\documentclass[a4paper,11pt]{article}

\usepackage[utf8]{inputenc}
\usepackage[UKenglish]{babel}
\usepackage[p,osf]{cochineal}
\usepackage[varqu,varl,var0]{inconsolata}
\usepackage[scale=.95,type1]{cabin}
\usepackage[cochineal,vvarbb]{newtxmath}
\usepackage[scr=boondoxo,cal=euler]{mathalfa}
\usepackage{comment}

\usepackage[top=3cm, bottom=4cm, left=3cm, right=3cm]{geometry}
\usepackage{setspace}
\makeatletter
\g@addto@macro\bfseries{\boldmath}
\makeatother
\usepackage{fancyhdr}
\usepackage{amsmath,amsthm}
\usepackage{amsfonts}
\usepackage{bm}
\usepackage{slashed}
\usepackage{mathtools}
\usepackage{thmtools}
\usepackage{amsbsy}
\numberwithin{equation}{section}
\allowdisplaybreaks

	\usepackage{pict2e}
	\makeatletter
		\newcommand{\loplus}{\mathbin{\mathpalette\dog@lsemi{+}}}
		\newcommand{\dog@lsemi}[2]{\dog@semi{#1}{#2}{270,90}}
		\newcommand{\dog@semi}[3]{%
  		\begingroup
  			\sbox\z@{$\m@th#1#2$}%
  			\setlength{\unitlength}{\dimexpr\ht\z@+\dp\z@\relax}%
  			\makebox[\wd\z@]{\raisebox{-\dp\z@}{%
    		\begin{picture}(1,1)
    			\linethickness{\variable@rule{#1}}
    			\roundcap
    			\put(0.5,0.5){\makebox(0,0){\raisebox{\dp\z@}{$\m@th#1#2$}}}
    			\put(0.5,0.5){\arc[#3]{0.5}}
    		\end{picture}%
  		}}%
  		\endgroup
		}
		\newcommand{\variable@rule}[1]{%
 		\fontdimen8  
  		\ifx#1\displaystyle\textfont3\else
    	\ifx#1\textstyle\textfont3\else
      	\ifx#1\scriptstyle\scriptfont3\else
        \scriptscriptfont3\relax
  		\fi\fi\fi
		}
	\makeatother

\usepackage{enumitem}
\usepackage{array}

\usepackage{tabularx}

\usepackage[textfont=it]{caption} 

\usepackage{hyperref}
\hypersetup{
    colorlinks,%
    citecolor=blue,%
    filecolor=blue,%
    linkcolor=blue,%
   	urlcolor=blue,
   	linktoc=page
}

\usepackage{cite}

\usepackage[toc,page]{appendix}

\usepackage{graphicx}
\usepackage{float}
\usepackage{subfig}
\usepackage{tikz}
\usepackage{tikz-layers}
\usetikzlibrary{calc}
\usetikzlibrary{decorations.pathreplacing}
\usetikzlibrary{decorations.pathmorphing}
\usetikzlibrary{decorations.markings}
\usetikzlibrary{arrows.meta}
\usetikzlibrary{positioning}
\usetikzlibrary{math}
\tikzset{every picture/.style={font issue=\footnotesize},
         font issue/.style={execute at begin picture={#1\selectfont}}
        }
\usepackage{xcolor}

\newcommand{\bfomega}{\boldsymbol{\upomega}}
\newcommand{\D}{\text{d}}

\newcommand{\bfupsilon}{{\boldsymbol{\upupsilon}}}
\newcommand{\bftau}{\boldsymbol{\uptau}}

\newcommand{\bfalpha}{{\boldsymbol{\upalpha}}}
\newcommand{\bfbeta}{{\boldsymbol{\upbeta}}}
\newcommand{\bfgamma}{{\boldsymbol{\upgamma}}}

\newcommand{\bfmu}
 {\boldsymbol{\upmu}}
\newcommand{\bk}[1]{\big[#1\big]}

\newcommand{\hframe}{{\text{\bf e}}}

\newcommand{\htheta}{{\boldsymbol{\uptheta}}{}}
\newcommand{\hupsilon}{{\boldsymbol{\upupsilon}}{}}
\newcommand{\bfg}{{\text{\bf g}}}

\newcommand{\vect}[1]{{\text{\bf #1}}}
\newcommand{\bfkappa}{{\boldsymbol{\upkappa}}}
\newcommand{\horbmnabla}{{\bar{\bm\nabla}}}

\newcommand{\gen}[1]{{\mathbb{#1}}}

\usepackage{accents}
\usepackage{adjustbox}
\renewcommand{\bar}[1]{\accentset{\rule[0.05ex]{0.4em}{0.42pt}}{#1}}
\renewcommand{\tilde}[1]{\accentset{\adjustbox{trim=0pt 0.2ex 0pt 0pt,clip}{\scalebox{0.75}[0.7]{$\sim$}}}{#1}}
 
\begin{document}

\setstretch{1.1}
\setcounter{tocdepth}{2}


\thispagestyle{empty}

\begin{flushright}
CPHT-RR021.082026
\end{flushright}
\vspace{0.5cm}

\begin{center}
\doublespacing
\begin{minipage}{0.8\textwidth}
\begin{center}
\begin{LARGE}
\textbf{\textsc{Carroll--Cotton Tensors and Gravitational Radiation at Null Infinity}}
\end{LARGE}
\end{center}
\end{minipage}
\end{center}

\vspace{7mm}
\begin{center} 
Adrien Fiorucci$^{a,b,}$\footnote{E-mail: \href{mailto:adrien.fiorucci@polytechnique.edu}{adrien.fiorucci@polytechnique.edu}}, 
Simon Pekar$^{c,d,}$\footnote{E-mail: \href{mailto:spekar@sissa.it}{spekar@sissa.it}}, 
P. Marios Petropoulos$^{a,}$\footnote{E-mail: \href{mailto:marios.petropoulos@polytechnique.edu}{marios.petropoulos@polytechnique.edu}},
and
Matthieu Vilatte$^{e,}$\footnote{E-mail: \href{mailto:matthieu.vilatte@umons.ac.be}{matthieu.vilatte@umons.ac.be}}
.

\normalsize
\bigskip\medskip
$^{a}$ \textit{Centre de Physique Th\'eorique, \'Ecole polytechnique, \\
Centre National de la Recherche Scientifique -- Unit\'e Mixte de Recherche 7644, \\
Institut Polytechnique de Paris, 91120 Palaiseau Cedex, France}
        
$^b$ \textit{Université Libre de Bruxelles and International Solvay Institutes \\
CP 231, Boulevard du Triomphe, 1050 Bruxelles, Belgium}

$^c$ \textit{International School for Advanced Studies (SISSA), \\
Via Bonomea 265, 34136 Trieste, Italy}\\
\bigskip
$^d$ \textit{Istituto Nazionale di Fisica Nucleare (INFN), \\
Sezione di Trieste, Via Valerio 2, 34127 Trieste, Italy} \\
\bigskip

$^e$ \textit{Service de Physique de l’Univers, Champs et Gravitation, Université de Mons \\
Place du Parc 20, 7000 Mons, Belgium}

\bigskip

\vspace{30pt}

\begin{abstract}\noindent
We provide a first-principles definition of Cotton tensors on conformal Carroll geometries by gauging the conformal Carroll algebra. On the mathematical side, this analysis yields a conformally covariant object designed to quantify deviations from conformal flatness in Carrollian geometries. On the physical side, it offers a powerful geometric tool to describe gravitational radiative degrees of freedom at the conformal boundary of four-dimensional asymptotically flat spacetimes. Unlike the Bondi news, whose fully-covariant definition requires amendment by supplementary data that we identify with the Geroch tensor, the Carroll--Cotton tensors offer the natural geometric objects to quantify deviations from stationarity due to the emission of gravitational waves.
\end{abstract}

\end{center}


\newpage

\setcounter{page}{2}

\begin{spacing}{0.85}

\tableofcontents

\end{spacing}

\vspace{15pt}
\noindent
\hrulefill

\setcounter{footnote}{0} 

\section{Introduction}

In the AdS/CFT correspondence, the holographic encoding of bulk gravitational degrees of freedom is naturally organised around two boundary tensors. The first is the holographic energy--momentum tensor $T^{\mu\nu}$, which captures the Coulombic data of the bulk gravitational field, such as mass and angular momentum \cite{Balasubramanian:1999re}. When the bulk Einstein equations hold, this symmetric tensor is covariantly conserved with respect to the boundary geometry and is interpreted as the expectation value of the energy--momentum tensor of the dual boundary theory. Its trace is fixed by local geometric data: it vanishes in odd boundary dimensions, while in even boundary dimensions it reproduces the conformal anomaly \cite{deHaro:2000vlm}.

The second tensor becomes relevant once one allows propagating gravitational degrees of freedom to interact with the conformal boundary. In asymptotically anti-de Sitter spacetimes, bulk gravitons can reach the boundary in finite affine time and leave an imprint on the boundary geometry. Describing this process requires relaxing the usual Dirichlet boundary conditions by allowing the boundary metric itself to fluctuate \cite{Compere:2008us}. These so-called \textit{leaky boundary conditions} \cite{Fiorucci:2020xto} promote the boundary metric to a dynamical mode in the dual description of Einstein gravity; in the standard holographic terminology, this is the non-normalisable, or Neumann, mode. More concretely, solving Einstein's equations with negative cosmological constant in radiative gauges such as Bondi or Newman--Unti reveals that specific components of the boundary metric act as sources for gravitational radiation reaching the conformal boundary of anti-de Sitter space \cite{Compere:2019bua,Campoleoni:2023fug}; see also \cite{Poole:2018koa,Ciambelli:2024kre,Arenas-Henriquez:2025rpt}. At the quantum level, local departures from conformal flatness on the boundary can be interpreted as insertions of gravitons, and boundary correlators are dual to bulk scattering amplitudes computed through AdS--Witten diagrams.

These observations point to the need for a boundary probe of local deviations from conformal flatness. In dimensions larger than three, such deviations are measured by the conformally invariant Weyl tensor, namely the tracefree part of the curvature tensor associated with the Levi-Civita connection of the boundary metric. For the case of primary physical interest, namely four-dimensional bulk spacetimes, the conformal boundary is three-dimensional and its Weyl tensor vanishes identically. The appropriate diagnostic tool is instead the \textit{Cotton tensor} \cite{Cotton1899}, built from third derivatives of the boundary metric:
\begin{equation}
C^\mu{}_\nu =\epsilon^{\mu\alpha\beta}\nabla_\alpha\big(R_{\beta\nu}-\tfrac14 R g_{\beta\nu}\big).
\end{equation}
The bracketed quantity is the Schouten tensor, with $R_{\mu\nu}$ the Ricci tensor of the boundary Levi-Civita connection and $R$ its trace. By construction, $C^{\mu\nu}$ is symmetric, traceless and covariantly conserved off-shell, the last property following from the boundary Bianchi identities.

When the bulk Einstein equations are imposed, $T^{\mu\nu}$ and $C^{\mu\nu}$ therefore share the same tensorial properties: both are symmetric, traceless tensors on the boundary and both obey conservation equations. This parallel is not accidental: the holographic energy--momentum tensor is the boundary value of the electric part of the bulk Weyl tensor, whereas the Cotton tensor captures its magnetic part. Gravitational electric--magnetic duality exchanges these two quantities, and self-dual solutions of Einstein's equations exhibit a direct proportionality between them \cite{deHaro:2007fg,Mansi:2008bs,deHaro:2008gp,Bakas:2008gz,Bakas:2009pbm,Miskovic:2009bm,Mukhopadhyay:2013gja,Petropoulos:2014yaa,Bakas:2014kfa}.\footnote{Strictly speaking, line elements solving the self-dual Einstein equations are real only in Euclidean or split signature.} Beyond its role as a probe of boundary interactions with bulk gravitational radiation, the Cotton tensor is also sensitive to topological data of the boundary, in particular to `nut' charges \cite{RamaswamySen1981,10.1063/1.525274} that can be obtained by smearing $C^{\mu\nu}$ against boundary conformal isometries.

Extending this geometric picture to asymptotically flat spacetimes is a crucial step towards a flat-space holographic correspondence. It is however technically subtle, because the conformal boundary, \textit{null infinity}, is then lightlike instead of timelike. Physics at null infinity is governed by the laws of Carroll physics \cite{Leblond}, which may be understood formally as the vanishing-speed-of-light limit of special relativity, naturally present on Lorentzian backgrounds. From the bulk viewpoint, Einstein's equations imply that the effective speed of light induced at the conformal boundary is proportional to the cosmological constant; it therefore vanishes in the flat limit. Consequently, the boundary Lorentzian metric degenerates along the null generators, and the appropriate intrinsic framework is that of null \cite{jankiewicz1954espaces,Vogel1965,Datcourt1967}, or Carrollian, manifolds \cite{Henneaux:1979vn,Ciambelli:2019lap,Henneaux:2021yzg,Chandrasekaran:2021hxc}.

This change of geometric setting has an immediate impact on symmetries. Carrollian conformal isometries arise as the vanishing-speed-of-light limit of ordinary Lorentzian conformal isometries; in particular, the three-dimensional conformal group gives rise in this limit to the infinite-dimensional BMS group \cite{Duval:2014uva}. Thus, the usual pseudo-Riemannian framework underlying AdS/CFT must be replaced by tools adapted to non-Lorentzian geometry \cite{Bekaert:2015xua,Bagchi:2019xfx,Henneaux:2021yzg,Herfray:2021qmp,Figueroa-OFarrill:2021sxz,Bergshoeff:2022eog,Freidel:2022vjq,Nguyen:2023vfz,Salzer:2023jqv,Vilatte:2024jjr,Nguyen:2025zhg,Ruzziconi:2026bix,Bekaert:2026cvx}. Despite these differences, the notion of a holographic energy--momentum tensor at null infinity has been investigated from several complementary perspectives since the early development of flat-space holography. It can be obtained, for instance, by taking asymptotically flat limits of AdS/CFT configurations \cite{Ciambelli:2018xat,Compere:2019bua,Fiorucci:2020xto,Campoleoni:2023fug,Alday:2024yyj,Lipstein:2025jfj}, or indirectly through twistor theory \cite{Adamo:2013tja,Adamo:2020yzi,Mason:2023mti,Kmec:2024nmu}, celestial amplitudes \cite{Kapec:2016jld,Donnay:2020lur,Pasterski:2022djr,Banerjee:2022wht,Ruzziconi:2024kzo}, and the fluid--gravity correspondence \cite{Ciambelli:2018wre,Ciambelli:2018ojf,Campoleoni:2018ltl}. In this sense, the holographic dictionary relating gravitational momenta to boundary energy--momentum data seems to survive even when the cosmological constant is set to zero: in both asymptotically anti-de Sitter and flat settings, it chiefly involves the Coulombic, third-subleading, contributions to the metric.

The encoding of radiative data is more delicate. Contrary to anti-de Sitter space, radiative degrees of freedom at null infinity do not enter the boundary background metric itself, but rather the boundary \textit{connection} \cite{Ashtekar:1981hw}. The latter has a richer structure than a Levi-Civita connection on a pseudo-Riemannian manifold, since compatibility with the Carrollian boundary structure and vanishing torsion, although both enforced by the bulk Einstein equations, do not determine it uniquely. The remaining connection data are precisely those needed to encode the shear of outgoing null congruences along which gravitational waves propagate towards null infinity. Their presence naturally leads to hypermomenta conjugate to independent fluctuations of the boundary connection. The resulting flux-balance equations govern the evolution of boundary energy and momenta in terms of the strength of gravitational radiation, encoded in the Bondi news tensor. Their holographic interpretation has recently been clarified in \cite{Fiorucci:2025twa}; see also \cite{Hartong:2025jpp,Hartong:2026rbr}. From a fluid--gravity viewpoint, these equations are Carrollian fluid equations evolving on a non-trivial background and sourced by the affine connection induced from the bulk Levi-Civita connection. Remarkably, this picture is consistent with the asymptotically flat limit of leaky AdS boundary conditions. The non-normalisable modes that fluctuate at the boundary of anti-de Sitter space are mapped, in the flat limit, to the Bondi news escaping to future null infinity \cite{Compere:2019bua,Fiorucci:2020xto,Fiorucci:2021pha,Campoleoni:2023fug}. This provides a direct bridge with the standard asymptotic analysis initiated by Bondi and Sachs more than half a century ago \cite{Bondi:1962px,Sachs:1962wk}; see also, for example, \cite{Wald:1999wa,Barnich:2010eb,Barnich:2011mi,Freidel:2021fxf,Geiller:2022vto,Rignon-Bret:2024wlu}.

Similarly to anti-de Sitter space, the manifestation of the gravitational field at the conformal boundary is therefore controlled by the asymptotic behaviour of the bulk Weyl tensor,\footnote{Since the work of Newman and Penrose \cite{Newman:1961qr,Newman:1962cia}, gravitational radiation in four-dimensional asymptotically flat spacetimes has been described covariantly by projecting the Weyl tensor on a null tetrad. The relevant radiative information is encoded in the Weyl scalars $\Psi^0_4$, $\Psi^0_3$ and $\mathrm{Im}(\Psi^0_2)$. Conditions for the absence of radiation are expressed by setting some of these quantities to zero and are closely related to the Petrov classification.} which can again be decomposed into an electric part encoding Coulombic degrees of freedom ultimately captured by the Carrollian energy--momentum tensor, and a magnetic part, which is both sensitive to radiative interactions and to topological properties of null infinity. In four-dimensional asymptotically flat spacetimes, it is then natural to expect a Carrollian counterpart of the Cotton tensor, dubbed \textit{Carroll--Cotton tensor}, to play a central role, and its derivation therefore constitutes an important step towards establishing the flat-space holographic correspondence. A first attempt to define such a tensor was proposed in \cite{Campoleoni:2023fug} and it was later rederived from a gravitational Chern--Simons action in \cite{Miskovic:2023zfz}. It relies on small-speed-of-light expansion of both the Lorentzian Cotton tensor and the aforementioned action, which is producing a finite tower of Carrollian replicas, together with their respective evolution equations. The purpose of this article is to give a different, intrinsically Carrollian derivation based on the gauging of boundary conformal isometries.\footnote{One hallmark of AdS/CFT is the dual role of $SO(d+1,2)$: it is both the conformal group of a $(d+1)$-dimensional conformally flat manifold and the isometry group of $\mathrm{AdS}_{d+2}$. In the flat case, $ISO(d+1,1)$ similarly acts both as the Poincar\'e group in $(d+2)$-dimensional bulk and as the group of conformal isometries of its $(d+1)$-dimensional Carrollian boundary.} 

As expected, the resulting Carroll--Cotton tensor encodes not only the conformal properties of the geometry at null infinity, but also the manifestation of gravitational waves at the boundary. This is made possible because the Bondi shear of four-dimensional asymptotically flat spacetimes is contained in the boundary connection. The distinctive feature of our approach is that it generalises the usual metric treatment to a \textit{Weyl--metric--affine} framework, in which the boundary frame and connection are kept independent and all quantities are manifestly Weyl covariant \cite{Campoleoni:2023fug,Fiorucci:2025twa}. Our construction through gauging procedures 
has three advantages. Firstly, it is intrinsic to null infinity; secondly, it does not rely on a limiting procedure, and finally, it yields a single geometrically defined set of tensors rather than several Carrollian replicas appearing at different orders in $c^2$. Our construction reproduces and extends the geometric treatment of radiation at null infinity initiated in \cite{Geroch:1977big} and further developed in \cite{Ashtekar:2014zsa,Compere:2018ylh,Freidel:2021fxf,Rignon-Bret:2024gcx,Fiorucci:2025twa}. In particular, we show that the radiative Newman--Penrose Weyl scalars are recovered exactly by expressing the Carroll--Cotton tensor in a Bondi frame. This consequently shows that our formalism provides the most general expression of these objects in arbitrary boundary Carrollian geometries. Moreover, no-radiation conditions are recast as conformal-flatness conditions at null infinity, which supplies a new entry in the flat-space holographic dictionary. Finally, our construction opens a geometric avenue to the study of electric--magnetic duality and self-dual solution spaces in four-dimensional asymptotically flat spacetimes from a purely boundary perspective, in the spirit of \cite{Mittal:2022ywl}, together with an intrinsic Carrollian-holographic characterisation of algebraically special asymptotically flat spacetimes \cite{Ciambelli:2018wre}.

The paper is organised as follows. We first review in Section \ref{sec:Lorentz} the relevant elements of conformal geometry on Lorentzian manifolds and introduce the gauging procedure in general dimensions. We then construct the three-dimensional Lorentzian Cotton tensor and discuss the gauge-fixing choices that have appeared in the literature. The remainder of the paper is devoted to the three-dimensional Carrollian case. Applying the same gauging strategy leads to a general expression of the Carroll--Cotton tensor with properties parallel to its Lorentzian counterpart, together with evolution equations following from Bianchi identities and is the subject of Section \ref{sec:Carroll}. We then relate its components to known expressions, analyse convenient boundary gauge fixings, and conclude by discussing the implications of our results for flat-space holography. In particular, we show how to use the gauging approach to enlighten the construction of a covariant version of the news tensor at null infinity in Section \ref{sec:News} and discuss ways of deriving the Carroll--Cotton tensors from variational principles involving Carrollian avatars of the Chern--Simons action in Section \ref{sec:ChernCarroll}. This article is supplemented by three appendices: Appendix \ref{sec:2d} discusses the gauging of the two-dimensional conformal algebra in both the Lorentzian and Carrollian cases; Appendix \ref{sec:invariantbilinear} reviews the computation of the invariant bilinear symmetric form on the conformal algebra in three dimensions; Appendix \ref{sec:ids} finally provides useful identities involving Riemann curvatures in the presence of non-trivial Weyl curvature.

\subsection*{Notations and conventions}

Spacetime manifolds are represented by calligraphic letters $\mathscr M$ and have $d+1$ dimensions, with $d$ spacelike dimensions and one timelike dimension. Tensor fields tangent to $\mathscr M$, including vector fields and differential forms, are denoted by boldface symbols, like $\vect V$, $\bfalpha$, \textit{etc.} In a basis $\{\htheta^A\}$ of $T^*\mathscr M$, covectors are expanded as $\vect W = W_A\htheta^A$. Tensor contractions and interior products are denoted by parentheses, \textit{i.e.}, $\iota_\vect{V}\vect W \equiv \vect W(\vect V)$. The dual basis is then defined as $\{\hframe_A\}\subset T\mathscr M$ that obeys $\htheta^A(\hframe_B) = \delta^A{}_B$. Capital Latin indices $A$ run from $0$ to $d$. Whenever a time-space split is explicit, $A = (0,a)$ where $0$ represents the privileged timelike direction and lowercase Latin indices $a=1,\dots,d$ are the transverse, hence spacelike, directions. Lie algebras and their elements are respectively denoted by gothic and blackboard-bold characters, \textit{i.e.}, $\mathfrak{g}$ and $\gen X$.

Koszul connections $\nabla : T\mathscr M\times T\mathscr M\to T\mathscr M$ are defined by their coefficients in a given basis as $\nabla_{\hframe_A}\hframe_B = \gamma^C{}_{AB}\hframe_C$. These coefficients are compiled into one-form fields $\bfgamma^A{}_B = \gamma^A{}_{CB}\htheta^C$. We reserve the symbol `$\nabla$' for spin connections. Namely, $\mathring\nabla$ is the Levi-Civita spin connection, $\bar{\nabla}$ is the projection of $\nabla$ transverse to a given timelike congruence, and $\hat{\nabla}$ is the $c\to 0$ limit of the latter. When discussing Weyl covariance, a spin connection $\nabla$ with coefficients $\gamma^A{}_{BC}$ can conveniently be traded for a non-metric but \textit{Weyl-metric} compatible connection, $\text D$, with coefficients $\omega^A{}_{BC} = \gamma^A{}_{BC} - \alpha_B\delta^A{}_C$ where $\alpha_B$ denotes the components of a Weyl-connection one-form. Weyl-covariant derivatives, equivalently constructed from $\nabla$ or the related $\text D$, are then denoted by calligraphic letters, $\mathscr D$. 
We have the relation: $\nabla_A V^B = (\nabla_{\hframe_A}\vect V)(\htheta^B)$, for a given vector field $\vect V = V^A\hframe_A$ for instance. Finally, our conventions for the curvatures are those of, \textit{e.g.}, \cite{Misner:1973prb}. In particular, AdS spacetimes have a negative Ricci curvature.

\section{Cotton tensor from gauging methods: the Lorentzian case} \label{sec:Lorentz}

In the following, we first review how the gauging procedure works for the usual Lorentzian conformal case \cite{Horne:1988jf,Garcia:2003bw} (see, \textit{e.g.}, \cite{Freedman:2012zz,Butter:2013goa} for a more physics-oriented review). 
In the next section, we shall then proceed in gauging the isotropic conformal Carroll algebra imposing similar equations of motion, which are crucially not the standard ones discussed when gauging the Poincar\'e algebra, even though the two algebras are isomorphic.

\subsection{Gauging the conformal algebra} \label{sec:gauging Lorentzian}

\subsubsection*{The conformal algebra}

The conformal algebra in $d+1$ dimensions, $\mathfrak{conf}(d,1) \simeq \mathfrak{so}(d+1,2)$, for any $d\in\mathbb N^+$, extends the Poincar\'e algebra $\mathfrak{iso}(d,1)$ into the Lie algebra of conformal transformations of $(d+1)$-dimensional Minkowski space $\mathscr M_0$. In Cartesian coordinates $(x^A)_{A=0,\ldots,d}$, the Minkowski metric takes the form
\begin{equation}
	\D s^2 = \eta_{AB}\D x^A\D x^B,\qquad \eta_{AB} = \text{diag}(-c^2,1,\dots,1), \label{eq:flat mink}
\end{equation}
where $c$ is a strictly positive real constant, explicitly introduced for latter convenience, and meant to represent the effective speed of light. Conformal isometries of \eqref{eq:flat mink} are vector fields $\vect V$ that obey $\mathscr L_{\vect V}\eta_{AB} = \tfrac{2}{d+1} (\bm\partial\cdot \vect V)\eta_{AB}$, which admits $\tfrac12 (d+2)(d+3)$ independent solutions, that we denote by
\begin{equation} \label{eq:diff realisation}
	\vect V_{\gen P_A} = \partial_A,\qquad \vect V_{\gen J_{AB}} = x_B\partial_A - x_A\partial_B,\qquad \vect V_{\gen D} = x^A\partial_A,\qquad \vect V_{\gen K_A} = 2 x_A x^B\partial_B - x^B x_B \partial_A,
\end{equation}
where the mapping $\mathfrak{conf}(d,1) \to \Gamma(T\mathscr M_0) : \gen X\mapsto \vect V_{\gen X}$ is a Lie algebra isomorphism. One says that the conformal algebra is realised by the vector fields \eqref{eq:diff realisation} seen as differential operators on $\mathscr C^\infty(\mathscr M_0)$. Translation and Lorentz generators, $\gen P_A$ and $\gen J_{AB}=\gen J_{[AB]}$, span the Poincaré algebra ($\bm\partial\cdot \vect V=0$) while $\gen D$ and $\gen K_A$ generate pure conformal isometries ($\bm\partial\cdot \vect V\neq 0$), namely, dilations and special conformal transformations. They form a complete basis of $\mathfrak{conf}(d,1)$ we shall work with, and satisfy the following commutation relations
\begin{equation}
	\begin{array}{lclcl}
		\bk{\gen D,\gen P_A} = \gen P_A, &\quad& \bk{\gen D,\gen K_A} = -\gen K_A, &\quad& \bk{\gen K_A,\gen P_B} = 2\eta_{AB} \gen D - 2 \gen J_{AB}, \\
		\bk{\gen J_{AB},\gen P_C} = 2\eta_{C[B}\gen P_{A]}, & & \bk{\gen J_{AB},\gen K_C} = 2\eta_{C[B}\gen K_{A]}, & & \\
		\multicolumn{3}{l}{\bk{\gen J_{AB},\gen J_{CD}} = 2\eta_{A[C}\gen J_{D]B} - 2\eta_{B[C}\gen J_{D]A},} & &
	\end{array} \label{eq:conformal alg}
\end{equation}
with all other commutators vanishing. This can be established by means of the aforementioned Lie-algebra isomorphism, which satisfies $\bk{\vect V_{\gen X},\vect V_{\gen Y}}_{\textrm{Lie}} \coloneqq \mathscr L_{\vect V_{\gen X}}\vect V_{\gen Y} = -\vect V_{[\gen X,\gen Y]}$, where the minus sign is up to conventions. For later convenience, we denote as $\{\gen X_I\}$ our chosen basis of $\mathfrak{conf}(d,1)$.

\subsubsection*{Gauge connections}

The \textit{gauging} of this algebra corresponds to represent it \textit{locally} on the tangent bundle of a generic curved Lorentzian manifold $(\mathscr M,\vect g)$. This first amounts to building a gauge connection $\bm\Gamma$, which is expanded on our chosen basis as\footnote{We have defined the gauge connection for local Lorentz transformations, $\bfgamma^{AB}$, with an extra factor of $\tfrac12$ to avoid double-counting arising from skew-symmetry. For consistency, this factor of $\tfrac12$ has also been carried over to the connection for local special conformal transformations, to avoid unusual normalisations, \textit{e.g.}, in Eq. \eqref{eq:kappa = schouten}. Furthermore, the connection $\bfalpha$ related to dilations has been defined with an additional minus sign to agree with our way to count Weyl weights, see, \textit{e.g.}, Eq. \eqref{eq:weyl metric lorentzian}, which establishes that the metric transforms as a Weyl primary with weight \textit{minus} two.}
\begin{equation}
	\bm\Gamma = \gen X_I \bm\Gamma^I = \gen P_A\htheta^A + \tfrac12 \gen J_{AB}\bfgamma^{AB} - \gen D\bfalpha + \tfrac12 \gen K_A\bfkappa^A \quad\in\quad \Omega^1\big(\mathscr M,\mathfrak{conf}(d,1)\big), \label{eq:Gamma}
\end{equation}
together with gauge transformations of generators $\Sigma\in \mathscr C^\infty\big(\mathscr M,\mathfrak{conf}(d,1)\big)$ under which the one-form $\bm\Gamma$ is meant to transform as a connection:
\begin{equation}
	\delta_\Sigma \bm\Gamma \coloneqq \D\Sigma + \bk{\bm\Gamma,\Sigma}. \label{eq:delta Gamma}
\end{equation}
The bracket in the right-hand side of Eq. \eqref{eq:delta Gamma} is the Lie bracket of the conformal algebra, given by Eq. \eqref{eq:conformal alg}. The gauge parameters also expand in the chosen basis of the conformal algebra as
\begin{equation}
	\Sigma = \Sigma^I\gen X_I \coloneqq -\tfrac{1}{2}L^{AB}\gen J_{AB} + B \gen D + \tfrac12 Z^A \gen K_A \label{eq:Sigma in basis}
\end{equation}
where the functions $L^{AB} = L^{[AB]}$, $B$ and $Z^A$ represent local point-isotropic conformal transformations, respectively local Lorentz, Weyl and special conformal transformations. Gauge transformations related to the gauging of the translation generators $\gen P_A$ are intentionally left out of the local gauge transformations generated by $\Sigma$. Indeed, they are infinitesimal diffeomorphisms acting over $\mathscr M$ and send points onto other neighbouring points. All the elements $\bm\Gamma^I$ entering in the gauge connection $\bm\Gamma$ are tensors and thus transform with Lie derivatives under infinitesimal diffeomorphisms.

Expanding also the gauge transformation \eqref{eq:delta Gamma} in the chosen basis, we extract the gauge transformation of each field as
\begin{subequations}\label{eq:delta gauge relat}
	\begin{align}
		\delta_\Sigma \htheta^A &= L^A{}_B\htheta^B - B\htheta^A, \label{eq:delta sigma thetaA} \\
		\delta_\Sigma \bfgamma^{AB} &= -\D L^{AB} + 2 L^{C[A}\bfgamma^{B]}{}_C + 2 Z^{[A}\htheta^{B]}, \label{eq:delta sigma gammaAB} \\
		\delta_\Sigma\bfalpha &= Z_A\htheta^A -\D B, \label{eq:Weyl connection variation} \\
		\delta_\Sigma\bfkappa^A &= L^A{}_B \bfkappa^B + B\bfkappa^A + \big(\D+\bfalpha\big)Z^A + Z^B\bfgamma^A{}_B. \label{eq:delta sigma kappaA}
	\end{align}
\end{subequations}
The $(d+1)$ one-form fields $\htheta^A$ provide a local basis of $\Gamma(T^*\mathscr M)$. The dual basis of $\Gamma(T\mathscr M)$, formed upon vector fields $\hframe_A$ that satisfy $\htheta^A(\hframe_B) = \delta^A{}_B$, is chosen to be orthonormal with respect to the metric, that is $\bfg(\hframe_A,\hframe_B) = \eta_{AB}$. This is equivalent to setting $\bfg\equiv \eta_{AB}\htheta^A\otimes\htheta^B$, where $\eta_{AB}$ is still given by the constant matrix as in Eq. \eqref{eq:flat mink}. Borrowing nomenclature from first-order formulation of general relativity, we shall refer to $\{\hframe_A\}$ and $\{\htheta^A\}$ respectively as the local Lorentz frame and coframe. Apart from flat space, such a frame is non-holonomic and we define its non-holonomy coefficients as the functions $c^C{}_{AB} = c^C{}_{[AB]}$ entering into
\begin{equation}
	\big[\hframe_A,\hframe_B\big]_{\textrm{Lie}} = c^C{}_{AB}\hframe_C\qquad\Leftrightarrow\qquad \D\htheta^A + \tfrac{1}{2}c^A{}_{BC}\htheta^B\wedge\htheta^C = \bm 0, \label{eq:non holonomy relativiste}
\end{equation}
which transform under local symmetries as
\begin{equation}
	\delta_\Sigma c^A{}_{BC} = L^A{}_D c^D{}_{BC} + 2 L^D{}_{[B} c^A{}_{C]D} - 2 \hframe_{[B}\big(L^A{}_{C]}\big) + B c^A{}_{BC} + 2\hframe_{[B}(B)\delta^A{}_{C]}. \label{eq:delta sigma cABC}
\end{equation}
Note that the integrability conditions $\D^2\htheta^A = \bm 0$ yield the non-trivial differential constraints:
\begin{equation}
	\hframe_{[B} \big(c^A{}_{CD]}\big) - c^A{}_{E[B}c^E{}_{CD]} = 0.
\end{equation}
From the transformation law \eqref{eq:delta sigma thetaA}, the coframe transforms as the component of a local Lorentz vector and belongs to the Weyl bundle of weight minus one. As a result, the metric transforms as
\begin{equation}
	\delta_\Sigma \bfg = -2B \bfg. \label{eq:weyl metric lorentzian}
\end{equation}
The gauging of the dilation operator endows the Lorentzian manifold $(\mathscr M,\bfg)$ with a Weyl structure, \textit{i.e.}, equivalence classes of metric fields with Weyl-weight minus-two: $\bfg'\in [\bfg]$ if and only if there exists a nowhere-vanishing function $\mathscr B$ such that $\bfg' = \mathscr B^{-2}\bfg$. Setting $\mathscr B = 1 + \epsilon B$ for some $\epsilon$ arbitrarily close to zero and expanding at first order in $\epsilon$ yields Eq. \eqref{eq:weyl metric lorentzian}. From now on in this Section, capital Latin indices will be lowered and raised by $\eta_{AB}$ and its inverse $\eta^{AB}$.

The gauge field $\bfgamma^{AB}$, which obeys $\bfgamma^{(AB)}=\bm 0$ by construction, transforms as a connection under the action of the local Lorentz algebra. It constitutes the spin connection whose action as a differential operator on fields expanded in the local coframe has been denoted by $\bm\nabla$. We have explicitly $\bm\nabla f = \D f$ on scalar fields, 
\begin{equation}
	\bm\nabla V^A = \D V^A + \bfgamma^A{}_B V^B,\qquad \bm\nabla W_A = \D W_A - \bfgamma^B{}_A W_B
\end{equation}
on vector fields $\vect V = V^A\hframe_A$ and one-form fields $\vect W = W_A\htheta^A$, and similarly on higher-rank tensors by means of the Leibniz rule.  Note that, contrary to the usual gauging of the Poincaré algebra, the spin connection also transforms inhomogeneously under local special conformal transformations. Furthermore, $\bfalpha = \alpha_A\htheta^A$ transforms inhomogeneously under Weyl transformations only and will be called Weyl connection for this reason. It is moreover pure-gauge for the local special conformal transformations that change it algebraically.\footnote{This instrumental feature is not at odds with all the non-explicitly Weyl-covariant analyses of boundary geometries in asymptotically AdS spacetime, see, \textit{e.g.}, \cite{FG1}. Indeed, the introduction of $\bfalpha$ is natural from the viewpoint of conformal compactification, which yields a Weyl geometry at the boundary, but is non-necessary. However, having a Weyl connection at our disposal allows one to manipulate manifestly Weyl-covariant expressions \cite{Bhattacharyya:2007vjd,Bhattacharyya:2008mz,Ciambelli:2018wre,Ciambelli:2019bzz}.} Finally, the last gauge field $\bfkappa^A = \kappa^A{}_B\htheta^B$ is a connection for local special conformal transformations and otherwise transforms as a local Lorentz vector field with Weyl-weight one.

\subsubsection*{Gauge curvatures}

The curvature associated with the connection \eqref{eq:Gamma} can be computed as
\begin{equation}
	\vect R[\bm\Gamma] = \D\bm\Gamma + \bm\Gamma\wedge\bm\Gamma \quad\in\quad \Omega^2\big(\mathscr M,\mathfrak{conf}(d,1)\big), \label{eq:def curvature conf}
\end{equation}
where the second term is a shorthand notation for the extension of the Lie bracket to algebra-valued one-forms like $\bm\Gamma = \bm\Gamma^I \gen X_I$, \textit{i.e.},
\begin{equation}
	\bm\Gamma\wedge\bm\Gamma \equiv \tfrac12 \bm\Gamma^I\wedge\bm\Gamma^J \bk{\gen X_I,\gen X_J}. \label{eq:wedge gamma gamma}
\end{equation}
In the chosen basis of the conformal algebra, the curvature $\vect R[\bm\Gamma]$ can be expanded as
\begin{equation}
	\vect R[\bm\Gamma] = \vect R[\bm\Gamma]^I\gen X_I \coloneqq \vect R[\gen P]^A \gen P_A + \tfrac12 \vect R[\gen J]^{AB}\gen J_{AB} - \vect R[\gen D]\gen D + \tfrac12 \vect R[\gen K]^A\gen K_A,
\end{equation}
and given in terms of the following two-forms:
\begin{subequations}\label{eq:gauge curvatures}
\begin{align}
	\vect R[\gen P]^A &= \big(\bm\nabla-\bfalpha\big)\wedge\htheta^A ,\\
	\vect R[\gen J]^{AB} &= \D\bfgamma^{AB} + \bfgamma^A{}_C\wedge\bfgamma^{CB} - 2 \bfkappa^{[A}\wedge\htheta^{B]}, \label{eq:RJdef} \\
	\vect R[\gen D] &= \D\bfalpha - \bfkappa_A\wedge\htheta^A, \\
	\vect R[\gen K]^A &= \big(\bm\nabla+\bfalpha\big)\wedge\bfkappa^A . \label{eq:RK}
\end{align}
\end{subequations}
The first one, $\vect R[\gen P]^A$, provides a Weyl-covariant evaluation of the torsion of the spin connection $\nabla$. The curvatures $\vect R[\gen J]^{AB}$ and $\vect R[\gen D]$ encode the Riemann curvature of the spin connection and the Weyl curvature, up to terms involving the new gauge field $\bfkappa^A$, rendering them covariant under special conformal transformations, see Eqs. \eqref{eq:deltaSigmaRJAB} and \eqref{eq:deltaSigmaRD}. We shall refer to them as \textit{composite Riemann and Weyl curvatures}, respectively. The last piece of curvature, $\vect R[\gen K]^A$, related to special conformal transformations, reduces to the Weyl-covariant derivative of $\bfkappa^A$ and is expanded as 
\begin{equation}
	\vect R[\gen K]^A \coloneqq \tfrac12 C^A{}_{BC}\htheta^B\wedge\htheta^C
\end{equation}
in the Cartan coframe. As we shall see later, under several assumptions that we disclose in the following, the functions $C^A{}_{BC} = C^A{}_{[BC]}$ provide the components of the Weyl-covariant Cotton tensor derived from the geometry. Therefore, besides endowing the initial manifold with a Weyl structure, the whole point of gauging the conformal algebra is actually to derive such an instance of the Cotton tensor from algebraic methods. Acting with transformations \eqref{eq:delta sigma thetaA}--\eqref{eq:delta sigma kappaA} on the above curvatures yields their respective transformation laws:
\begin{subequations}\label{eq:transfo curvatures}
	\begin{align}
		\delta_\Sigma \vect R[\gen P]^A &= L^A{}_B\vect R[\gen P]^B - B \vect R[\gen P]^A, \\
		\delta_\Sigma \vect R[\gen J]^{AB} &= 2 L^{[A}{}_C\vect R[\gen J]^{C|B]} + 2 Z^{[A}\vect R[\gen P]^{B]}, \label{eq:deltaSigmaRJAB} \\
		\delta_\Sigma \vect R[\gen D] &= Z_A\vect R[\gen P]^A, \label{eq:deltaSigmaRD} \\
		\delta_\Sigma \vect R[\gen K]^A &= L^A{}_B\vect R[\gen K]^B + B \vect R[\gen K]^A + \vect R[\gen J]^{AB}Z_B + Z^A\vect R[\gen D].
	\end{align}
\end{subequations}
One can notice that the curvature $\vect R[\gen J]^{AB}$ of the spin connection transforms tensorially under local Lorentz action if and only if there is no curvature in the direction of the translation generators. The same remark applies to the Weyl curvature, which is in that case fully invariant. Demanding $\vect R[\gen P]^A = \bm 0$ is indeed equivalent to an absence-of-torsion requirement. In $d \geq 2$, curvatures obey differential `Bianchi identities', coming from integrability conditions involving the curvatures. Indeed, computing the exterior derivative of Eq. \eqref{eq:def curvature conf} yields $\D\vect R[\bm\Gamma] = [\vect R[\bm\Gamma],\bm\Gamma]$, which is expanded as
\begin{subequations} \label{eq: Bianchi Lorentz}
\begin{align}
	\big(\bm\nabla-\bfalpha\big)\wedge\vect R[\gen P]^A &= \vect R[\gen J]^A{}_B \wedge \htheta^B - \vect R[\gen D] \wedge \htheta^A, \label{eq:bianchi for RP} \\
	\bm\nabla \wedge \vect R[\gen J]^{AB} &= - 2 \vect R[\gen P]^{[A} \wedge \bfkappa^{B]} - 2 \vect R[\gen K]^{[A} \wedge \htheta^{B]}, \label{eq:bianchi for RJ} \\
	\bm\nabla \wedge \vect R[\gen D] &= - \vect R[\gen K]_A \wedge \htheta^A + \vect R[\gen P]_A \wedge \bfkappa^A, \label{eq:bianchi for RD} \\
	\big(\bm\nabla+\bfalpha\big)\wedge \vect R[\gen K]^A &= \vect R[\gen J]^A{}_B \wedge \bfkappa^B + \vect R[\gen D] \wedge \bfkappa^A . \label{eq:bianchi for RK}
\end{align}
\end{subequations}
in the chosen basis.

Up to this point, all the connection coefficients are utterly free. Given the transformation laws \eqref{eq:transfo curvatures} and the identities \eqref{eq: Bianchi Lorentz}, the constraints that we can impose among them without restricting neither the underlying geometry nor the symmetries must fall into one of the following categories:
\begin{enumerate}
	\item\label{item:condition1} \textit{Torsion-free:} $\vect R[\gen P]^A=\bm 0$, which completely fixes $\bfgamma^{AB}$ in terms of $\htheta^A$ and $\bfalpha$;
	\item\label{item:condition2} \textit{Torsion-free and Weyl-curvature-free:} $\vect R[\gen P]^A=\bm 0$ and $\vect R[\gen D]=\bm 0$, which further fixes $\kappa_{[AB]}$ in terms of $\bfalpha$ and its first derivatives.
	\item\label{item:condition3} \textit{Torsion-free and Ricci flat:} $\vect R[\gen P]^A=\bm 0$ and $\vect{Ric}[\gen J]^A=\bm 0$, where the Ricci tensor of the composite Riemann curvature is defined as $\vect{Ric}[\gen J]^A \coloneqq \vect R[\gen J]^{AC}(\cdot,\hframe_C)$. For any $d>1$, these conditions completely fix $\bfkappa^A$ in terms of $\bfgamma^A$ and $\bfalpha$ and their first derivatives, and $\vect R[\gen D] = \bm 0$ is automatically satisfied.\footnote{When $d\leq 2$, imposing Ricci flatness is equivalent to imposing Riemann flatness, \textit{i.e.}, $\vect R[\gen J]^{AB} = \bm 0$, since there is no Weyl tensor in low dimensions. For $d=1$ specifically, Riemann flatness only fixes the trace of $\bfkappa^A$, since the composite Riemann tensor has only one independent component, $R[\gen J]^{01}{}_{01}$.  Therefore, $\vect R[\gen D] = \bm 0$ must be imposed manually in addition to the other conditions, leading to a fourth category, specific to two dimensions, which we refer to as the \textit{torsion-free and Riemann--Weyl-flat} case. Note that the traceless part of $\kappa_{(AB)}$ is left unconstrained by the aforementioned conditions, see Appendix \ref{sec:2d} for more details.}
\end{enumerate}
In the language of gauging methods \cite{Horne:1988jf}, one usually refers to such systems of constraints as \textit{equations of motion}. In the above enumeration, they are displayed from the weakest to the strongest. Importantly, one cannot impose more general equations of motion like $\vect R[\gen J]^{AB} = \bm 0$ or $\vect R[\gen K]^A = \bm 0$, as these would place constraints on the geometry. In particular, imposing that all curvatures vanish is equivalent to conformal flatness.\footnote{Indeed, assuming that we are in the \hyperref[item:condition3]{third} case, \textit{i.e.}, torsion-free and Ricci flat, the only non-vanishing curvatures are what remains of $\vect R[\gen J]^{AB}$, which is nothing but the Weyl tensor related to the spin connection, and $\vect R[\gen K]^A$. In dimensions strictly greater than three, the Weyl tensor is non-trivial. Assuming that it vanishes implies conformal flatness, since it vanishes in flat space and is Weyl-covariant. Therefore, the Cotton tensor $\vect R[\gen K]^A$ automatically vanishes by virtue of the identity \eqref{eq:identity weyl cotton}. In three dimensions, the Weyl vanishes identically. Therefore, assuming that $\vect R[\gen K]^A$ vanishes as it does in flat space also implies conformal flatness because the Cotton tensor is Weyl-covariant.}

\subsection{The Weyl--Levi-Civita connection}
\label{sec:WLC relat}

On Lorentzian geometries, there is a fundamental theorem that implies the existence of a unique torsion-free connection that is compatible with the metric, the Levi-Civita connection $\mathring\nabla$. The related coefficients coincide with the Christoffel symbols of the second kind,
\begin{equation}
	\mathring\gamma{}^A{}_{BC} = \mathring\gamma{}^A{}_{BC}(\bfg,\bm\partial\bfg) = \tfrac12\big(c^A{}_{BC} + c_B{}^A{}_C + c_C{}^A{}_B\big), \label{eq:levi civita}
\end{equation}
expressed in the orthonormal local frame $\hframe_A$. Whenever a Weyl structure is present, no particular metric $\bfg$ in a given Weyl class $[\bfg]$ should be preferred and the definition of the Levi-Civita connection, for this particular representative $\bfg$ breaks Weyl symmetry. Thanks to the one-form field $\bfalpha$ that transforms as a connection under Weyl transformations, \textit{i.e.}, $\delta_B\bfalpha = -\D B$ by virtue of Eq. \eqref{eq:Weyl connection variation}, one is able to define a connection $\text D$ preserving the Weyl class $[\bfg]$, or in other words, is compatible with the metric $\bfg$ up to Weyl rescalings. This is the strongest compatibility condition one can require between $\text D$ and $\bfg$ in the presence of a Weyl structure \cite{Weyl:1918ib,Weyl:1919fi,folland1970weyl,hall1992weyl,Ciambelli:2019bzz}. We show here how this construction emerges from the gauging procedure of $\mathfrak{conf}(d,1)$.

For that purpose, we assume that the gauging of the conformal algebra is done in the \hyperref[item:condition3]{third} case above, \textit{i.e.}, that the gauge curvatures \eqref{eq:gauge curvatures} obey the following constraints:
\begin{equation}
	\boxed{
	\vect R[\gen P]^A = \bm 0,\qquad \vect{Ric}[\gen J]^A = \bm 0,
	} \label{eq:constraints gauging conformal}
\end{equation}
which have the property of determining completely the spin connection from non-holonomy coefficients and the Weyl connection and expressing the additional gauge field $\bfkappa^A$ in terms of the geometry. We start by solving the condition $\vect R[\gen P]^A=\bm 0$, which controls the torsion of the spin connection as
\begin{equation}
	\D\htheta^A + \bfgamma^A{}_B\wedge\htheta^B - \bfalpha\wedge \htheta^A = \bm 0 \qquad\Leftrightarrow\qquad \gamma^A{}_{[BC]} = \tfrac{1}{2}c^A{}_{BC} + \tfrac12 \alpha_{B}\delta^A{}_{C} - \tfrac12 \alpha_C\delta^A{}_B . \label{eq:torsion of spin co}
\end{equation}
Because $\gamma^{(A|B|C)} = 0$ by construction, the unique solution to the above constraint is
\begin{equation}
	\gamma^A{}_{BC} = \tfrac12(c^A{}_{BC}+c_B{}^A{}_C+c_C{}^A{}_B) - \delta^A{}_B\alpha_C + \alpha^A\eta_{BC}. \label{eq:gamma for WLC}
\end{equation}
By virtue of Eqs. \eqref{eq:delta sigma gammaAB} and \eqref{eq:Weyl connection variation}, we can check that the coefficients $\gamma^A{}_{BC}$ are Weyl-covariant with weight one. However, it differs in two respects from the usual notion of Levi-Civita connection in the presence of a Weyl connection. Firstly, being always metric-compatible by design, it is tied to a particular representative of the Weyl class $[\bfg]$, thus breaking Weyl invariance. Secondly, it is not torsion-free when the Weyl connection is non-trivial, as evidenced by Eq. \eqref{eq:torsion of spin co}. The examination of the issue naturally offers the solution: one can use the Weyl connection $\bfalpha$ itself to build an auxiliary connection $\text D$, with coefficients
\begin{equation}
		\bfomega^A{}_B = \bfgamma^A{}_B - \delta^A{}_B\bfalpha\quad\Leftrightarrow\quad \omega^A{}_{BC} = \tfrac12\big(c^A{}_{BC} + c_B{}^A{}_C + c_C{}^A{}_B\big) - \delta^A{}_B \alpha_C + \alpha^A\eta_{BC} - \delta^A{}_C \alpha_B,
	 \label{eq:omega in terms of gamma and alpha}
\end{equation}
hence uniquely defined in terms of the geometry through non-holonomy coefficients and a choice of Weyl connection. This auxiliary connection is now torsion-free in the usual sense, since
\begin{equation}
	\vect D\wedge\htheta^A = \D\htheta^A + \bfomega^A{}_B\wedge\htheta^B = \bm 0 \label{eq:no torsion WLC}
\end{equation} 
from Eq. \eqref{eq:torsion of spin co}. Furthermore, it obeys the following crucial property:
\begin{equation}
	\boxed{\vect D \bfg - 2 \bfalpha \otimes \bfg = \bm 0 \quad\Leftrightarrow\quad \bfomega_{(AB)} = -\eta_{AB}\bfalpha,} \label{eq:weyl LC}
\end{equation}
which is referred to as \textit{Weyl-metricity}, since it generalises the concept of metric compatibility to the whole Weyl class instead of a particular representative. Indeed, recalling that $\delta_B\bfgamma^A{}_B=\bm 0$ and $\delta_B \bfalpha = -\D B$, see Eqs. \eqref{eq:delta sigma gammaAB} and \eqref{eq:Weyl connection variation}, we have that
\begin{equation}
	\delta_B\bfomega^A{}_B = -\delta^A{}_B\D B\quad\Rightarrow\quad \delta_B\big(\vect D\bfg - 2\bfalpha\otimes\bfg\big) = -2B\big(\vect D\bfg - 2\bfalpha\otimes\bfg\big) = \bm 0,
\end{equation}
and imposing \eqref{eq:weyl LC} does not break Weyl invariance. The torsion-free and Weyl-metric connection $\text D$ shall therefore be referred to as the \textit{Weyl--Levi-Civita connection} built from the choice $\bfalpha$ of Weyl connection, as it offers the closest analogue to the Levi-Civita connection that agrees with Weyl connection. Furthermore, in the frame where $\bfalpha = \bm 0$, which can be achieved through a special conformal transformation, see Section \ref{sec:gauges relat}, the Weyl--Levi-Civita reduces to the ordinary Levi-Civita connection for the selected representative $\bfg$ of $[\bfg]$, $\omega^A{}_{BC}=\mathring\gamma^A{}_{BC}$, see Eq. \eqref{eq:levi civita}.

For later convenience, we build a Weyl-covariant derivative operator $\mathscr D$. More specifically, we shall require that it acts internally in the space of sections of the Weyl bundle with weight $w$, \textit{i.e.},
\begin{equation}
	\delta_B\vect A = w B \vect A \qquad\Rightarrow\qquad \delta_B \pmb{\mathscr D}\vect A = w B \pmb{\mathscr D}\vect A
\end{equation}
for any weight-$w$ tensor $\vect A$. It is field-dependent by design as the Weyl-weight of the field in argument appears explicitly in the above definition. In practice, when working in local frames, it is more convenient to define weights with respect to components rather than full tensors, because the frame itself transforms non-trivially under Weyl rescaling. Therefore, we shall define the Weyl-covariant derivative acting on weight-$w$ scalars $f$, vector components $V^A$ and one-form components $W_A$ as
\begin{subequations}\label{eq:curly D relat def}
	\begin{align}
		\mathscr D_A f &\coloneqq \text D_A f + w \alpha_A f,\\ \mathscr D_A V^B &\coloneqq \text D_A V^B + (w+1)\alpha_A V^B,\\ \mathscr D_A W_B &\coloneqq \text D_A W_B + (w-1)\alpha_A W_B.
	\end{align}
\end{subequations}
In components, each of these quantities transforms as Weyl-covariant fields with weights $(w+1)$. In the gauging approach, the spin connection $\nabla$ is of a more fundamental nature than the connection $\text D$ which is a composite operator built up from the former and the gauge connection related to local dilations. In terms of $\nabla$, the definitions of Eq. \eqref{eq:curly D relat def} simplify to\footnote{In \cite{Fiorucci:2025twa}, the $\mathscr D$ operator was naturally built upon $\omega^A{}_{BC}$ instead of $\gamma^A{}_{BC}$.}
\begin{equation}
	\mathscr D_A f = \nabla_A f + w\alpha_A f,\qquad \mathscr D_A V^B = \nabla_A V^B + w \alpha_A V^B,\qquad \mathscr D_A W_B = \nabla_A W_B + w \alpha_A W_B. \label{eq:curly D in terms of gamma nabla}
\end{equation}
The correcting weight-dependent term is now universal, in the sense that it is insensitive to the rank of the tensor in argument. This simplification is due to the fact that the spin connection is always Weyl-covariant, as evidenced by Eq. \eqref{eq:delta sigma gammaAB}, and the only inhomogeneous term that can be induced by a Weyl transformation comes from the transport term, \textit{e.g.}, $\hframe_A(V^B)$ in the vector case. Finally, in terms of the Levi-Civita connection $\mathring\nabla$ derived from the same representative, the above formulas develop into \cite{Campoleoni:2023fug}
\begin{subequations}\label{eq:curly D on various objects}
	\begin{align}
		\mathscr D_A V^B &= \mathring\nabla_A V^B + w\alpha_A V^B + \alpha^B V_A - \delta^B{}_A \alpha_C V^C,\\
		\mathscr D_A W_B &= \mathring\nabla_A W_B + w\alpha_A W_B + \alpha_B W_A - \eta_{AB} \alpha_C W^C.
	\end{align}
\end{subequations}
The generalisation of Eqs. \eqref{eq:curly D relat def}--\eqref{eq:curly D on various objects} to higher-rank tensors is direct thanks to Leibniz rule.\footnote{For instance, one can show that $\mathscr D_A A_{BC} = \mathring \nabla_A A_{BC} + w\alpha_A A_{BC} + \alpha_B A_{AC} + \alpha_C A_{BA} - \alpha^D(A_{DC}\eta_{AB} + A_{BD}\eta_{AC})$ for any rank-two tensor $\vect A = A_{BC}\htheta^B\otimes\htheta^C$ with weight-$w$ components.} Importantly, one cannot associate connection coefficients for any field to $\mathscr D_A$ as one usually does for $\nabla_A$, simply because the weight of the field appears explicitly in the definition, see either Eq. \eqref{eq:curly D relat def} or Eq. \eqref{eq:curly D in terms of gamma nabla}. Therefore, although the connection induced by $\mathscr D_A$ is metric-compatible, $\pmb{\mathscr D}\bfg=\bm 0$, on account of Eq. \eqref{eq:weyl LC}, it is non-affine by construction.

We now turn our attention to solving the second condition in Eq. \eqref{eq:constraints gauging conformal}. In the absence of composite torsion, $\vect R[\gen P]^A = \bm 0$, the Bianchi identity \eqref{eq:bianchi for RP} yields
\begin{equation}
	\vect R[\gen J]^A{}_B\wedge\htheta^B = \vect R[\gen D]\wedge\htheta^A\qquad\Rightarrow\qquad (d-1)\vect R[\gen D] = \vect{Ric}[\gen J]_A\wedge\htheta^A = \bm 0. \label{eq:RD vanishes auto}
\end{equation}
If $d>1$, the absence of composite Weyl curvature is therefore automatically satisfied. We shall assume that $d>1$ in what follows and comment on the $d=1$ case in Appendix \ref{sec:2d}. Performing a suitable contraction of Eq. \eqref{eq:RJdef} and imposing $\vect{Ric}[\gen J]^A = \bm 0$ leads to
\begin{equation}
	\vect{Ric}[\bfgamma]^A \coloneqq \vect{R}[\bfgamma]^{BA}(\hframe_B,\cdot) = (d-1)\bfkappa^A + \kappa^B{}_B \htheta^A \label{eq:Ricgamma in terms of kappa}
\end{equation}
where $\vect{R}[\bfgamma]^{AB}\coloneqq \D\bfgamma^{AB} + \bfgamma^A{}_C\wedge\bfgamma^{CB}$ denotes the Riemann tensor deriving from the spin connection $\bfgamma^{AB}$ and $\vect{Ric}[\bfgamma]^A$ the related Ricci tensor. When broken into (skew-)symmetric parts, Eq. \eqref{eq:Ricgamma in terms of kappa} first implies, on the one hand, that the skew-symmetric part of the Ricci tensor is algebraically and identically locked to the skew-symmetric part of the gauge field $\bfkappa^A$ as
\begin{equation}
	\kappa_{[AB]} = \tfrac{1}{d-1}\text{Ric}[\bfgamma]_{[AB]}, \label{eq:Ric antisym}
\end{equation}
and, on the other hand, that its symmetric part is determined by the \textit{Schouten tensor} of the spin connection, \textit{i.e.},
\begin{equation}
	\kappa_{(AB)} = \text{Sch}[\bfgamma]_{AB} \coloneqq \tfrac{1}{d-1}\big(\text{Ric}[\bfgamma]_{(AB)} - \tfrac{1}{2d} R[\bfgamma]\eta_{AB}\big),  \label{eq:kappa = schouten}
\end{equation}
where $R[\bfgamma] \coloneqq \eta^{AB}\text{Ric}[\bfgamma]_{(AB)}$ denotes the Ricci scalar of the spin connection. Crucially, the Ricci tensor $\text{Ric}[\bfgamma]_{AB}$ is not symmetric in the presence of a non-trivial Weyl curvature. Indeed, the vanishing of the $\gen D$-curvature fixes the skew-symmetric part of $\bfkappa^A$ in terms of the Weyl curvature as
\begin{equation}
	\bfkappa_A\wedge\htheta^A = \D\bfalpha \quad\Leftrightarrow\quad \kappa_{[AB]} = -\nabla_{[A}\alpha_{B]}, \label{eq:antisymm schouten}
\end{equation}
which has a direct impact on Eq. \eqref{eq:Ric antisym}. This matter of fact has already been observed, \textit{e.g.}, in \cite{Ciambelli:2019bzz}, see also references therein.

\subsection{The Cotton tensor}
When the conditions \eqref{eq:constraints gauging conformal} are obeyed, the components $C^A{}_{BC}$ of the $\gen K$-curvature are identified with those of the \textit{Cotton tensor} \cite{Cotton1899}. Indeed, thanks to the Weyl-covariant operator $\mathscr D$, we can rewrite Eq. \eqref{eq:RK} in components in a compact way as
\begin{equation}
	\boxed{ C^A{}_{BC} = 2\mathscr D_{[B}\kappa^A{}_{C]}, }\label{eq:cotton explicit}
\end{equation}
which is the expected definition of the Cotton tensor as the curl of the Schouten tensor, displayed here in a manifestly Weyl-covariant fashion.\footnote{Note that here, we consider that the full $\kappa_{AB}$ is the Schouten tensor, which is a slight abuse of nomenclature. Indeed, since we are working with a Weyl-covariant connection, such a definition of the Schouten tensor must contain a skew-symmetric piece, which is related to the Weyl curvature, see Eq. \eqref{eq:antisymm schouten}.} As a consistency check, we can derive the Bianchi identities involving the Cotton tensor in order to get the algebraic and differential constraints this tensor shall obey without relying on the explicit expression \eqref{eq:cotton explicit}. We can form the composite Weyl tensor
\begin{equation}
	\vect{W}[\gen J]^{AB} = \tfrac12 W^{AB}{}_{CD}\htheta^C\wedge\htheta^D \coloneqq \vect R[\gen J]^{AB} + 2\, \vect{Sch}[\gen J]^{[A}\wedge\htheta^{B]}, \label{eq:introd of weyl}
\end{equation}
where the second term involves the composite Schouten tensor
\begin{equation}
	\vect{Sch}[\gen J]^{A} = \tfrac{1}{d-1}\big(\vect{Ric}[\gen J]^A - \tfrac{1}{2d}R[\gen J]\htheta^A\big)
\end{equation}
and $R[\gen J]\coloneqq \vect{Ric}[\gen J]^A(\hframe_A)$ denotes the composite Ricci scalar curvature. On account of Eq. \eqref{eq:bianchi for RP}, the Weyl tensor enjoys the usual symmetry properties:
\begin{equation}
	W_{(AB)CD} = 0 = W_{AB(CD)},\quad W^A{}_{[BCD]}=0\Rightarrow W_{ABCD} = W_{CDAB},\quad W^A{}_{BAC} = 0, \label{eq:symmetries of the weyl}
\end{equation}
which, in particular, make it vanish if $d\leq 2$. It is also Weyl invariant: $\delta_B\vect W[\gen J]^{AB} = \bm 0$.

Since the composite Ricci tensor is supposed to vanish by virtue of Eq. \eqref{eq:constraints gauging conformal}, we have that $\vect{W}[\gen J]^A{}_B = \vect R[\gen J]^A{}_B$ when the equations of motion hold. Acting with an exterior derivative and using \eqref{eq:bianchi for RJ} yields a Bianchi identity for the Cotton tensor:
\begin{equation}
	\pmb{\mathscr D}\wedge \vect W[\gen J]^{AB} + 2 \vect R[\gen K]^{[A}\wedge \htheta^{B]} = \bm 0 \quad\Leftrightarrow\quad \mathscr D_{[E} W^{AB}{}_{CD]} = 2 \delta^{[A}{}_{[E} C^{B]}{}_{CD]}. \label{eq:bianchi weyl C}
\end{equation}
In dimensions strictly greater than three ($d > 2$), it fixes the Cotton tensor as the divergence of the Weyl tensor. Indeed, taking a double trace over the index pairs $(A,E)$ and $(B,D)$ in Eq. \eqref{eq:bianchi weyl C} first implies
\begin{equation}
	C^A{}_{AB} = 0 = C^A{}_{BA}, \label{eq:tracefree cotton}
\end{equation}
\textit{i.e.}, the tracelessness of the Cotton tensor, and therefore, we obtain
\begin{equation}
	(d-2)C^B{}_{CD} = \mathscr D_A W^{AB}{}_{CD} \label{eq:identity weyl cotton}
\end{equation}
by taking only one trace over the index pair $(A,E)$. The symmetries of the Weyl tensor in the right-hand side imply also that the fully skew-symmetric part of the Cotton tensor vanishes, $C_{[ABC]} = 0$. This property, together with Eq. \eqref{eq:tracefree cotton}, can be termed as the fact that the Cotton tensor lies in the irreducible mixed-symmetry representation of the Lorentz group. Furthermore, its curl is controlled by a last (differential) Bianchi identity which is obtained from Eq. \eqref{eq:bianchi for RK}:
\begin{equation}
	\pmb{\mathscr D}\wedge\vect R[\gen K]^A = \vect R[\gen J]^A{}_B \wedge \bfkappa^B = \vect W^A{}_B\wedge\bfkappa^B, \label{eq:curl of cotton}
\end{equation}
applied for $\vect R[\gen D] = \bm 0$. Therefore, the Cotton tensor is curl-free if and only if the composite Weyl tensor vanishes. In particular, the Cotton tensor is identically curl-free in three dimensions.

\subsection{Metric and rheotactic gauges}
\label{sec:gauges relat}
The most important feature of the gauging procedure is that it brings about more fields than in the usual case: even on-shell, by which we mean when the equations of motion \eqref{eq:constraints gauging conformal} are satisfied, the local coframe $\htheta^A$ has to be supplemented by a Weyl connection $\bfalpha$ in order to make the formalism explicitly Weyl-covariant. However, this additional field transforms algebraically under special conformal transformations $Z^A$: it can therefore be completely gauge-fixed at the expense of restricting these supplementary transformations.\footnote{Whether the local special conformal transformations with parameter $Z^A$ carry a non-vanishing charge, thereby implying that these gauge fixings are not made for free, particularly in a holographic setting \cite{Ciambelli:2019bzz}, is an interesting issue and will be addressed elsewhere.} We now elaborate on two convenient choices that have appeared in previous literature.

One can choose $\bfalpha = \bm 0$. In this case, the Weyl--Levi-Civita connection coincides with the usual Levi-Civita connection for a given representative $\bfg$ in the Weyl class $[\bfg]$. This gauge, also called the $K$-gauge in \cite{Freedman:2012zz}, is preserved by local Weyl transformations if and only if the gauge parameter for special conformal transformations $Z^A$ is fixed in terms of the Weyl rescaling function $B$ as $Z_A = \hframe_A[B]$. Imposing this gauge condition and the equations of motion means that the gauge connection $\bm\Gamma$ is completely determined in terms of the local Lorentz frame $\hframe_A$. In a holographic setting, the fact that one can cancel the effect of a Weyl rescaling by a transformation leaving the boundary frame invariant is reminiscent of the so-called Penrose--Brown--Henneaux diffeomorphisms \cite{Brown:1986nw,Imbimbo:1999bj}. They appear in the context of asymptotically anti-de Sitter gravity at conformal infinity: whenever the action of radial diffeomorphisms causes a Weyl rescaling of the boundary metric, it can always be compensated by a subleading diffeomorphism in the directions tangent to the boundary and completely determined by the gradient of the Weyl factor \cite{Ciambelli:2019bzz}. In summary, choosing the gauge $\bfalpha = \bm 0$ and imposing the equations of motion, one needs only to work with the metric $\bfg$ which transforms trivially under local Lorentz and is sensitive only to diffeomorphisms and Weyl. We therefore refer to it as the \textit{metric gauge}.

In some contexts, such as physics at conformal boundaries of gravitational theories, it is suitable to keep Weyl covariance throughout. This is the case, for instance, in fluid/gravity correspondence, where bulk gravity dynamics can be reconstructed from conformal fluids defined at the conformal boundary, see \cite{Loganayagam:2008is,Bhattacharyya:2008mz,Leigh:2011au,Leigh:2012jv,Mukhopadhyay:2013gja,Ciambelli:2017wou} for applications in AdS/CFT holography, \cite{Ciambelli:2018xat,Ciambelli:2018wre,Campoleoni:2018ltl,Ciambelli:2020eba,Fiorucci:2025twa} for applications in flat-space holography and \cite{Hubeny:2011hd,Petropoulos:2014yaa} for reviews. In this setting, the `fluid' flow induces a natural and privileged timelike vector field over $(\mathscr M,\bfg)$, which is the velocity field $\bfupsilon$. Assuming that it is Weyl-invariantly normalised as $\bfg(\bfupsilon,\bfupsilon) = -c^2$, it must transform as $\delta_B\bfupsilon = B\bfupsilon$ under infinitesimal Weyl rescalings. As a result, one can define the Weyl connection as \cite{Loganayagam:2008is,Bhattacharyya:2008mz}
\begin{equation}
	\bfalpha_\bfupsilon \coloneqq \bm\upvarphi + \frac{\theta}{d}\bftau \label{eq:rheotactic}
\end{equation}
where $\bftau \coloneqq -c^2\bfg(\bfupsilon,\cdot)$ is the canonical clock form related to the congruence $\bfupsilon$, the one-form field
\begin{equation}
	\bm\upvarphi \coloneqq - \big(\bfupsilon\cdot \mathring{\bm\nabla}\big)\bftau, \label{eq:varphi from u}
\end{equation}
transverse to $\bfupsilon$ as the latter is geodesic, is the acceleration computed thanks to the Levi-Civita connection $\mathring{\nabla}$ related to $\bfg$ and $\theta = \mathring{\bm\nabla}\cdot\bfupsilon$ is the expansion. It is not difficult to show from Eq. \eqref{eq:rheotactic} that $\delta_B \bfalpha_\bfupsilon = -\D B$ as required. By construction, the Weyl--Levi-Civita connection built as \eqref{eq:weyl LC} from $\bfalpha_\bfupsilon$ satisfies the following constraints:
\begin{equation}
	{\pmb{\mathscr D}}\cdot \bfupsilon = 0,\qquad \big(\bfupsilon\cdot{\pmb{\mathscr D}}\big) \bfupsilon  = \bm 0,
\end{equation}
We refer to this gauge fixing as \textit{rheotactic}, as it is built upon the choice of a particular velocity field $\bfupsilon$. It is therefore not preserved by local Lorentz boosts that affect the velocity ($\delta_\lambda \bfupsilon = \lambda^a{}_0 \hframe_a$) and change $\bm\upvarphi$ and $\theta$ inhomogeneously. There is however no contradiction with the general transformation \eqref{eq:Weyl connection variation} if the $Z$-transformation is tuned to take this inhomogeneous shift into account.

\subsection{Three dimensions: time-space split and expressions in rheotactic gauge}
\label{sec:rheotactic lorentz}
To conclude this introductory discussion, let us focus on the three-dimensional case, for which the Cotton tensor can be dualised into a symmetric rank-two tensor. Indeed, if $d=2$, a Levi-Civita symbol with three indices exists and one can define
\begin{equation}
	\vect R[\gen K]^A = \tfrac12 C^A{}_{BC}\htheta^B\wedge\htheta^C \coloneqq \tfrac12 C^{AD}\varepsilon_{BCD}\htheta^B\wedge\htheta^C. \label{eq:Cotton dualised}
\end{equation}
Since the Weyl tensor identically vanishes, so does the $\gen J$-curvature by virtue of the equations of motion and thanks to the related Bianchi identity \eqref{eq:bianchi for RJ}, one establishes that the two-index instance of the Cotton tensor is symmetric:
\begin{equation}
	\vect R[\gen K]^{[A}\wedge\htheta^{B]} = \bm 0\quad\Leftrightarrow\quad C^{[AB]} = 0. \label{eq:cotton is symm}
\end{equation}
Unlike in higher dimensions, there is nothing more to be extracted from this Bianchi identity. Looking at the Bianchi identity \eqref{eq:bianchi for RD} related to the $\gen D$-curvature, we recover the fact that the dualised Cotton tensor is tracefree:
\begin{equation}
	\vect R[\gen K]_A\wedge\htheta^A = \bm 0\quad\Leftrightarrow\quad C^A{}_{A} = \eta_{AB}C^{AB} = 0, \label{eq:cotton is TF}
\end{equation}
where we used that $\vect D$ is torsionless, \textit{i.e.}, $\vect R[\gen P]^A=\bm 0$. Finally, expanding Eq. \eqref{eq:curl of cotton}, we obtain
\begin{equation}
	\mathscr D_{[A} C^D{}_{BC]}\htheta^A\wedge\htheta^B\wedge\htheta^C = 0\qquad\Leftrightarrow\qquad \mathscr D_A C^{AB} = 0,
\end{equation}
by virtue of which the Cotton tensor is covariantly conserved.

The three-dimensional Cotton tensor is instrumental in four-dimensional Einstein gravity with asymptotically locally AdS boundary conditions, where it provides a geometric encoding, on the boundary, of bulk gravitational radiation \cite{Ciambelli:2017wou,Ciambelli:2018wre,Campoleoni:2023fug,Ciambelli:2024kre,Fernandez-Alvarez:2025qqx} and magnetic gravitational charges \cite{deHaro:2008gp,Bakas:2008gz,Mukhopadhyay:2013gja,Bakas:2015opa,Mittal:2022ywl}. It therefore represents the magnetic counterpart of the holographic energy--momentum tensor \cite{Balasubramanian:1999re,deHaro:2000vlm}, which emanates from the electric part of the boundary value of the bulk Weyl tensor \cite{Ashtekar:1984zz,Ashtekar:1999jx}. 
In order to prepare the Carrollian limit, we display the detailed expressions of the Cotton tensor in three dimensions in the rheotactic gauge, which was used in \cite{Campoleoni:2023fug} in the gravitational context and discuss some of its properties. Recall that the rheotactic gauge choice $\bfalpha \equiv \bfalpha_\bfupsilon$ consists in deriving the Weyl connection one-form onto from a given timelike congruence $\bfupsilon$. We shall use notations that allow for the immediate comparison between the Lorentzian and Carrollian cases, up to suitable $c\to 0$ limits. To this aim, we split the geometry into longitudinal and transverse parts with respect to the congruence, which is in fact mandatory to deal with non-Lorentzian situations with the required amount of care and precision.

We adapt the Cartan frame $\{\hframe_A\}$ to the congruence by setting $\hframe_0 \equiv \bfupsilon$. Therefore, $\{\hframe_a\}$ define a basis of vectors transverse to $\bfupsilon$. The dual basis is chosen such that $\htheta^0 = \bftau$, the canonical clock form related to $\bfupsilon$ and the $\{\htheta^a\}$ span the dual basis of $\{\hframe_a\}$ on transverse spaces. We split the non-holonomy coefficients accordingly as
\begin{equation} \label{eq:nonholo}
\begin{array}{ccc}
	[\hframe_a, \hframe_b] = 2\varpi_{ab} \bfupsilon + c^c{}_{ab} \hframe_c ,&\quad& [\bfupsilon, \hframe_a] = \varphi_a \bfupsilon - c^b{}_a \hframe_b, \\
	\D\htheta^a + \frac12 c^a{}_{bc} \htheta^b \wedge \htheta^c + c^a{}_b \htheta^b\wedge \bftau = 0 , &\quad& \D\bftau + \varpi_{ab} \htheta^a \wedge \htheta^b - \varphi_a \htheta^a\wedge\bftau = 0 .
\end{array}
\end{equation}
Vertical non-holonomy coefficients $\varphi_a$ and $\varpi_{ab}$ transform as $SO(2)$ tensors. The one-form $\bm \upvarphi = \varphi_a\htheta^a$ is the acceleration of $\bfupsilon$, that already appeared in Eq. \eqref{eq:varphi from u}. The two-form of components $\varpi_{ab}$ is the rotational of $\bfupsilon$, \textit{i.e.}, its vorticity. It represents the obstruction for $\bftau$ to define an integrable distribution of horizontal spaces. Both transform as connections under Lorentz boosts. Furthermore, horizontal non-holonomy coefficients are sensitive to the particular choice of basis in the distribution of $\bftau$ and thus transform as connections under $SO(2)$ local rotation. The symmetric part
\begin{equation}
	c_{(ab)} = \frac{\theta}{2}\delta_{ab} + \xi_{ab}
\end{equation}
however transforms as a tensor under the aforementioned local symmetry: it quantifies the dragging of the horizontal geometry on the flow of $\bfupsilon$. The trace part involves the expansion $\theta$ that has already appeared above, while the tracefree part $\xi_{ab}$ has been coined as the shear.\footnote{In a holographic setting, one is interested in multiple shears that have different physical and geometrical meaning. Therefore, to avoid any confusion, one shall refer to $\xi_{ab}$ as the intrinsic (or geometric) shear, by opposition to the physical shear $\mathscr C_{ab}$ of outgoing null geodesics supporting gravitational radiation.}

Next, the selection of a particular timelike congruence $\bfupsilon$ allows us to define a transverse projector $\bm\Pi \coloneqq \textbf{id} - \bfupsilon\otimes\bftau$ onto horizontal spaces. At any point $P$ on the flow of $\bfupsilon$, the image of $\bm\Pi$ provides the local hyperplane of simultaneity of the observer $\bfupsilon$ at $P$. Vectors that are stabilised by the action of $\bm\Pi$ are referred to as transverse with respect to $\bfupsilon$. Moreover, transverse one-forms are stabilised by the action of the transposed projector $\bm\Pi^\text{t} = \textbf{id} - \bftau\otimes\bfupsilon$. Any spin connection $\bfgamma^A{}_B$ on $\mathscr M$ can therefore be projected onto transverse space to define a \textit{horizontal connection} as
\begin{equation}
	\bar\nabla_\bot \vect A \coloneqq \bm\Pi\circ(\nabla_\bfupsilon \vect A)\circ \bm\Pi^\text{t},\qquad \bar\nabla_{\hframe_a} \vect A \coloneqq \bm\Pi\circ(\nabla_{\hframe_a} \vect A)\circ \bm\Pi^\text{t}
\end{equation}
on any transverse rank-two tensor $\vect A = A^a{}_b\hframe_a\otimes\htheta^b$. In components, we find\footnote{In practice, since the connection $\bar\nabla$ is projected onto transverse spaces, it is therefore blind to `$0$' indices. For instance, $A^0{}_a$ will be seen as a transverse one-form only and the upper `$0$' index as a scalar.} \begin{equation}
	\bar\nabla_\bot A^a{}_b \coloneqq \bfupsilon(A^a{}_b) + \gamma^a{}_c A^c{}_b - A^a{}_c\gamma^c{}_b,\qquad
	\bar\nabla_a A^b{}_c \coloneqq \hframe_a(A^b{}_c) + \gamma^b{}_{ad} A^d{}_c - A^b{}_d\gamma^d{}_{ac} .
\end{equation}
As expected, the relevant spin connection coefficients for $\bar\nabla$ are the horizontal one-forms $\bfgamma^a{}_b$. By construction, the horizontally projected connection $\bar\nabla$ acts internally on horizontal spaces. In general, it has torsion even though the parent connection is torsion-free.\footnote{Indeed, the horizontal torsion is given by $\bar{\vect T}{}^a = \D\htheta^a + \bfgamma^a{}_b\wedge\htheta^b = -(\gamma^a{}_b + c^a{}_b)\htheta^b\wedge\bftau + \big(\gamma^a{}_{[bc]}-\tfrac12 c^a{}_{bc}\big)\htheta^b\wedge\htheta^c.$ If the three-dimensional connection has no torsion, $\gamma^a{}_{[bc]} = \tfrac12 c^a{}_{bc}$ and the purely spatial components vanish. However, on the basis of Eq. \eqref{eq:projconn}, one can easily check that there will remain non-trivial pieces in the mixed components whenever the flow of $\bfupsilon$ does not preserve the horizontal metric. see, \textit{e.g.}, \cite{Ciambelli:2018xat,Campoleoni:2023fug}.} Furthermore, it has fiducial torsion on the vertical direction whenever $\bftau$ is not integrable \cite{Campoleoni:2023fug}.\footnote{See, \textit{e.g.}, Eqs. \eqref{eq:D0Db Phi} and \eqref{eq:DaDbVc} for a concrete manifestation of this matter of fact.}

Reported in terms of a time-space split related to the congruence $\bfupsilon$, the Weyl--Levi-Civita spin connection $\bfgamma^A{}_B$ associated with the rheotactic-gauge fixing takes the following explicit expression:
\begin{equation}
	\boxed{
	\bfgamma^0{}_0 = \bm 0,\quad \bfgamma^a{}_0 = \big(\xi^a{}_b-c^2\varpi^a{}_b\big)\htheta^b,\quad \bfgamma^0{}_a = \big(\tfrac{1}{c^2}\xi_{ba} + \varpi_{ba}\big)\htheta^b,\quad \bfgamma^a{}_b = \gamma^a{}_b\bftau + \gamma^a{}_{cb}\htheta^c,
	} \label{eq:gamma split rheo}
\end{equation}
that we obtain by merging information from Eqs. \eqref{eq:gamma for WLC}, \eqref{eq:rheotactic} and \eqref{eq:nonholo}. In this case, the connection coefficients of the horizontally projected connection are given by
\begin{equation}
	\gamma_{ab} = \gamma_{[ab]} = -c_{[ab]} - c^2\varpi_{ab},\qquad \gamma^a{}_{bc} = \tfrac12 \big( c^a{}_{bc} + c_b{}^a{}_c + c_c{}^a{}_b\big) - \delta^a{}_b\varphi_c + \varphi^a \delta_{bc}. \label{eq:projconn}
\end{equation} 
Namely, the purely horizontal spin-connection coefficients are those of the Weyl--Levi-Civita connection on the transverse space, for the particular choice $\alpha_a=\varphi_a$ of horizontal Weyl connection. Following the same steps as explained in Section \ref{sec:WLC relat}, the horizontal spin connection $\bar\nabla$ can be promoted into a Weyl-class compatible derivative operator $\bar{\mathscr D}$, acting on Weyl-weight-$w$ horizontal tensors like $\vect A = A^a{}_b\hframe_a\otimes\htheta^b$ as
\begin{equation}
	\bar{\mathscr D}_\bot A^a{}_b \coloneqq \bar\nabla_\bot A^a{}_b + \tfrac12 w \theta A^a{}_b,\qquad \bar{\mathscr D}_a A^b{}_c \coloneqq \bar\nabla_a A^b{}_c + w \varphi_a A^b{}_c. \label{eq:curly D hat acting on Aab}
\end{equation}
Again, the supplementary term ensuring Weyl covariance of the resulting derivative is universal, in the sense that it does not depend on the rank of $\vect A$, when the first term only involves the spin connection. By construction, this operator is compatible with the Weyl class of the projected metric $\bar{\bfg} = \bm\Pi\circ \bfg\circ\bm\Pi^\text{t} = \delta_{ab}\htheta^a\otimes\htheta^b$ onto spacelike hypersurfaces orthogonal to $\bfupsilon$, \textit{i.e.},
\begin{equation}
	\big(\bar{\mathscr D}_\bot\bar{\bfg}\big)_{ab} = 0,\qquad \big(\bar{\mathscr D}_a\bar{\bfg}\big)_{bc} = 0.
\end{equation}
Evaluating the commutators of such derivatives reveals the pieces of curvature and torsion of $\bar\nabla$ in a Weyl-covariant fashion. We find
\begin{subequations}\label{eq:commu curly D bar}
	\begin{align}
		\big[\bar{\mathscr D}_a,\bar{\mathscr D}_b\big]\Phi &= 2\varpi_{ab}\bar{\mathscr D}_\bot\Phi + w \Omega_{ab}\Phi, \label{eq:DaDb Phi} \\
		\big[\bar{\mathscr D}_\bot,\bar{\mathscr D}_a\big]\Phi &= -\big(\xi^b{}_a - c^2\varpi^b{}_a\big)\bar{\mathscr D}_b\Phi + w \bar{\mathscr R}_a \Phi, \label{eq:D0Db Phi} \\
		\big[\bar{\mathscr D}_a,\bar{\mathscr D}_b\big] V^c &= \bar{\mathscr R}{}^c{}_{dab}V^d + 2\varpi_{ab}\bar{\mathscr D}_\bot V^c + w \Omega_{ab} V^c, \label{eq:DaDbVc} \\
		\big[\bar{\mathscr D}_\bot,\bar{\mathscr D}_a\big]V^b &= -\bar{\mathscr R}{}^b{}_{ac}V^c - \big(\xi^c{}_a - c^2\varpi^c{}_a\big)\bar{\mathscr D}_c V^b + w \bar{\mathscr R}_a V^b, \label{eq:D0DaVb}
	\end{align}
\end{subequations}
for weight-$w$ scalar field $\Phi$ and vector field $V^a$, where the tensor components $\bar{\mathscr R}{}^a{}_{bc}$, $\bar{\mathscr R}{}^a{}_{bcd}$, $\bar{\mathscr R}_a$ and $\Omega_{ab}$ are given by Eqs. \eqref{eq:Rabc weyl}, \eqref{eq:Rabcd split relat} and \eqref{eq:weyl curv rheotactic} respectively. Purely horizontal torsion is manifestly absent since we are projecting a torsion-free connection. However, torsion may appear in the vertical direction depending on how the spacelike hypersurfaces are stacked along the flow of $\bfupsilon$, in particular the integrability of the distribution induced by $\bftau$. The components of the horizontal curvature tensor with respect to the orthonormal frame adapted to $\bfupsilon$ are given by
\begin{subequations}
	\begin{align}
		\bar{\pmb{\mathscr R}}{}^a{}_b &= \D\bfgamma^a{}_b + \bfgamma^a{}_c\wedge\bfgamma^c{}_{b} = \bar{\mathscr R}{}^a{}_{bc}\bftau\wedge\htheta^c + \tfrac12 \bar{\mathscr R}{}^a{}_{bcd}\htheta^c\wedge\htheta^d, \label{eq:horizontal curvature} \\
		\bar{\mathscr R}{}^a{}_{bc} &= \bfupsilon(\gamma^a{}_{cb}) - (\hframe_c+\varphi_c)\gamma^a{}_b + \gamma^a{}_{db}c^d{}_c + \gamma^a{}_d\gamma^d{}_{cb} - \gamma^a{}_{cd}\gamma^d{}_b \nonumber \\
		&= \bar{\mathscr D}{}^a\big(\xi_{bc}-c^2\varpi_{bc}\big) - \bar{\mathscr D}_c\big(\xi^a{}_b - c^2\varpi^a{}_b\big) + \delta^a{}_b\bar{\mathscr R}_c - \delta_{bc}\bar{\mathscr R}{}^a, \label{eq:Rabc weyl} \\
		\bar{\mathscr R}{}^a{}_{bcd} &= 2\hframe_{[c}(\gamma^a{}_{d]b}) + 2 \gamma^a{}_{[c|e}\gamma^e{}_{d]b} - c^e{}_{cd}\gamma^a{}_{eb} - 2\varpi_{cd}\bar\gamma^a{}_b,
	\end{align}
\end{subequations}
and are Weyl primary fields of weight $2$. To establish the second equality in Eq. \eqref{eq:Rabc weyl}, we use the Bianchi identities arising from the integrability conditions $\D^2\bftau = \bm 0 = \D^2\htheta^a$, in particular
\begin{equation}
	\bfupsilon(c^a{}_{bc}) + 2 \hframe_{[b}(c^a{}_{c]}) + 2 \varphi_{[b}c^a{}_{c]} - 2 c^a{}_{d[b}c^d{}_{c]} - c^a{}_d c^d{}_{bc} = 0.
\end{equation}
Moreover, the purely horizontal curvature components reduce to one degree of freedom, the Gauss mean curvature $\bar{\mathscr K}$, since horizontal spaces are two-dimensional:
\begin{equation}
	\bar{\mathscr R}_{abcd} = \bar{\mathscr K}(\delta_{ac}\delta_{bd} - \delta_{ad}\delta_{bc}),\qquad \bar{\mathscr R} = 2 \bar{\mathscr K}, \qquad \bar{\mathscr K} \coloneqq \mathring K + \mathring\nabla_a \varphi^a + 2c^2(\ast\varpi)^2, \label{eq:Rabcd split relat}
\end{equation}
where $\mathring K$ represents the Gauss mean curvature of the Levi-Civita connection, with coefficients $\mathring\gamma^a{}_{bc}$, deriving from the horizontal metric, and $\ast\varpi \coloneqq \tfrac12\varepsilon^{ab}\varpi_{ab}$ is the Hodge dual of the transverse two-form $\bm\upvarpi$ through the two-dimensional Levi-Civita symbol $\varepsilon_{ab}$, which satisfies $\varepsilon_{ac}\varepsilon_b{}^c = \delta_{ab}$. More on the transverse dualisation will be said later, around Eq. \eqref{eq:Cab bar}.

In the time-space split, the full curvature tensor is subsequently computed from the knowledge of the horizontal curvature $\bar{\mathscr R}{}^a{}_b$ and the second fundamental form $\bfgamma^a{}_0$ of the spacelike hypersurfaces orthogonal to $\bfupsilon$ thanks to Gauss--Codazzi equations:
\begin{equation}
\begin{split}
	\pmb{\mathscr R}^a{}_0 &= c^2 \delta^{ab}\pmb{\mathscr R}^0{}_b = \bm\nabla\wedge\bfgamma^a{}_0 = \tfrac12 \mathscr R^a{}_{0bc}\htheta^b\wedge\htheta^c + \mathscr R^a{}_{0b0}\htheta^b\wedge\bftau,\\
	\pmb{\mathscr R}^a{}_b &= \bar{\pmb{\mathscr R}}{}^a{}_b + \bfgamma^a{}_0\wedge\bfgamma^0{}_b = \tfrac12 \mathscr R^a{}_{bcd}\htheta^c\wedge\htheta^d + \mathscr R^a{}_{bc0}\htheta^c\wedge\bftau.
\end{split}
\end{equation}
In the presence of a non-trivial Weyl curvature, weight-dependent terms appear as well in the commutators, and involve the components
\begin{equation}\label{eq:weyl curv rheotactic}
	\begin{split}
		\D \bfalpha &\coloneqq \tfrac12 \Omega_{ab}\htheta^a\wedge\htheta^b - \bar{\mathscr R}_a\htheta^a\wedge\bftau, \\
		\bar{\mathscr R}_a &= \bar\nabla_\bot\varphi_a + \big(\xi^b{}_a - c^2\varpi^b{}_a\big)\varphi_b -\tfrac12 \bar\nabla_a\theta, \qquad
		\Omega_{ab} = - \varepsilon_{ab}\bar{\mathscr A}
	\end{split}
\end{equation}
expressed here for the rheotactic gauge choice, where the scalar field
\begin{equation}
	\bar{\mathscr A} \coloneqq \ast\varpi\,\theta - \varepsilon^{ab}\bar\nabla_a\varphi_b = -2\bar{\mathscr D}_\bot(\ast\varpi) \label{eq:weyl curv rheo 2}
\end{equation}
corresponds to (minus two times) the vertical derivative of the non-integrability two-form, again by virtue of integrability conditions $\D^2\bftau=\bm 0$. We have $\bar{\mathscr R}{}^a = \bar{\mathscr R}{}^{ab}{}_b$ by taking a trace in Eq. \eqref{eq:Rabc weyl}. In the time-space split viewpoint, the components of the Schouten tensor \eqref{eq:kappa = schouten}--\eqref{eq:antisymm schouten} read
\begin{subequations}\label{eq:kappa for rheo relat}
    \begin{alignat}{6}
        \kappa^0{}_0
        &= \tfrac12 \mathscr R^{0a}{}_{0a}-\tfrac14 \mathscr R^{ab}{}_{ab},
        &\qquad \kappa^0{}_a
        &= \mathscr R^{0b}{}_{ab},
        &\qquad \kappa^a{}_0
        &= \bar{\mathscr R}{}^a-c^2 \mathscr R^{0ba}{}_b,
        \\
        \kappa^a{}_a
        &= \tfrac12 \mathscr R^{ab}{}_{ab},
        &\qquad \kappa_{\langle ab\rangle}
        &= \mathscr R^0{}_{\langle a|0|b\rangle},
        &\qquad \kappa_{[ab]}
        &= -\tfrac12\Omega_{ab},
    \end{alignat}
\end{subequations}
in terms of the components of the curvature tensor and the rheotactic Weyl curvature. The relevant contractions of the spin-connection curvature tensor are found to be \begin{subequations}\label{eq:contractions riemann rheotactic}
\begin{align}
	\mathscr R^{0a}{}_{0a} &= \tfrac{1}{c^2} \xi_{ab} \xi^{ab} - 2c^2 (\ast\varpi)^2, \\
	\mathscr R^{0b}{}_{ab} &= - \tfrac{1}{c^2} \bar{\mathscr D}{}^b \xi_{ab} - \varepsilon_{ab} \bar{\mathscr D}{}^b (\ast\varpi),\\
	\mathscr R^{ab}{}_{ab} &= 2\bar{\mathscr K} - \tfrac{1}{c^2} \xi_{ab} \xi^{ab} + 2 c^2 (\ast\varpi)^2, \\
	\mathscr R^0{}_{\langle a|0|b\rangle} &= \tfrac{1}{c^2} \bar{\mathscr D}_\bot \xi_{ab}.
\end{align}
\end{subequations} 
We can now compute the Cotton tensor, presented in a manifestly Weyl-covariant fashion as
\begin{subequations}
\begin{align}
	\vect R[\gen K]^0 &= \bar{\pmb{\mathscr D}} \kappa^0{}_0 \wedge \bftau + \bar{\pmb{\mathscr D}} \kappa^0{}_a \wedge \htheta^a + \big( \kappa^a{}_b - \delta^a{}_b \kappa^0{}_0 \big) \bfgamma^0{}_a \wedge \htheta^b + 2 \kappa_{(0a)} \bfgamma^{0a} \wedge \bftau, \\
	\vect R[\gen K]^a &= \bar{\pmb{\mathscr D}} \kappa^a{}_0 \wedge \bftau + \bar{\pmb{\mathscr D}} \kappa^a{}_b \wedge \htheta^b - c^2 \big( \kappa^a{}_b - \delta^a{}_b \kappa^0{}_0 \big) \bfgamma^{0b} \wedge \bftau - \kappa_{0b} \bfgamma^{0a} \wedge \htheta^b - \kappa^a{}_0 \bfgamma^0{}_b \wedge \htheta^b.
\end{align}
\end{subequations}
Identifying the relevant coefficients and dualising as in Eq. \eqref{eq:Cotton dualised}, we recover the scalar $C^{00}$, vector $C^{a0}$ and tracefree-tensorial $C^{\langle ab\rangle}$ components of the Cotton according to the transverse decomposition with respect to the congruence $\bfupsilon$. We find
\begin{equation}
\begin{split}
	C^{00} &= 4c^2 (\ast\varpi)^3 + \big(\bar{\mathscr D}_a\bar{\mathscr D}{}^a+2\bar{\mathscr K}\big)(\ast\varpi) \\
	&\quad + \frac{1}{c^2}\left( \bar{\mathscr D}_a\bar{\mathscr D}_b\big({\ast}\xi^{ab}\big) - 2 (\ast\varpi) \xi_{ab}\xi^{ab}\right) + \frac{1}{c^4}\big({\ast}\xi^{ab}\big)\bar{\mathscr D}_\bot \xi_{ab} 
\end{split} \label{eq:C00 bar}
\end{equation}
for the scalar component,
\begin{equation}
\begin{split}
	C^{a0} &= 2c^2\,{\ast}\bar{\mathscr D}{}^a (\ast\varpi)^2 + \frac{1}{2} {\ast} \bar{\mathscr D}{}^a\bar{\mathscr K} + \frac{1}{2}\bar{\mathscr D}{}^a\bar{\mathscr A} - 3(\ast\varpi)\bar{\mathscr D}_b\xi^{ab} + \xi^{ab}\bar{\mathscr D}_b(\ast\varpi) - 2(\ast\varpi)\bar{\mathscr R}{}^a \\
	&\quad -\frac{1}{c^2}\left(\bar{\mathscr D}_b\bar{\mathscr D}_\bot(\ast\xi^{ab}) + (\ast\xi^{ab})\bar{\mathscr D}{}^c\xi_{bc} + \frac{1}{4}{\ast}\bar{\mathscr D}{}^a(\xi_{bc}\xi^{bc})\right)
\end{split} \label{eq:Ca0 bar}
\end{equation}
for the vector components, and
\begin{equation}
\begin{split}
	C^{\langle ab\rangle} &= c^2\big(2{\ast}\xi^{ab}(\ast\varpi)^2 - \bar{\mathscr D}{}^{\langle a}\bar{\mathscr D}{}^{b\rangle}(\ast\varpi)\big) + (\ast\varpi)\bar{\mathscr D}_\bot \xi^{ab} + {\ast}\xi^{ab}\bar{\mathscr K} + \frac{1}{2}\xi^{ab}\bar{\mathscr A} \\ 
	&\quad - {\ast}\bar{\mathscr D}{}^{\langle a}\bar{\mathscr R}{}^{b\rangle} - \varepsilon^{c\langle a}\bar{\mathscr D}_c\bar{\mathscr D}_d \xi^{b\rangle d} + \frac{1}{c^2}\big(\bar{\mathscr D}_\bot^2 {\ast}\xi^{ab} - {\ast}\xi^{ab}\xi_{cd}\xi^{cd}\big)
\end{split} \label{eq:Cab bar}
\end{equation}
for the tracefree-tensor components. In the above expressions, when acting on tensor quantities, the operator $\ast$ denotes \textit{transverse dualisation}, that is, the following endomorphism of spaces traceless transverse tensors of arbitrary rank:
\begin{equation}
	\ast v^a = \varepsilon^b{}_a v_b,\quad \ast w_{ab} = \varepsilon^c{}_a w_{cb},
\end{equation}
and similarly for higher-rank tensors. By construction, this endomorphism is anti-involutive and anti-self-adjoint for the horizontal Euclidean scalar product, as it obeys
\begin{equation}
	{\ast} {\ast} v^a = -v^a,\qquad (\ast v^a)w_a = -v^a(\ast w_a),
\end{equation}
upon direct evaluation. 

In the perspective of investigating the Carrollian $c\to 0$ limit of the expressions \eqref{eq:C00 bar}--\eqref{eq:Cab bar}, we note that there are hidden factors of $c^2$ in the projected derivatives $\bar{\mathscr D}_\bot$ because the connection coefficients ${\gamma}^a{}_b$ themselves contain such powers of $c^2$ involving the non-integrability two-form, see Eq. \eqref{eq:gamma split rheo}. To remedy this and offer an explicit expansion in powers of $c^2$ of the Cotton tensor, we can trade the projected connection for the horizontal spin connection $\hat\nabla$ defined in \cite{Campoleoni:2023fug}, such that
\begin{equation}
	\hat{\gamma}{}^a{}_b \coloneqq {\gamma}^a{}_b + c^2\varpi^a{}_b,\qquad \hat{\gamma}{}^a{}_{bc} \coloneqq {\gamma}^a{}_{bc},
\end{equation}
up to Weyl-improvement terms in the purely spatial part to ensure Weyl covariance of the spin connection. We see that only the vertical derivative gets modified and the two horizontally projected connections obviously coincide in the $c\to 0$ if the latter is taken while keeping $\varpi_{ab}$ fixed. In particular, $\hat{\mathscr K} = \bar{\mathscr K} - 2c^2(\ast\varpi)^2$, but $\hat{\mathscr A} = \bar{\mathscr A}$ and $\hat{\mathscr R}_a = \bar{\mathscr R}_a$ since the Weyl connection has not been changed. The properties of the horizontal connection $\hat{\nabla}$ have been extensively discussed in the above reference and shall not be repeated here for the sake of brevity. Moreover, they can easily be deduced from the various expressions above. Replacing the time-derivative operator in the explicit expressions of the Cotton tensor yields the following fragmentation:
\begin{equation}
	\begin{split}
		C^{00} &\coloneqq c^2 C_{(-1)} + C_{(0)} + \frac{1}{c^2} C_{(1)} + \frac{1}{c^4} C_{(2)}, \\
		 C^{a0} &\coloneqq c^2  \psi^a +  \chi^a + \frac{1}{c^2} z^a,\qquad  C^{\langle ab\rangle} \coloneqq -c^2  \Psi^{ab} -  X^{ab} - \frac{1}{c^2} Z^{ab},
	\end{split}\label{eq:cotton in c2}
\end{equation}
which defines four scalars:
\begin{equation}
\begin{aligned}
	C_{(-1)} &= 8(\ast\varpi)^3,\quad &C_{(0)} &= \big(\hat{\mathscr D}_a\hat{\mathscr D}{}^a + 2\hat{\mathscr K}\big)(\ast\varpi),\\ 
	C_{(1)} &= \hat{\mathscr D}_a\hat{\mathscr D}_b\big({\ast}\xi^{ab}\big),\quad &C_{(2)} &= {\ast}\xi^{ab}\hat{\mathscr D}_\bot \xi_{ab},
\end{aligned} \label{eq:Cotton Cs}
\end{equation}
three vectors:
\begin{equation}
\begin{split}
	\psi^a &= 3\, {\ast}\hat{\mathscr D}^a(\ast\varpi)^2,\\
	\chi^a &= \frac12 \big({\ast}\hat{\mathscr D}^a \hat{\mathscr K} + \hat{\mathscr D}^a \hat{\mathscr A}\big) - 2(\ast\varpi)\big(\hat{\mathscr R}^a + 2 \hat{\mathscr D}_b\xi^{ab}\big) + 3 \hat{\mathscr D}_b\big({\ast}\varpi\xi^{ab}\big), \\
	z^a &= -\hat{\mathscr D}_b\hat{\mathscr D}_\bot \big({\ast}\xi^{ab}\big) - ({\ast}\xi^a{}_b)\hat{\mathscr D}_c\xi^{bc} - \frac14 {\ast}\hat{\mathscr D}{}^a\big(\xi_{bc}\xi^{bc}\big),
\end{split}\label{eq:Cotton psi chi z}
\end{equation}
and three tracefree rank-two tensors:
\begin{equation}
	\begin{split}
		\Psi^{ab} &= \hat{\mathscr D}^{\langle a}\hat{\mathscr D}{}^{b\rangle} (\ast\varpi) - 2 (\ast\varpi)^2 {\ast}\xi^{ab}, \\
		X^{ab} &= \frac12 \varepsilon^{ca}\hat{\mathscr D}_c \big( \hat{\mathscr R}{}^b + \hat{\mathscr D}_d \xi^{bd}\big) + \frac12 \varepsilon^{cb}\hat{\mathscr D}{}^a \big(\hat{\mathscr R}_c + \hat{\mathscr D}{}^d \xi_{cd}\big) - \frac32 \xi^{ab}\hat{\mathscr A} - {\ast}\xi^{ab}\hat{\mathscr K} + 3 (\ast\varpi) \hat{\mathscr D}_\bot\xi^{ab}, \\
		Z^{ab} &= -\hat{\mathscr D}_\bot^2 \xi^{ab} + {\ast}\xi^{ab}\big(\xi_{cd}\xi^{cd}\big).
	\end{split}\label{eq:Cotton Psi Chi Z}
\end{equation}
Therefore, we verify that our gauging procedure reproduces the expressions of the Cotton established in Ref.~\cite{Campoleoni:2023fug} by direct evaluation, up to a conventional sign difference in defining the tracefree transverse tensor components.

\begin{description}
	\item[Remark.] We have only displayed a set of independent components of the Cotton tensor, but all of them can be computed in our time-split formalism and one is able to check the Bianchi identities \eqref{eq:cotton is symm} and \eqref{eq:cotton is TF} explicitly. Namely, one can show that $C^a{}_a = c^2 C^{00} - (\bar{\mathscr D}_\bot \bar{\mathscr A} + {\ast}\bar{\mathscr D}_a\bar{\mathscr R}{}^a)$, and the two last terms cancel out by virtue of the Bianchi identities $\D^2\bfalpha = \bm 0$ related to the Weyl curvature. Therefore, $C^a{}_a + C^0{}_0 = 0$ as expected. Furthermore, one can also compute $C^{[a,b]}$ and show that it vanishes on account of the Bianchi identities for the projected curvature $\bar{\pmb{\mathscr R}}{}^a{}_b$ and the Bianchi identities involving the Weyl curvature. Finally, to show that $C^{a0} = C^{0a}$ requires using all these identities and the commutation properties \eqref{eq:commu curly D bar}.
\end{description}

\subsection{Gravitational Chern--Simons action}
\label{sec:relat chern simons}

In this last part, we review how to derive the three-dimensional Cotton tensor from variational methods through the gravitational Chern--Simons action.

First, it is a well-known story that one can compute the covariant or Einstein--Hilbert energy--momentum tensor of any given theory of fields by coupling it minimally to a curved background and varying with respect to this background. Let $\phi$ be the dynamical fields and $(\mathscr M,\bfg)$ a given background manifold. Varying the action $S[\phi,\bfg]$, resulting from the minimal coupling, gives
\begin{equation}
	\delta S = \int_{\mathscr M} \D^{d+1} x\,\sqrt{-g} \left(\mathscr E_\phi \delta\phi + \tfrac12 T^{\mu\nu} \delta g_{\mu\nu}\right)
\end{equation}
after discarding possible boundary terms, where $\{x^\mu\}$ a given coordinate patch. Setting the dynamical fields $\phi$ on-shell, \textit{i.e.}, assuming that they solve $\mathscr E_\phi = 0$, the first variation of the action provides the energy--momentum tensor $T^{\mu\nu}$ of the theory. By construction, it is symmetric and conserved by virtue of diffeomorphism invariance of the minimally coupled action. Furthermore, it is also traceless by virtue of Weyl invariance if the theory is conformal. It is crucial to distinguish dynamical fields from the background ones for this procedure to be well-defined. 

We now apply similar considerations to the Chern--Simons action gauging the full conformal algebra in three dimensions, $\mathfrak{so}(3,2)$:
\begin{equation}
	S_\text{CS}[\bm\Gamma] = \frac{1}{2c} \int_{\mathscr M} \text{CS}[\bm\Gamma],\qquad \text{CS}[\bm\Gamma]\coloneqq \text{Tr}\big[\bm\Gamma \wedge \D \bm\Gamma + \tfrac23 \bm\Gamma \wedge \bm\Gamma \wedge \bm\Gamma \big]. \label{eq:CS}
\end{equation}
In this equation, $c$ is again the speed of light, $\bm\Gamma$ was defined in Eq.~\eqref{eq:Gamma} and $\text{Tr}$ denotes the non-degenerate symmetric bilinear form associated with the conformal algebra. Since $\mathfrak{so}(3,2)$ is simple, this bilinear form is unique, up to normalisation, and coincides with the Killing form $G : \mathfrak{so}(3,2)\times \mathfrak{so}(3,2)\to \mathbb R$. The only non-vanishing pairings through the Killing form are found to be \cite{Horne:1988jf}
\begin{equation}
	G(\gen J_{AB},\gen J_{CD}) = 2(\eta_{AD}\eta_{BC}-\eta_{AC}\eta_{BD}),\qquad G(\gen P_A,\gen K_B) = -4\eta_{AB},\qquad G(\gen D,\gen D) = 2, \label{eq:GforConformal}
\end{equation}
see Appendix \ref{sec:invariantbilinear} for more details. In particular, the first relation showcases the canonical Killing form on the Lorentz algebra. Recalling that the wedge product in Eq. \eqref{eq:CS} has been defined in Eq. \eqref{eq:wedge gamma gamma}, this action reads
\begin{equation} \label{eq: Chern-Simons conformal}
	S_{\text{CS}}[\bm\Gamma] = \frac{1}{2c} \int_{\mathscr M} \big[ \text{CS}[\bfgamma] + 2 \bfalpha \wedge \D \bfalpha - 4 \bfkappa_A \wedge \big( \D \htheta^A + \bfgamma^A{}_B \wedge \htheta^B - \bfalpha \wedge \htheta^A \big)  \big] 
\end{equation}
in terms of the different gauge fields, where $\text{CS}[\bfgamma] =  \bfgamma^A{}_B \wedge \D \bfgamma^B{}_A + \tfrac23 \bfgamma^A{}_B \wedge \bfgamma^B{}_C \wedge \bfgamma^C{}_A$ through the canonical Killing form on the Lorentz algebra. In the above, we discarded the boundary term
\begin{equation}
	\frac{1}{c}\oint_{\partial\mathscr M} \htheta^A\wedge\bfkappa_A. \label{eq:boundary term lorentz}
\end{equation}
Equations of motion obtained by variation with respect to the fields $\htheta^A$, $\bfgamma^A{}_B$, $\bfalpha$ and $\bfkappa^A$ are respectively $\vect R[\gen K]^A = \bm 0$, $\vect R[\gen J]^{AB} = \bm 0$, $\vect R[\gen D] = \bm 0$ and $\vect R[\gen P]^A = \bm 0$. These are the equations of motion of three-dimensional conformal gravity \cite{Horne:1988jf}. Therefore, when all the Chern--Simons equations of motion are satisfied, the manifold $\mathscr M$ is endowed with a conformally flat connection. 

In order to derive the canonical momentum as above, we need to distinguish between background and dynamical fields. We shall therefore impose the equations of motion for the latter and report their solution back into the Chern--Simons action to find out what remains. We refer to the action resulting from this procedure as the reduced action. For consistency, we consider again the geometry as background, which means that we shall not impose equations of motion obtained by variation of $S_{\text{CS}}$ with respect to $\htheta^A$. Requiring strong Lorentz invariance, by which we mean that $\delta_L S_{\text{CS}} = 0$ irrespective of the boundary conditions on $\partial\mathscr M$, the reduced action will in fact depend on the metric $\bfg = \eta_{AB}\htheta^A\otimes\htheta^B$ built up from the Cartan coframe $\htheta^A$. Moreover, by Weyl invariance, it will only depend on the Weyl class of the metric $\bfg$. Equivalently, it will be assumed to depend on the couple $(\bfg,\bfalpha)$, which encompasses the same pieces of information. We now recall that imposing only the equations of motion for $\bfgamma^A{}_B$ and $\bfkappa^A$, \textit{i.e.}, $\vect R[\gen J]^{AB}=\bm 0$ and $\vect R[\gen P]^A = \bm 0$, has the effect of fixing the spin connection $\bfgamma^{AB}$ as in Eq. \eqref{eq:gamma for WLC} and the tensor $\bfkappa^A$ as
\begin{equation}
	\kappa_{(AB)} = \text{Sch}[\bfgamma]_{AB} = \text{Ric}[\bfgamma]_{(AB)} - \tfrac14 R[\bfgamma]\eta_{AB},\qquad \kappa_{[AB]} = \text{Ric}[\bfgamma]_{[AB]}.
\end{equation}
in terms of $\htheta^A$ and $\bfalpha$, see Eqs. \eqref{eq:Ric antisym} and \eqref{eq:kappa = schouten}. Moreover, we recall that $\vect R[\gen D]=\bm 0$ is automatically satisfied, and thus $\kappa_{[AB]} = -\mathring \nabla_{[A}\alpha_{B]}$. This is in line with the fact that $\bfalpha$ shall remain completely free.\footnote{We stress that imposing the whole set of Chern--Simons equations of motion still leaves $\bfalpha$ completely free. This has to be expected, since the latter is the only gauge field that transforms \textit{algebraically} under a gauge transformation, namely the special conformal transformations, and cannot therefore acquire an on-shell value without breaking this symmetry sector. Of course, gauge fields transforming with derivatives of the gauge parameters are still allowed to have non-trivial equations of motion since the underlying symmetry is protected by the existence of Noether identities.} The Weyl one-form $\bfalpha$ therefore appears as a redundant pure-gauge degree of freedom, that is however instrumental to ensure manifest Weyl covariance. Considering it as background or dynamical field is left to one's appreciation and the two viewpoints are in fact equivalent. Here, we have chosen to treat $\bfalpha$ as dynamical, in the sense that $\vect R[\gen D]=\bm 0$ is imposed, albeit as an identity. This allows us to retain only $\htheta^A$ as the background geometry.

Consequently, the Chern--Simons action \eqref{eq:CS} then reduces to
\begin{equation}
	S_{\text{CS}}^{\text{red}}[\htheta^A] = S_{\text{gCS}}[\htheta^A] - \frac{1}{2c}\int_{\partial\mathscr M} \alpha_A\D\htheta^A, \label{eq:redCS relat}
\end{equation}
and therefore corresponds, up to an irrelevant boundary term, to the \textit{gravitational Chern--Simons action} for the Levi-Civita spin connection:
\begin{equation}
	S_{\text{gCS}}[\htheta^A] = \frac{1}{2c}\int_\mathscr{M} \text{CS}[\mathring\bfgamma] = \frac{1}{2c}\int_{\mathscr M} \big[\mathring\bfgamma^A{}_B\wedge\D\mathring\bfgamma^B{}_A + \tfrac23 \mathring\bfgamma^A{}_B\wedge\mathring\bfgamma^B{}_C\wedge\mathring\bfgamma^C{}_A\big]. \label{eq:grav CS}
\end{equation}
The latter is a third-order action for the geometry, \textit{i.e.}, for the Cartan coframe $\htheta^A$, since $\mathring\bfgamma^A{}_B$ has to be considered as a function of (derivatives of) $\htheta^A$, given in components by Eq. \eqref{eq:levi civita}. The demonstration of Eq. \eqref{eq:redCS relat} relies on the \textit{transgression formula} that relates two Chern--Simons forms:
\begin{equation}
	\begin{split}
		\text{CS}[\mathring\bfgamma+\bm\Phi] - \text{CS}[\mathring\bfgamma] &= 2 \bm\Phi^A{}_B\wedge \vect R[\mathring\bfgamma]^B{}_A + \bm\Phi^A{}_B\wedge\mathring{\bm\nabla}\bm\Phi^B{}_A \\
		&\quad +\tfrac23 \bm\Phi^A{}_B\wedge\bm\Phi^B{}_C\wedge\bm\Phi^C{}_A - \D(\mathring\bfgamma^A{}_B\wedge\bm\Phi^B{}_A),
	\end{split}
\end{equation}
where $\bm\Phi^{AB} = \alpha^A\htheta^B - \eta^{AB}\bfalpha$ in the present case. The first term vanishes because $\vect R[\mathring\bfgamma]^{AB}$ is skew-symmetric in $(A,B)$ and obeys the Bianchi identity $ \vect R[\mathring\bfgamma]^A{}_B\wedge\htheta^B = \bm 0$. Next, a direct computation shows that $\bm\Phi^A{}_B\wedge\mathring{\bm\nabla}\bm\Phi^B{}_A = (d-2)\bfalpha\wedge\D\bfalpha = \bm 0$ in three dimensions. Finally, the cubic term in $\bm\Phi^A{}_B$ vanishes algebraically and the boundary term is reworked using the fact that the Levi-Civita spin connection is torsion-free by definition, $\D\htheta^A + \mathring{\bfgamma}^A{}_B\wedge\htheta^B = \bm 0$.

Now that the sequence of equations of motion that can be imposed has been discussed and the end point of the reduction has been clarified, we can go back to the original action \eqref{eq: Chern-Simons conformal}, impose stationarity of the action with respect to dynamical fields, by which we mean all gauge fields except $\htheta^A$, and then perform a variation with respect to the background fields, to finally obtain:
\begin{equation}
	\boxed{\delta S_\text{CS}^{\text{red}}[\htheta^A] = -\frac{2}{c}\int_{\mathscr M} \delta \htheta^A \wedge \vect R[\gen K]_A = -\frac1c\int_{\mathscr M} \varepsilon_{ABD}C_{C}{}^D \htheta^A\wedge\htheta^B\wedge\delta\htheta^C,} 
\end{equation}
providing a variational definition of the Cotton tensor. This observation motivates our discussions about the Carrollian avatars of the gravitational Chern--Simons action in Section \ref{sec:ChernCarroll}, see also \cite{Miskovic:2023zfz}. It is worth noticing that the boundary term \eqref{eq:boundary term lorentz} does not play any role since it reduces to a total derivative when $\bfkappa^A$ is on-shell, by virtue of Eq. \eqref{eq:antisymm schouten}. It therefore leads to a vanishing integral if $\partial\mathscr M$ is assumed to have no boundary. In passing, we have recovered the well-known fact that the energy--momentum tensor deriving from the gravitational Chern--Simons action is the Cotton tensor, since $\delta S_\text{CS}^{\text{red}}[\htheta^A] = \delta S_{\text{gCS}}[\htheta^A]$ after discarding a boundary term, which does not affect the definition of the Cotton tensor. This concludes our discussion of the Lorentzian conformal case: our attention now turns to the Carrollian case. 

\section{Cotton tensor from gauging methods: the Carrollian case} \label{sec:Carroll}

In this section, we derive the Carroll--Cotton tensor by gauging the conformal Carroll algebra, which is isomorphic to the Poincar\'e algebra in one dimension higher. This observation might suggest that little can be learnt from such a gauging, since the corresponding procedure for the Poincar\'e algebra was developed long ago and provides a straightforward derivation of the first-order formulation of gravity. However, interpreting the Poincar\'e algebra as a conformal algebra in one dimension lower imposes entirely different conditions on the gauging procedure.

\subsection{Gauging the conformal Carroll algebra}
\label{sec:gauging Carrollian}

\subsubsection*{The conformal Carroll algebra}

Conformal Carroll algebras have been defined by Duval, Gibbons and Horvathy in their seminal works \cite{Duval:2014lpa,Duval:2014uva} as the algebra of conformal isometries of the flat Carroll manifold $(\mathscr C_0,\bfg,\bfupsilon)$. The latter is defined as the $c\to 0$ limit of the special-relativistic flat manifold $(\mathscr M_0,\bfg)$ defined by Eq. \eqref{eq:flat mink}, that is
\begin{equation}
	\D s^2 = 0\times\D t^2 + \delta_{ab}\D x^a\D x^b,\qquad \bfupsilon = \partial_t,\qquad \bfg(\bfupsilon,\cdot)= \bm 0, \label{eq:flat carroll}
\end{equation}
in Cartesian coordinates $(x^A) = (t,x^a)$, where $a=1,\dots,d$. Owing to the degeneracy of the metric $\bfg$ in the direction of the field of observers $\bfupsilon$, these two quantities may scale independently under conformal transformations. The conformal isometries of \eqref{eq:flat carroll} form a one-parameter family of algebras, denoted by $\mathfrak{confcarr}_z^\star(d,1)$, generated by infinitesimal diffeomorphisms satisfying
\begin{equation}
	\mathscr L_{\vect V}\bfg = \frac{2}{d}(\bm\partial\cdot\vect V)\bfg,\qquad \mathscr L_{\vect V}\bfupsilon = -\frac{z}{d}(\bm\partial\cdot\vect V)\bfupsilon, \label{eq:conf carroll general}
\end{equation}
where $z\in\mathbb R$ is sometimes referred to as the \textit{dynamical exponent}. Since the flat background is an invariant Carroll structure, $\mathscr L_{\bfupsilon}\bfg = \bm 0$, the set of solutions of \eqref{eq:conf carroll general} is infinite-dimensional, as it contains the so-called supertranslations: $\vect V = f\bfupsilon$ with $\mathscr L_\bfupsilon f = 0$. Requiring that the flow of $\vect V$ also preserves the flat connection on $\Gamma(T\mathscr C_0)$ \textit{up to a trace} breaks supertranslations down to a finite subgroup, \textit{i.e.}, invariant functions $f$ that are at most quadratic in $x$.\footnote{In the non-conformal case, vertical diffeomorphisms generated by $\vect V = f\bfupsilon$, with $\mathscr L_\bfupsilon f = 0$, that preserve a choice of connection on $\Gamma(T\mathscr C_0)$ are at most \textit{linear} in $x$. In the conformal case, one must allow homogeneous rescalings of the metric in general, which induces non-trivial trace terms in the transformation of purely spatial connection coefficients, as we will see later on. Therefore, the function $f$ can be at most quadratic in $x$.} We denote $\mathfrak{confcarr}_z(d,1)$ these finite-dimensional algebras \cite{Afshar:2024llh}. Furthermore, the $z=1$ case is special as it represents the isotropic configuration in which the field of observers transforms with weight one under conformal transformations. This occurs both on null hypersurfaces\footnote{Every null hypersurface carries a Carrollian geometry, but only exceptional null hypersurfaces realise its full isotropic conformal symmetry. Such conformal symmetry arises on distinguished geometries, notably the light cone and null infinity in conformal compactifications of spacetimes satisfying suitable asymptotic conditions. In these settings, the conformal Carroll algebra is closely intertwined with the BMS algebra \cite{Duval:2014uva,Ciambelli:2019lap}.} embedded into Lorentzian spacetimes and on Carrollian limits of Lorentzian conformal manifolds. 

In the following, we shall focus on the finite-dimensional isotropic Carroll conformal algebra $\mathfrak{confcarr}_1(d,1)$, which is realised on $\mathscr C_0$ by the following vector fields
\begin{equation}
\begin{array}{lclcl}
	\vect V_{\gen P_a} = \partial_a, \ \vect V_{\gen H} = \partial_t , &\qquad& \vect V_{\gen J_{ab}} = x_b\partial_a - x_a\partial_b, &\qquad& \vect V_{\gen B_a} = -x_a \partial_t, \\
	\vect V_{\gen D} = x^a\partial_a + t \partial_t, &\qquad& \vect V_{\gen K_a} = 2 x_a x^b\partial_b - x^b x_b \partial_a  + 2 x_a t \partial_t , &\qquad& \vect V_{\gen K_0} = -x^a x_a \partial_t .
\end{array} \label{eq:carr diff realisation}
\end{equation}
The generators displayed in the first line represent space and time translations, spatial rotations and Carroll boosts, respectively. They are isometries of the flat Carroll manifold, \textit{i.e.}, for which Eqs. \eqref{eq:conf carroll general} have no right-hand side. Those displayed in the second line are pure conformal extensions of the Carroll isometry algebra, in the sense that $\bm\partial\cdot\vect V\neq\bm 0$ in Eqs. \eqref{eq:conf carroll general}. They generate respectively dilations, space and time special conformal transformations. Borrowing notations from the previous Section, the mapping $\mathfrak{confcarr}_1(d,1) \to \Gamma(T\mathscr C_0) : \gen X\mapsto \vect V_{\gen X}$ is a Lie algebra isomorphism. The non-trivial Lie brackets among our chosen basis generators, $\{\gen X_I\} = \{\gen P_a,\gen H,\gen J_{ab},\gen B_a,\gen D,\gen K_a,\gen K_0\}$, are given by
\begin{equation}
\begin{array}{lclcl}
	\bk{\gen D,\gen P_a} = \gen P_a, &\quad& \bk{\gen D,\gen K_a} = -\gen K_a, &\quad& \bk{\gen K_a,\gen P_b} = 2\delta_{ab}\gen D - 2 \gen J_{ab}, \\
	\bk{\gen J_{ab},\gen P_c} = 2 \delta_{c[b}\gen P_{a]}, &\quad& \bk{\gen J_{ab},\gen K_c} = 2 \delta_{c[b}\gen K_{a]}, &\quad& \bk{\gen J_{ab},\gen B_c} = 2 \delta_{c[b}\gen B_{a]}, \\
	\bk{\gen D,\gen H} = \gen H, &\quad& \bk{\gen D,\gen K_0} = - \gen K_0, &\quad& \bk{\gen J_{ab},\gen J_{cd}} = 2\delta_{a[c} \gen J_{d]b} - 2\delta_{b[c} \gen J_{d]a}, \\
	\bk{\gen B_a,\gen P_b} = \delta_{ab} \gen H , &\quad& \bk{\gen B_a,\gen K_b} = \delta_{ab} \gen K_0, &\quad& \bk{\gen K_0,\gen P_a} = -\bk{\gen K_a,\gen H} = -2 \gen B_a.
\end{array} \label{eq:conformal carroll algebra}
\end{equation}
The special-relativistic conformal algebra $\mathfrak{conf}(d,1)$ and the Carrollian instance we are interested in, $\mathfrak{confcarr}_1(d,1)$, have the same dimensions, $\frac{1}{2}(d+2)(d+3)$. Furthermore, the latter is the \.In\"on\"u--Wigner contraction of the former when $c$, seen as a defining parameter of $\mathfrak{conf}(d,1)$ through the background Minkowski metric \eqref{eq:flat mink}, is sent to zero.\footnote{Comparing Eqs. \eqref{eq:diff realisation} and \eqref{eq:carr diff realisation}, one readily prove that the $c\to 0$ limit only affects the Lorentz-boost generators $\vect V_{\gen J_{0a}}$ and the temporal special conformal generator $\vect V_{\gen K_0}$, transforming them into the Carroll-boost generators $\vect V_{\gen B_a}$ and the corresponding special conformal generator in the Carroll limit. Alternatively, one can replace $\gen J_{0a}\to c \gen B_a$ in the commutators \eqref{eq:conformal alg} and then perform the $c\to 0$ limit to get the commutators \eqref{eq:conformal carroll algebra}.} The isotropic conformal Carroll algebra is also isomorphic to the Poincaré algebra in one dimension higher \cite{Duval:2014uva,Duval:2014lpa}:
\begin{equation}
	\lim_{c\to 0}\mathfrak{conf}(d,1) \equiv \mathfrak{confcarr}_1(d,1)\simeq \mathfrak{iso}(d+1,1),
\end{equation}
which decomposes into Lorentz transformations $\mathfrak{so}(d+1,1)$ and translations $\mathfrak t\simeq \mathbb R^{d+1,1}$ as
\begin{equation}
	\mathfrak{so}(d+1,1) = \text{Span}\lbrace \gen J'{}_{MN}\rbrace = \text{Span}\lbrace \gen P_a,\gen J_{ab},\gen D,\gen K_a\rbrace,\quad \mathfrak{t} = \lbrace \gen P'{}_M\rbrace = \text{Span}\lbrace \gen H,\gen B_a,\gen K_0\rbrace. \label{eq:decomposotion poincare}
\end{equation}
where $M,N = 0,\dots,d+1$. The precise isomorphism is given by the following linear combinations \cite{Donnay:2022wvx,Nguyen:2023vfz}
\begin{equation}
	\begin{split}
		\gen P'{}_0 &= \tfrac{1}{\sqrt 2}(\gen K_0+\gen H),\quad \gen P'{}_a = \sqrt 2 \gen B_a,\quad \gen P'{}_{d+1} = \tfrac{1}{\sqrt 2}(\gen K_0 - \gen H), \\
		\gen J'{}_{ab} &= \gen J_{ab},\quad \gen J'{}_{a0} = \tfrac12(\gen P_a+\gen K_a),\quad \gen J'{}_{a(d+1)} = \tfrac12(\gen P_a-\gen K_a),\quad \gen J'{}_{0(d+1)} = \gen D,
	\end{split} \label{eq:carroll to poin gens}
\end{equation}
and the generators $\gen J'{}_{MN}$ and $\gen P'{}_M$ reproduce the usual non-vanishing commutation relations
\begin{equation}
	\big[\gen J'{}_{MN},\gen P'{}_P\big] = 2 \eta_{P(M}\gen P'{}_{N)}, \quad \big[\gen J'{}_{MN},\gen J'{}_{PQ}\big] = 2\eta_{M[P}\gen J'{}_{Q]N} - 2\eta_{N[P}\gen J'{}_{Q]M}.
\end{equation}
of the $(d+1)$-dimensional Poincaré algebra.

\subsubsection*{Gauge connections}

We now discuss the gauging of the isotropic finite-dimensional conformal Carroll algebra. Since a substantial amount of details on the special-relativistic case has already been given in Section \ref{sec:gauging Lorentzian} and many technical steps are essentially the same, the presentation will be shorter, although crucial differences and technical subtleties shall be duly stressed. From now on, we consider a generically curved Carrollian manifold $(\mathscr C,\bfg, \bfupsilon)$, with the only constitutive relation that $\bfg(\bfupsilon,\cdot)=\bm 0$. Due to the universal character of the field of observers $\bfupsilon$, such a manifold is endowed with the structure of a fibre bundle $\mathscr C\to\mathscr S$, where $\mathscr S$ is a $d$-dimensional Riemannian manifold \cite{Bekaert:2015xua,Ciambelli:2019lap}. 

On $T\mathscr C$, we introduce a gauge connection $\bm\Gamma$ for $\mathfrak{confcarr}_1(d,1)$ as
\begin{equation}
	\bm\Gamma \coloneqq \gen P_a\htheta^a + \gen H \bftau + \tfrac12 \gen J_{ab}\bfgamma^{ab} + \gen B_a \bfbeta^a - \gen D\bfalpha + \tfrac12 \gen K_a\bfkappa^a +  \tfrac12 \gen K_0 \bfkappa^0. \label{eq:Gamma carroll}
\end{equation}
The $(d+1)$-dimensional set $\{\bftau,\htheta^a\}$ of one-form fields is the Carrollian analogue of the Cartan coframe \cite{Campoleoni:2023fug}. In particular, $\bftau$ is an Ehresmann connection that allows for the split of local tangent spaces of the fibre bundle $\mathscr C\to\mathscr S$ into vertical and horizontal components. Its `normalisation' is conventionally chosen such that $\bftau(\bfupsilon) = 1$. At each point $P\in\mathscr C$, one defines $\text H_P(\mathscr C)\coloneqq \ker(\bftau_P)$, the local horizontal space. This is the closest analogue to the local space of simultaneity encountered in Lorentzian physics. A basis of horizontal vectors is provided by a complete set of linearly independent vectors $\hframe_a$ that obey $\bftau(\hframe_a) = 0$, which we will choose as orthonormal with respect to the ambient metric: $\bfg(\hframe_a,\hframe_b)=\delta_{ab}$. Therefore, an orthonormal basis of $\Gamma(T\mathscr C)$ is given by $\{\bfupsilon,\hframe_a\}$, which we shall refer to as a Carroll--Cartan frame. By construction, its dual coframe is given by $\{\bftau,\htheta^a\}$ provided we impose that $\htheta^a(\hframe_b) = \delta^a{}_b$. The other gauge fields introduced in Eq. \eqref{eq:Gamma carroll} are easy to identify. On the one hand, the Lorentzian spin connection $\bfgamma^{AB}$ splits in the limit into a purely spatial spin connection $\bfgamma^{ab}$ and a Carroll-boost connection $\bfbeta^a$. The latter plays an important role when (conformal) Carroll geometries are considered for applications in gravitational theories. In particular, expanding
\begin{equation}
	\bfbeta^a = \beta^a\bftau + \beta^a{}_b\htheta^b, \qquad \beta_{(ab)} \coloneqq \tfrac1d \beta \delta_{ab} - \tfrac12 \mathscr C_{ab}, \label{eq:separation beta}
\end{equation}
in the Carroll--Cartan coframe, the tracefree tensor $\mathscr C_{ab}$, $\delta^{ab}\mathscr C_{ab} = 0$, represents, in the $d=2$ case, the transverse shear induced by gravitational waves crossing $\mathscr C$ \cite{Fiorucci:2025twa} (see also \cite{Geroch:1977big,Ashtekar:1981hw} for earlier references), seen as a null hypersurface embedded into a higher-dimensional Lorentzian manifold. On the other hand, purely conformal generators come with their own connections: $\bfalpha$ is the Weyl connection while $\bfkappa^0$ and $\bfkappa^a$ are resulting from the time-space split of the Lorentzian special conformal connection $\bfkappa^A$.

Gauge transformations are again defined as the point-isotropic transformations of the isotropic finite-dimensional conformal Carroll algebra realised locally as
\begin{equation}
	\delta_\Sigma \bm\Gamma = \D\Sigma + \bk{\bm\Gamma,\Sigma},\qquad \Sigma \coloneqq -\tfrac{1}{2} \lambda^{ab} \gen J_{ab} + \lambda^a \gen B_a + B \gen D + \tfrac12 Z^a \gen K_a + \tfrac12 Z^0 \gen K_0 .\label{eq:Sigma carr}
\end{equation}
Under these gauge transformations, the various gauge fields that we have introduced transform as
\begin{subequations}\label{eq:transfo gauge fields carroll}
\begin{align}
	\delta_\Sigma \htheta^a &= \lambda^a{}_b \htheta^b - B\htheta^a , \label{eq:transfo theta a}\\
	\delta_\Sigma \bftau &= -\lambda_a \htheta^a - B\bftau , \label{eq:transfo tau}\\
	\delta_\Sigma \bfgamma^{ab} &= -\horbmnabla \lambda^{ab} + 2 Z^{[a}\htheta^{b]}, \label{eq:transfo gamma carr} \\
	\delta_\Sigma \bfbeta^a &= \horbmnabla \lambda^a + \lambda^a{}_b \bfbeta^b + Z^0 \htheta^a - Z^a \bftau , \label{eq:transfo beta}\\
	\delta_\Sigma \bfalpha &= Z_a\htheta^a -\D B, \label{eq:transfo alpha carr} \\
	\delta_\Sigma \bfkappa^a &= (\horbmnabla + \bfalpha)Z^a + \lambda^a{}_b \bfkappa^a + B \bfkappa^a \label{eq:transfo kappa carr},\\
	\delta_\Sigma \bfkappa^0 &= (\horbmnabla + \bfalpha)Z^0 + B \bfkappa^0 - \lambda_a \bfkappa^a + Z_a \bfbeta^a.
\end{align}
\end{subequations}
In the above equations, $\horbmnabla$ denotes the action of the spin connection on the horizontal space $\text H(\mathscr C)$ built up from $\bfgamma^{ab}$ only, \textit{i.e.},
\begin{equation}
	\horbmnabla f = \D f,\qquad \horbmnabla V^a = \D V^a + \bfgamma^a{}_bV^b,\qquad \horbmnabla W_a = \D W_a - \bfgamma^b{}_a W_b \label{eq:projected carroll conn}
\end{equation}
on any scalar $f$, horizontal vector $V^a$ and one-form $W_a$. Generalisation to higher-rank tensors is straightforward thanks to Leibniz rule. In the $c$-to-zero-limit viewpoint, it corresponds to the projected connection on horizontal spaces, hence the same notation; in particular,
\begin{equation}
	\bfgamma^a{}_b = \gamma^a{}_b \bftau + \gamma^a{}_{cb}\htheta^c.
\end{equation}
The connections on $\Gamma(T\mathscr S)$ built up from $\gamma^a{}_b$ and $\gamma^a{}_{bc}$, will again be denoted as $\bar\nabla_\bot$ and $\bar\nabla_a$ respectively. Transformations \eqref{eq:transfo gauge fields carroll} justify \textit{a posteriori} the qualification of each gauge field present in Eq. \eqref{eq:Gamma carroll}. The gauge parameters $\lambda^a{}_b$ represents an infinitesimal local $\mathfrak{so}(d)$ rotation of the horizontal orthonormal coframe $\htheta^a$. Moreover, the $\lambda_a$ are the components of a transverse one-form field that can induce infinitesimal local Carroll boosts, or Ehresmann gauge transformations, as
\begin{equation}
	\delta_\lambda \bftau = -\lambda_a\htheta^a,\quad \delta_\lambda \htheta^a = \bm 0\qquad\Leftrightarrow\qquad \delta_\lambda \bfupsilon = \bm 0,\quad \delta_\lambda \hframe_a = \lambda_a\bfupsilon.
\end{equation}
These transformations preserve the underlying Carroll structure and only appear when a horizontal slicing has to be defined, hence $\bftau$ to be introduced. Furthermore, $B$ generates a Weyl transformation: Eqs. \eqref{eq:transfo theta a} and \eqref{eq:transfo tau} imply that the Carroll structure transforms as
\begin{equation}
	\delta_\Sigma \bfg = -2B\bfg,\qquad \delta_\Sigma\bfupsilon = B\bfupsilon.
\end{equation}
As in the Lorentzian case, gauging the dilations, and thereby introducing the gauge field $\bfalpha$, endows the background Carroll manifold with a Weyl structure:
\begin{equation}
	\bfg' \in [\bfg]\ \Leftrightarrow\ \bfg' = \mathscr B^{-2}\bfg,\qquad \bfupsilon'\in[\bfupsilon]\ \Leftrightarrow\ \bfupsilon' = \mathscr B\bfupsilon \label{eq:weyl structure carr}
\end{equation}
for a given non-vanishing $\mathscr B\in\mathscr F(\mathscr C)$. The fact that we are gauging the conformal Carroll algebra with $z=1$ imposes that the resulting Weyl structure is also isotropic, \textit{i.e.}, that the relative factor between the Weyl weights of the metric and the field of observers is exactly minus two, as in Eq. \eqref{eq:weyl structure carr}. Finally, functions $Z^0$ and $Z^a$ generate the Carrollian instances of local special conformal transformations that we have already encountered in the Lorentzian case. 

\begin{description}
	\item[Remark.] As in the Lorentzian case, local special conformal transformations will again be central to the algebraic transformation of the Weyl connection $\bfalpha$ in what follows. There is, however, a crucial difference between Eqs.~\eqref{eq:Weyl connection variation} and \eqref{eq:transfo alpha carr}: while the Lorentzian Weyl connection can be fully gauged away by a suitable choice of special-conformal frame, encoded in the parameters $Z^A$, only the transverse part of its Carrollian counterpart can be gauged away by choosing the parameters $Z^a$. Indeed, the transverse projection of Eq. \eqref{eq:transfo alpha carr} yields
\begin{equation}
	\delta_\Sigma \alpha_a = \lambda_a{}^b\alpha_b + B\alpha_a - \hframe_a(B) + Z_a, \label{eq:delta alpha a carroll}
\end{equation}
and its longitudinal projection gives
\begin{equation}
	\delta_\Sigma \alpha_0 = B\alpha_0 - \bfupsilon(B). \label{eq:delta alpha 0 carroll}
\end{equation}
The vertical component $\alpha_0$ is therefore not pure gauge, since $Z^0$ no longer enters its transformation. This important discrepancy is closely related to the construction of the (Weyl) Levi-Civita connection on conformal Carrollian backgrounds, which, as we shall see below, locks $\alpha_0$ to the geometry and therefore forbids to set the form $\bfalpha=\bm 0$ in the presence of non-trivial expansion.
\end{description}

Except in flat space, Carroll--Cartan frames $\{\bftau,\htheta^a\}$, like their Lorentzian counterpart, are not expected to be holonomic. In Section \ref{sec:rheotactic lorentz}, we have prepared the Carrollian limit by performing a time-space split along a timelike congruence $\bfupsilon$. In the limit, the latter becomes the unique field of observers $\bfupsilon$ of the resulting Carroll structure. Therefore, we keep the same notations as Eq. \eqref{eq:nonholo} for the non-holonomy functions of the Carroll--Cartan basis. First, the vertical non-holonomy coefficients reveal the Carrollian acceleration $\varphi_a$ and vorticity $\varpi_{ab}$: the former quantifies the Lie-dragging of $\bftau$ on the flow of the field of observers, while the latter represents the obstruction for $\bftau$ to define an integrable distribution of horizontal spaces in $T\mathscr C$. They transform under local symmetries as
		\begin{equation}
			\begin{split}
				\delta_\Sigma \varphi_a &= \lambda_a{}^b \varphi_b + B\varphi_a - \hframe_a(B) + \bfupsilon(\lambda_a) + c^b{}_a\lambda_b, \\
				\delta_\Sigma \varpi_{ab} &= 2\lambda_{[a}{}^c\varpi_{c|b]} + B \varpi_{ab} + \hframe_{[a}(\lambda_{b]}) + \lambda_{[a}\varphi_{b]} - \tfrac12 \lambda_c c^c{}_{ab}.
			\end{split}
		\end{equation}
		They thus behave like connections under local Carroll boosts. Second, the horizontal non-holonomy coefficients $c^a{}_b$ and $c^a{}_{bc}$ are covariant under Carroll boosts, in the sense that they transform as a multiplet, but sensitive to the particular choice of basis in the distribution of $\bftau$ and thus transform as connections under local rotations. We have indeed
\begin{equation}
	\begin{split}
		\delta_\Sigma c^a{}_b &= \lambda^a{}_c c^c{}_b - \lambda^c{}_b c^a{}_c - \bfupsilon(\lambda^a{}_b) + B c^a{}_b - \delta^a{}_b\bfupsilon(B), \\
		\delta_\Sigma c^a{}_{bc} &= \lambda^a{}_d c^d{}_{bc} - 2\lambda^d{}_{[b}c^a{}_{d|c]} -2 \hframe_{[b}(\lambda^a{}_{c]}) + 2 c^a{}_{[b}\lambda_{c]} + B c^a{}_{bc} - 2\delta^a{}_{[b}\hframe_{c]}(B).
	\end{split}
\end{equation}		
The symmetric part $c_{(ab)}$ is again a fully-covariant tensor under local rotations and boosts, with tracefree part $\xi_{ab}$ also a tensor under Weyl transformations.

\subsubsection*{Gauge curvatures}

The curvature associated with the gauge connection \eqref{eq:Gamma carroll} is again given by
\begin{equation}
	\vect R[\bm\Gamma] = \D\bm\Gamma + \bm\Gamma\wedge\bm\Gamma\quad\in\quad \Omega^2\big(\mathscr C,\mathfrak{confcarr}_1(d,1)\big)
\end{equation}
with the same notations as in Eqs. \eqref{eq:def curvature conf} and \eqref{eq:wedge gamma gamma}. It is expanded in the basis we have opted for as $\vect R[\bm\Gamma]=\vect R[\bm\Gamma]^I\gen X_I$, or
\begin{equation}
	\vect R[\gen P]^a \gen P_a + \vect R[\gen H]\gen H + \tfrac12 \vect R[\gen J]^{ab}\gen J_{ab} + \vect R[\gen B]^a\gen B_a - \vect R[\gen D]\gen D + \tfrac12 \vect R[\gen K]^a\gen K_a + \tfrac12 \vect R[\gen K]^0\gen K_0,
\end{equation}
where each term is expressed as follows
\begin{subequations}
\begin{align}
	\vect R[\gen P]^a &= \big(\horbmnabla-\bfalpha\big)\wedge\htheta^a , \label{eq:RPa carroll}\\
	\vect R[\gen H] &= \big(\horbmnabla -\bfalpha\big)\wedge\bftau + \bfbeta_a \wedge \htheta^a \label{eq:RH carroll},\\
	\vect R[\gen J]^{ab} &= \D\bfgamma^{ab} + \bfgamma^a{}_c\wedge\bfgamma^{cb} - 2\bfkappa^{[a}\wedge\htheta^{b]}, \label{eq:RJab carroll} \\
	\vect R[\gen B]^a &= \horbmnabla \wedge \bfbeta^a + \htheta^a\wedge\bfkappa^0 - \bftau \wedge \bfkappa^a , \label{eq:RBa carroll}\\
	\vect R[\gen D] &= \D\bfalpha - \bfkappa_a\wedge\htheta^a, \label{eq:RD carroll} \\
	\vect R[\gen K]^a &= \big(\horbmnabla+\bfalpha\big)\wedge\bfkappa^a , \label{eq:RKa carroll} \\
	\vect R[\gen K]^0 &= \big(\horbmnabla+\bfalpha\big)\wedge\bfkappa^0 + \bfbeta_a \wedge \bfkappa^a \label{eq:RK0 carroll},
\end{align}\label{eq:curvatures carroll}%
\end{subequations}
in terms of the various connection coefficients. It is worth paying attention to the crucial role of $\bfalpha$, which again teams up with the horizontal derivative to display manifestly Weyl-invariant combinations. We shall come back on this point soon. The curvatures $\vect R[\gen K]^a$ and $\vect R[\gen K]^0$ offer natural candidates for the Carrollian counterpart of the Lorentzian-relativistic Cotton tensor $\vect R[\gen K]^A$, which can be understood from a simple $c$-to-zero-limit argument. Finding the expressions for these curvatures and discussing their geometric properties and interpretation are our principal motivations to consider the gauging of the conformal Carroll algebra. 

By taking an extra exterior derivative, we can derive the following Bianchi identities:
\begin{subequations} \label{eq: Bianchi Carroll}
\begin{align}
	\big(\horbmnabla-\bfalpha\big) \wedge \vect R[\gen P]^a &= \vect R[\gen J]^a{}_b \wedge \htheta^b - \vect R[\gen D] \wedge \htheta^a , \label{eq:bianchi carroll RPa}\\
	\big(\horbmnabla-\bfalpha\big) \wedge \vect R[\gen H] &= \vect R[\gen B]_a \wedge \htheta^a - \vect R[\gen D] \wedge \bftau + \vect R[\gen P]^a \wedge \bfbeta_a , \label{eq:bianchi carroll RH}\\
	\horbmnabla \wedge \vect R[\gen J]^{ab} &=   2 \vect R[\gen P]^{[a} \wedge \bfkappa^{b]} - 2 \vect R[\gen K]^{[a} \wedge \htheta^{b]} , \label{eq:bianchi carroll RJab}\\
	\horbmnabla \wedge \vect R[\gen B]^a &=  \vect R[\gen J]^a{}_b \wedge \bfbeta^b + \vect R[\gen K]^a \wedge \bftau - \vect R[\gen K]^0 \wedge \htheta^a - \vect R[\gen P]^a \wedge \bfkappa^0 + \vect R[\gen H] \wedge \bfkappa^a , \label{eq:bianchi carroll RBa} \\
	\horbmnabla \wedge \vect R[\gen D] &= -\vect R[\gen K]_a \wedge \htheta^a + \vect R[\gen P]_a \wedge \bfkappa^a, \label{eq:bianchi carroll RD} \\
	\big(\horbmnabla+\bfalpha\big)\wedge \vect R[\gen K]^a &= \vect R[\gen J]^a{}_b \wedge \bfkappa^b + \vect R[\gen D] \wedge \bfkappa^a , \label{eq:bianchi carroll RKa} \\
	\big(\horbmnabla+\bfalpha\big)\wedge \vect R[\gen K]^0 &= \vect R[\gen B]_a \wedge \bfkappa^a + \vect R[\gen K]^a \wedge \bfbeta_a + \vect R[\gen D] \wedge \bfkappa^0. \label{eq:bianchi carroll RK0}
\end{align}
\end{subequations}
The explicit expressions in Eqs. \eqref{eq:curvatures carroll} allow us to derive the gauge transformations of the various pieces of curvature under \eqref{eq:Sigma carr} thanks to \eqref{eq:transfo gauge fields carroll}. We find
\begin{subequations} 
\begin{align}
	\delta_\Sigma \vect R[\gen P]^a &= \lambda^a{}_b\vect R[\gen P]^b - B\vect R[\gen P]^a ,\label{eq:delta RPa}\\
	\delta_\Sigma \vect R[\gen H] &= -\lambda_a \vect R[\gen P]^a - B\vect R[\gen H] , \label{eq:delta RH}\\
	\delta_\Sigma \vect R[\gen J]^{ab} &= 2 \lambda^{[a}{}_c\vect R[\gen J]^{c|b]} + 2 Z^{[a}\vect R[\gen P]^{b]}, \label{eq:delta RJab} \\
	\delta_\Sigma \vect R[\gen B]^a &= \lambda^a{}_b \vect R[\gen B]^b - \lambda_b \vect R[\gen J]^{ba} + Z^0\vect R[\gen P]^a - Z^a\vect R[\gen H] , \label{eq:delta RBa} \\
	\delta_\Sigma \vect R[\gen D] &= Z_a\vect R[\gen P]^a, \label{eq:delta RD} \\
	\delta_\Sigma \vect R[\gen K]^a &= \lambda^a{}_b\vect R[\gen K]^b + B \vect R[\gen K]^a + \vect R[\gen J]^{ab}Z_b + Z^a\vect R[\gen D] , \label{eq:delta RKa}\\
	\delta_\Sigma \vect R[\gen K]^0 &= - \lambda_a\vect R[\gen K]^a + B \vect R[\gen K]^0 + \vect R[\gen B]^{a} Z_a + Z^0 \vect R[\gen D]. \label{eq:delta RK0}
\end{align}\label{eq: transfo Carrollian curvatures}%
\end{subequations}
As in the Lorentzian case, we now derive an interesting invariant subset of equations of motion, in the sense that they are covariant under all the local symmetries. As we shall see, we cannot require anymore that the conditions do not constrain the background geometry: we shall only require that they impose the minimal amount of geometric constraints to ensure the absence of torsion. The fibre structure naturally splits constraints and curvatures into horizontal and vertical components, which can be dealt with separately in accordance with the symmetries. First, we can impose that the horizontal composite torsion vanishes, \textit{i.e.}, $\vect R[\gen P]^a = \bm 0$, since it transforms covariantly, by virtue of Eq. \eqref{eq:delta RPa}. Then, Eq. \eqref{eq:delta RH} implies that $\vect R[\gen H] = \bm 0$, \textit{i.e.}, the cancellation of the vertical torsion, can be consistently imposed. These two constraints fix the coefficients $\gamma^a{}_b$ and $\gamma^a{}_{bc}$ of the horizontal connection, and \textit{some part} of the Carroll--boost connection, in terms of the coframe $\bftau,\htheta^a$ and the Weyl connection $\bfalpha$. The mixed vertical-horizontal component of $\vect R[\gen P]^a=\bm 0$ further imposes a constraint on the geometry: it restricts the flow of $\bfupsilon$ to be shear-free. In line with \cite{Bekaert:2015xua}, we refer to this situation as a \textit{conformally invariant Carrollian structure}, which can therefore be endowed with a torsion-free gauge connection. For this reason, there remain degrees of freedom in the gauge connection.

Next, the transformation laws \eqref{eq:delta RJab} and \eqref{eq:delta RBa} can be massaged to find that the linear combination $\vect{Ric}[\gen J]^a + \vect R[\gen B]^a(\bfupsilon,\cdot)$ involving the Ricci tensor $\vect{Ric}[\gen J]^a\coloneqq \vect R[\gen J]^{ab}(\cdot,\hframe_b)$ associated with the composite Riemann curvature and the vertical leg of the Carroll-boost curvature are invariant under Carroll-boosts. With the torsion-free assumption, this combination further covariantly transforms as a $\mathfrak{so}(d)$ vector, and can therefore be assumed to vanish,
\begin{equation}
	\vect{Ric}[\gen J]^a + \vect R[\gen B]^a(\bfupsilon,\cdot) = \bm 0. \label{eq:ricciflat carroll}
\end{equation}
By analogy with the Lorentz-relativistic case, we refer to gauge connections obeying this constraint as \textit{horizontally Carroll--Ricci flat connections}. Bianchi identities \eqref{eq:bianchi carroll RPa} and \eqref{eq:bianchi carroll RH} then reduce to
\begin{equation}
	\vect R[\gen J]^a{}_b\wedge\htheta^b = \vect R[\gen D]\wedge\htheta^a,\qquad \vect R[\gen B]^a\wedge\htheta_a = \vect R[\gen D]\wedge\bftau,
\end{equation}
which can be repackaged into the following simple condition:
\begin{equation}
	(d-1)\vect R[\gen D] = \left[\vect{Ric}[\gen J]^a + \vect R[\gen B]^a(\bfupsilon,\cdot)\right]\wedge\htheta_a = \bm 0.
\end{equation}
For any $d>1$, the composite Weyl curvature automatically drops out on account of the Bianchi identities. Again, we shall temporarily discard the $d=1$ case from the discussion and give further comments on this particular configuration later on. Assuming now that all the previous constraints are satisfied, the remaining components $\vect{Ric}[\gen B] \coloneqq \vect R[\gen B]^a(\cdot,\hframe_a)$ of the composite Ricci tensor, namely those purely related to the Carroll-boost curvature, are left invariant by local symmetries: $\delta_\Sigma \vect{Ric}[\gen B] = \bm 0$. Therefore, we can also require that it vanishes to define our suitable gauge connection. One can show that any stronger restriction, for any $d\geq 3$, would impose more constraints on the background geometry than conformal invariance on the flow of $\bfupsilon$. For $d=2$, demanding that the Ricci curvatures vanish is equivalent to requiring that the Riemann curvatures vanish since Weyl tensors are absent in three dimensions. All in all, we are left with the following conditions:
\begin{equation}\label{eq:EOM carroll}
\boxed{\begin{array}{lclcl}
	\vect R[\gen P]^a = \vect R[\gen H] = \bm 0 = \vect R[\gen D], &\qquad& \vect R[\gen J]^{ab}(\hframe_a,\hframe_c) + \vect R[\gen B]^b(\hupsilon,\hframe_c) = 0, \\
	\vect R[\gen J]^a{}_b(\hframe_a,\hupsilon) = 0 = \vect R[\gen B]^{a}(\hframe_a,\hframe_b) , &\qquad& \vect R[\gen B]^a(\hframe_a,\hupsilon) = 0,
\end{array}
}
\end{equation}
expanded onto the local Carroll--Cartan basis, and which we shall refer to as the \textit{Carroll--Ricci flat conditions}, for which the gauge fields $\bfkappa^0$ and $\bfkappa^a$ are completely determined in terms of the geometry and the Weyl connection. The conditions \eqref{eq:EOM carroll} constitute the Carrollian equivalent of the torsion-free and completely Ricci-curvature-free conditions, Eq. \eqref{eq:constraints gauging conformal}. Moreover, it is straightforward to show that the $c\to 0$ limit of the Lorentzian equations of motion \eqref{eq:constraints gauging conformal} yields \eqref{eq:EOM carroll}.

\subsection{Weyl--Carroll--Levi-Civita connections}
We now solve the constraints \eqref{eq:EOM carroll} and derive the Carrollian counterpart of Weyl--Levi-Civita connection, Eq. \eqref{eq:weyl LC}, which we shall refer to as \textit{Weyl--Carroll--Levi-Civita connections}. First, let us recall that contrary to the Lorentzian case, the no-torsion conditions do not fix uniquely the boost connection $\bfbeta^a$, which now contains degrees of freedom encoded into an arbitrary symmetric rank-two transverse tensor, but impose instead a constraint on the background Carrollian geometry $(\mathscr C,\bfg,\bfupsilon)$. In the purely Carrollian context, by which we mean without any Weyl extension, it has been known for a long time \cite{jankiewicz1954espaces,Vogel1965,Datcourt1967} in the manifestation of gravitational fields on null hypersurfaces, and recently re-discovered \cite{Bekaert:2015xua,Hartong:2015xda} in a Carroll-intrinsic approach. In the conformal context, it has been mainly investigated through the perspective of intrinsic geometry of null infinity \cite{Geroch:1977big,Ashtekar:1981hw,Ciambelli:2019lap}. Let us recover these features through our construction.

Imposing that the horizontal part of the composite torsion tensor vanishes yields
\begin{equation}
	\vect R[\gen P]^a = \bm 0 \quad\Rightarrow\quad \gamma^a{}_{[bc]} = \tfrac12 c^a{}_{bc} - \alpha_{[b} \delta^a{}_{c]} , \quad \gamma^a{}_{b} = \alpha_0 \delta^a{}_b - c^a{}_b.
\end{equation}
The solution to the first equation is
\begin{equation}
	\gamma^a{}_{bc} = \tfrac12 \big(c^a{}_{bc} + c_b{}^a{}_c + c_c{}^a{}_b\big) - \delta^a{}_b\alpha_c + \delta_{bc}\alpha^a, \label{eq:gammasol}
\end{equation}
recalling that $\gamma_{(a|b|c)}=0$. Next, upon separating the second equation into trace, symmetric tracefree and skew-symmetric parts, we find
\begin{equation}
	\alpha_0 = \tfrac{1}{d}\theta,\qquad \xi_{ab} = 0,\qquad \gamma_{[ab]} =- c_{[ab]}, \label{eq:constraints WCLC}
\end{equation}
since $\bfgamma_{(ab)}=\bm 0$ by construction. The first and last equations can always be imposed at no expense by adapting the choice of temporal Weyl connection $\alpha_0$ and spin connection $\gamma^a{}_{b}$ to the geometry. However, the central equation imposes that the flow of $\bfupsilon$ preserves the Weyl class, \textit{i.e.}, $\mathscr L_\bfupsilon\bfg \propto \bfg$. In plain words, it enforces the background Carroll structure to be conformally invariant.

\begin{description}
	\item[Remark.] There is again a crucial difference with the Lorentzian case. Indeed, in the latter, one can define a Weyl--Levi-Civita connection for any choice of Weyl connection $\bfalpha$. In the Carrollian case, even when the constraint $\xi_{ab}=0$ is satisfied by the geometry, thereby ensuring the existence of a Carroll--Weyl--Levi-Civita connection, the absence of torsion is not compatible with arbitrary Weyl connections. Rather, it forces the vertical component $\alpha_0$ to be fixed by the geometry, namely by the expansion $\theta$ of $\bfupsilon$. This is consistent with the fact that $\alpha_0$ is not pure gauge in the Carrollian case, as stressed earlier.
\end{description}

Whenever the background geometry is endowed with a conformally invariant Carrollian structure, the system of equations $\vect R[\gen P]^a=\bm 0$ admits solutions for the spin-connection coefficients but is under-determined. Indeed, as three components of the torsion identically vanish by virtue of the conditions $\alpha_0 = \tfrac{1}{d}\theta$ and $\xi_{ab} = 0$, there shall remain $\frac{d(d+1)}{2}$ degrees of freedom in the final gauge connection after all the constraints \eqref{eq:EOM carroll} are solved. Indeed, the only constraints to be imposed on $\bm\upbeta{}_{a}$ are found as
\begin{equation}
	\vect R[\gen H] = \bm 0 \quad\Rightarrow\quad \beta_a = \varphi_a - \alpha_a,\quad \beta_{[ab]} = - \varpi_{ab}
\end{equation}
by imposing the absence of vertical torsion, meaning that its transverse symmetric part, $\beta$ and $\mathscr C_{ab}$ in Eq. \eqref{eq:separation beta}, are utterly free. We shall see later that the gauge fields $\bfkappa^0$ and $\bfkappa^a$ are uniquely determined from the previous geometric data, namely the Carroll--Cartan frame and the Weyl connection thanks to the remaining equations of motion.

At this stage, it is instructive to have a closer look at the transformation law \eqref{eq:transfo beta}, which can be split into longitudinal and transverse (trace and symmetric tracefree) parts as
\begin{subequations}
	\begin{align}
		\delta_\Sigma \beta_{a} &= \lambda_a{}^b\beta_{b} + B\beta_{a} + \bar\nabla_\bot\lambda_a - Z_a, \label{eq:delta beta a}\\ 
		\delta_\Sigma\beta &= \big(\bar\nabla_a+\beta_{a}\big)\lambda^a + B\beta + d Z^0, \label{eq:delta beta trace} \\ 
		\delta_\Sigma \mathscr C_{ab} &= 2 \lambda_{\langle a}{}^c\mathscr C_{b\rangle c} + B\mathscr C_{ab} - 2\big(\bar\nabla_{\langle a}+\beta_{\langle a}\big)\lambda_{b\rangle}. \label{eq:delta Cab}
	\end{align}
\end{subequations}
Clearly, both $\beta$ and $\beta_{a}$ are pure-gauge fields under special conformal transformations parameterised by $Z^0$ and $Z^a$. Indeed, there always exist special-conformal-transformation parameters $Z^0$ and $Z^a$ allowing for a gauge-fixing like $\beta = 0$ and $\beta_a=0$. After fixing this gauge, the residual gauge parameters are locked to the Carroll-boost parameters as $Z_a = \bar{\nabla}_\bot \lambda_a$ and $Z^0 = -\tfrac{1}{d} (\bar\nabla_a+\beta_a)\lambda^a$. Setting $\beta_a = 0$ for a torsion-free connection, \textit{i.e.}, $\vect R[\gen H] = \bm 0$, amounts to reach the \textit{rheotactic gauge}
\begin{equation}
	\beta_a = 0\qquad\Rightarrow\qquad \alpha_a = \varphi_a, \label{eq:rheotactic carroll}
\end{equation}
since we automatically have $\alpha_0 = \tfrac{1}{d}\theta$ by virtue of the constraint $\vect R[\gen P]^a = \bm 0$. This is the gauge chosen in, \textit{e.g.}, \cite{Campoleoni:2023fug}. Setting further $\beta = 0$ amounts to impose the \textit{rheotactic tracefree gauge}.

Alternatively, we recall that the transverse part of the Weyl connection also transforms purely algebraically under transverse special-conformal transformations, see Eq. \eqref{eq:delta alpha a carroll}. Therefore, there always exists a special-conformal local frame such that $\alpha_a = 0$, which amounts to reach the \textit{transverse metric gauge}. In this case, the information that would be encoded by $\alpha_a$, namely the value of the Carrollian acceleration $\varphi_a$, as required by Eq. \eqref{eq:rheotactic carroll}, is passed on to $\beta_a$. This is the gauge chosen in, \textit{e.g.}, \cite{Fiorucci:2025twa} and usual analyses in Bondi--Sachs coordinates, see \cite{Bondi:1962px,Sachs:1962wk,Tamburino:1966zz,Barnich:2010eb,Barnich:2011mi,Freidel:2021fxf,Geiller:2022vto,Rignon-Bret:2024wlu,Hartong:2025jpp,Hartong:2026rbr} and references therein. Again, it is always reachable by adjusting the special-conformal frame, and is preserved by residual gauge parameters locked to the Weyl parameter as $Z_a = \hframe_a[B]$. Let us also stress that there is no notion of a full metric gauge, \textit{i.e.}, $\alpha_0 = 0$, for a torsion-free connection over a Carroll manifold with $\theta\neq 0$, as evidenced by the transformation law \eqref{eq:delta alpha 0 carroll}. Enforcing a vanishing $\alpha_0$ would either impose a non-trivial constraint on the geometry, $\theta=0$, or induce transverse torsion, $\vect R[\gen P]^a\neq \bm 0$. In the present case, the only field that supports the inhomogeneous transformation under longitudinal special-conformal transformation is the trace $\beta$. A convenient gauge fixing amounts to set $\alpha_a = 0$ and $\beta=0$, which we refer to as the \textit{transverse tracefree metric gauge}.

Therefore, the true degrees of freedom lie in the tracefree part, $\mathscr C_{ab}$, which has $\frac{d(d+1)}{2}-1$ independent components. Unlike $\beta$ and $\beta_a$, there is no gauge choice allowing to cancel $\mathscr C_{ab}$ unless it derives from a given one-form field. This happens in particular situations which are of instrumental interest for gravitational-wave analysis: indeed, in a holographic perspective, recognising null infinity as an invariant conformal Carroll manifold with $d=2$, $\mathscr C_{ab}$ is nothing but the shear of outgoing gravitational waves, sometimes referred to as the \textit{Bondi shear}, since its derivative along the null generators of null infinity is the Bondi news tensor.\footnote{See Sections \ref{sec:Carroll--Schouten tensors} and \ref{sec:covariantnews} for our Weyl-covariant definition of the news tensor and a discussion on its fundamental character with respect to the Carroll--Cotton tensors.} In particular, we shall see that the transformation law \eqref{eq:delta Cab} coincides with that of the Bondi shear \cite{Fiorucci:2025twa}, see also \cite{Baulieu:2025itt}. In his seminal work \cite{Ashtekar:1981hw}, Ashtekar discards the trace $\beta$ of the transverse Carroll-boost connection at null infinity thanks to a subleading Weyl rescaling in the bulk. The effect of such transformations has recently been understood in the covariant-phase-space approach to null infinity in \cite{Barnich:2011ty,Geiller:2022vto}. Our analysis proves that discarding the trace, in order to be left with the only degrees of freedom of the gravitational field encoded in $\mathscr C_{ab}$ only, can be performed in a purely Carroll-intrinsic way. However, it crucially requires to consider the gauging of the full conformal Carroll algebra at null infinity, not only local Carroll transformations and dilations. In our viewpoint, special conformal transformations are sensitive to the embedding of the conformal boundary into the bulk,\footnote{More specifically, the unphysical spacetime in the parlance of Penrose's conformal compactification \cite{Penrose:1962ij,Penrose:1965am}, see \cite{Fiorucci:2025twa} for a discussion in our formalism.} and keep track of how outgoing (resp. incoming) null rays approach future (resp. past) null infinity.

\begin{description}
	\item[Remark.] Let us stress that, unlike the Weyl connection itself, the Weyl curvature is left to a gauge choice. Indeed, although the former cannot be set to zero by gauge transformations if the expansion $\theta$ is non-trivial, the latter transforms as a connection under the transverse special-conformal transformations:
\begin{equation}
	\delta_\Sigma \D\bfalpha = \big(\bar{\bm\nabla}+\bfalpha\big)Z_a\wedge\htheta^a.
\end{equation}	
Therefore, the \textit{Weyl-flat gauge} $\D\bfalpha=\bm 0$ can always be reached by adjusting the special-conformal frame. This demonstrates \textit{a posteriori} the soundness of our hypotheses in \cite{Fiorucci:2025twa}, where the gravitational flux-balance laws have been derived at null infinity assuming no Weyl curvature. Residual gauge transformations are then restricted to curl-free transverse covectors that are parallel transported covectors along the fibre: $(\bar\nabla_\bot +\tfrac{1}{d}\theta)Z_a = 0$ and $(\bar\nabla_{[a}+\alpha_{[a})Z_{b]}=0$. This gauge is always compatible with the metric gauge; however, it is only compatible with the rheotactic gauge in particular configurations.
\end{description}

Armed with the so-determined connection coefficients $\bfgamma^a{}_b$, $\bfbeta^a$ and $\bfalpha$, for a given choice of the related degrees of freedom in $\bfbeta^a$, we can build the Carrollian avatar of the connection \eqref{eq:weyl LC}, which is defined as a particular torsion-free connection $\text D$ preserving the Carroll conformal class $[(\bfg,\bfupsilon)]$:
\begin{equation}
	\boxed{
		\vect D\bfg - 2\bfalpha\otimes\bfg = \bm 0,\quad \vect D\bfupsilon + \bfalpha\otimes\bfupsilon = \bm 0,\quad \vect D\wedge\bftau = \bm 0,\quad \vect D\wedge\htheta^a = \bm 0.
	} \label{eq:def WCLC}
\end{equation}
It naturally goes under the name of \textit{Weyl--Carroll--Levi-Civita connection} and belongs, up to gauge choice, to an affine space modeled on transverse symmetric traceless tensors $\mathscr C_{ab}$. The last two conditions above are equivalent to $\vect R[\gen H] = \bm 0$ and $\vect R[\gen P]^a=\bm 0$, while the first two conditions give
\begin{equation}
	\bfomega_{(ab)} = -\bfalpha \delta_{ab},\quad \bfomega^0{}_0 = -\bfalpha,\quad \bfomega^a{}_0 = \bm 0,\quad \bfomega^0{}_a = \bfbeta{}_a.
\end{equation}
Denoting as before spacetime components as $A = 0,a$ for $a=1,\ldots,d$, the solution to Eq. \eqref{eq:def WCLC} can be written compactly as
\begin{equation}
	\boxed{\bfomega^A{}_B = \bfgamma^a{}_b \delta^A{}_a\delta^b{}_B - \bfalpha \delta^A{}_B + \bfbeta_b\delta^A{}_0\delta^b{}_B,}\label{eq:WCLC omegaAB}
\end{equation}
where the various coefficients are given by
\begin{equation}
	\boxed{
	\begin{aligned}
		\bfgamma^a{}_b &= -\delta^{ac}c_{[cb]} \bftau + \gamma^a{}_{cb}\htheta^c,\\ 
		\bfalpha &= \tfrac{1}{d} \theta \bftau + \alpha_a\htheta^a,\\
		\bfbeta_a &= (\D-\bfalpha)\wedge\bftau(\hframe_a,\cdot) + \left(\tfrac{1}{d}\beta \delta_{ab} - \tfrac12 \mathscr C_{ab}\right)\htheta^b,
	\end{aligned}
	} \label{eq:WCLC detailed}
\end{equation}
in terms of the transverse connection coefficients $\gamma^a{}_{bc}$, displayed in Eq. \eqref{eq:gammasol}, pure-gauge fields $\alpha_a$ and $\beta$, and the extra degrees of freedom $\mathscr C_{ab}$. In particular, the purely horizontal coefficients are those of the Weyl--Levi-Civita connection on the base manifold $\mathscr S$, 
\begin{equation}
	\omega^a{}_{bc} = \tfrac12 \big(c^a{}_{bc} + c_b{}^a{}_c + c_c{}^a{}_b\big) - \delta^a{}_b\alpha_c - \delta^a{}_c\alpha_b + \delta_{bc}\alpha^a,
\end{equation}
which reinforces the justification of our proposed terminology. Note crucially that, unlike the Weyl--Levi-Civita connection, the given of a Weyl--Carroll structure does not fully prescribe $\text D$, as there are additional degrees of freedom in $\bfomega^0{}_a$. The only way to fix this ambiguity is to prescribe how $\text D$ transports the Ehresmann connection $\bftau$. Finally, passing from the connection $\text D$ to the field-dependent Weyl-covariant derivative operator ${\mathscr D}$ proceeds exactly as before.

\begin{description}
	\item[Remark.] Reduced in longitudinal and transverse components relative to a timelike congruence $\bfupsilon$, with related clock form $\bftau$, the coefficients \eqref{eq:gamma for WLC} of the Weyl--Levi-Civita spin connection are
	\begin{equation}
		\begin{split}
			\bfgamma^a{}_0 &= c^2\big(\varphi^a - \alpha^a\big)\bftau + \big(\tfrac{1}{d}\theta - \alpha_0\big) \htheta^a + \big(\xi^a{}_b - c^2\varpi^a{}_b\big)\htheta^b, \\
			\bfgamma^0{}_a &= (\D-\bfalpha)\wedge\bftau (\hframe_a,\cdot) + \tfrac{1}{c^2}\big(\tfrac{1}{d}\theta - \alpha_0\big)\htheta_a + \tfrac{1}{c^2}\xi_{ab}\htheta^b, \\
			\bfgamma^a{}_b &= \delta^{ac}\big(c_{[cb]} - c^2\varpi_{cb}\big)\bftau + \gamma^a{}_{cb}\htheta^c,
		\end{split}
	\end{equation}
where the $\gamma^a{}_{bc}$ are those of the Weyl--Levi-Civita connection on $\mathscr S$. The $c\to 0$ limit of the Lorentzian Weyl--Levi-Civita connection is well-defined if and only if 
\begin{equation}\label{eq:hypotheses limit}
	\alpha_0 \stackrel{!}{=} \tfrac{1}{d}\big(\theta - c^2\beta\big),\qquad \xi_{ab} \stackrel{!}{=} -\tfrac12 c^2 \mathscr C_{ab},
\end{equation}
where $\beta$ and the symmetric traceless transverse tensor $\mathscr C_{ab}$ are assumed to be finite in the limit. As a result, the Weyl--Carroll structure is conformally invariant in the limit and the longitudinal component of the Weyl connection is locked by the expansion. In this case, the resulting connection coincides with the Carroll--Weyl--Levi-Civita connection with degrees of freedom encoded by the arbitrary fields $\beta$ and $\mathscr C_{ab}$. It is worth noting that the limit taken within the rheotactic gauge fixing, Eq. \eqref{eq:rheotactic}, yields the Carrollian tracefree rheotactic gauge fixing ($\beta = 0$). The second equation in \eqref{eq:hypotheses limit} has been understood in the holographic setting as the absence of extrinsic curvature of the conformal boundary inside the conformally compactified spacetime \cite{Compere:2019bua,Campoleoni:2023fug,Poole:2018koa}. On the conformal-spacetime boundary, it allows one to start from a completely fixed Weyl--Levi-Civita connection\footnote{In this viewpoint, $c$ is the effective speed of light on the boundary, related to the bulk cosmological constant $\Lambda\in\mathbb R^-$ as $c^2 = -\frac{\Lambda}{3}$. The bulk asymptotically flat limit $\Lambda\to 0$ therefore induces a Carroll $c\to 0$ limit on the boundary, see, \textit{e.g.}, \cite{Ciambelli:2018xat,Donnay:2022aba,Campoleoni:2023fug,Alday:2024yyj}.} and to land on a Carroll--Weyl--Levi-Civita connection that contains degrees of freedom. Although the first equation, which relates arbitrary functions $\alpha_0$ and $\beta$, is harmless to the definition of Weyl--Levi-Civita since both are, a priori, free data for any finite $c$ and then merely induces a renaming of variables, the second equation unlocks $\xi_{ab}$ from its expression in terms of derivatives of the metric, which induces a clear deviation from the (Weyl--)Levi-Civita connection. In an intrinsic approach of the limit and the gravitational flux-balance laws, including these degrees of freedom, hence positing \eqref{eq:hypotheses limit} for any finite $c$, therefore requires to give up either the Weyl-metric compatibility of the parent connection or allow for torsion.

\end{description}

\subsection{Carroll--Schouten tensors}
\label{sec:Carroll--Schouten tensors}

Finally, we prove that the extra fields $\bfkappa^0 = \kappa^0{}_0 \bftau + \kappa^0{}_a\htheta^a$ and $\bfkappa^a = \kappa^a{}_0 \bftau + \kappa^a{}_b\htheta^b$ are uniquely determined in terms of the previous geometric data, namely the Carroll--Riemann curvature tensor, fragmented in the time-space decomposition as
\begin{subequations}
	\begin{align}
		\vect R[\bfgamma]^a{}_b &\coloneqq \D \bfgamma^a{}_b + \bfgamma^a{}_c \wedge \bfgamma^c{}_b = \tfrac12 {\mathscr R}^a{}_{bcd} \htheta^c \wedge \htheta^d + {\mathscr R}^a{}_{bc} \bftau \wedge \htheta^c, \\
		\vect R[\bfbeta]_a &\coloneqq \bar{\bm\nabla}\wedge \bfbeta_a = \tfrac12 {\mathscr R}^0{}_{abc} \htheta^b \wedge \htheta^c + {\mathscr R}^0{}_{ab} \bftau \wedge \htheta^b,
	\end{align}
\end{subequations}
where $\mathscr R^a{}_{bcd} = \bar{\mathscr R}{}^a{}_{bcd}$ and $\mathscr R^a{}_{bc} = \bar{\mathscr R}{}^a{}_{bc}$ are the components of the projected connection transverse to $\bfupsilon$, and the Weyl curvature
\begin{equation}
	\D\bfalpha \coloneqq \tfrac12\bar\Omega_{ab}\htheta^a\wedge\htheta^b - \bar{\mathscr R}_a\htheta^a\wedge\bftau,\quad \bar\Omega_{ab} \coloneqq -\varepsilon_{ab}\bar{\mathscr A}.
\end{equation}
In the above equations, $\bar\Omega_{ab}$, $\bar{\mathscr R}_a$ and $\bar{\mathscr A}$ denote the same quantities as in the relativistic case, see Eqs. \eqref{eq:weyl curv rheotactic} and \eqref{eq:weyl curv rheo 2} for explicit expressions in terms of the non-holonomy coefficients.

Inspired by the Lorentzian case, we shall call the special conformal gauge fields on-shell the \textit{Carroll--Schouten tensors}. Let us derive their expression in arbitrary dimensions $d\geq 2$. From Bianchi identities \eqref{eq:bianchi carroll RPa} and \eqref{eq:bianchi carroll RH}, we find that the Riemann curvatures of the spin and boost connections obey the constraints $\vect R[\gen J]^a{}_b\wedge\htheta^b = \bm 0$ and $\vect R[\gen B]_a\wedge\htheta^a = \bm 0$ in the absence of torsion and composite Weyl curvature. In the Carroll--Cartan basis, they expand as
\begin{equation}
	\bar{\mathscr R}{}^a{}_{[bcd]} = -2\kappa_{[bc}\delta^a{}_{d]},\quad \bar{\mathscr R}{}^a{}_{[bc]} = \delta^a{}_{[b}\kappa_{c]0},\quad \mathscr R^0{}_{[abc]} = 0,\quad \mathscr R^0{}_{[ab]} = \kappa_{[ab]}. \label{eq:Bianchi carroll detail}
\end{equation}
Taking a trace in the first two identities implies that
\begin{equation}
	(d-2)\kappa_{[ab]} = \bar{\mathscr R}{}^c{}_{[a|c|b]},\qquad (d-1)\kappa^a{}_0 = \bar{\mathscr R}{}^{ab}{}_b.
\end{equation}
If $d = 2$, the first equation above is a tautology; otherwise, and similarly to the Lorentzian case, the skew-symmetric part of the Carroll--Schouten tensor is controlled by the skew-symmetric part of the Ricci tensor, which is non-vanishing whenever there is Weyl curvature. Indeed, the vanishing of the curvature $\vect R[\gen D]$ fixes
\begin{equation}
	\kappa_{[ab]} = \tfrac12\varepsilon_{ab}\bar{\mathscr A},\qquad \kappa^a{}_0 = \bar{\mathscr R}{}^a, \label{eq:antisym schouten carroll}
\end{equation}
and we recover the identity $\bar{\mathscr R}{}^a = \frac{1}{d-1}\bar{\mathscr R}{}^{ab}{}_b$. We solve the remaining constraints in \eqref{eq:EOM carroll} to find
\begin{equation}
	\kappa^0{}_a = \tfrac{1}{d-1}\mathscr R^{0b}{}_{ab},\qquad \kappa^0{}_0 = \tfrac1d \big( \mathscr R^{0a}{}_a - \kappa^a{}_a\big). \label{eq:kappa0a and kappa00 carroll}
\end{equation}
The components of the Carroll--Schouten tensors that remain to be fixed are $\kappa_{(ab)}$. Again by virtue of the Bianchi identity \eqref{eq:bianchi carroll RPa}, the transverse composite Ricci tensor $\vect{Ric}[\gen J]^a$ is symmetric, in the sense that $\text{Ric}[\gen J]^a{}_0 = 0$ and $\text{Ric}[\gen J]_{[ab]} = 0$. The only non-trivial constraint that remains to be solved in Eq. \eqref{eq:EOM carroll} is thus $\text{Ric}[\gen J]_{(ab)} = \vect R[\gen B]_{(a}(\hframe_{b)},\bfupsilon)$. Owing to Eq. \eqref{eq:kappa0a and kappa00 carroll}, we obtain
\begin{equation}
	\kappa^a{}_a = \tfrac{1}{2(d-1)}\bar{\mathscr R}{}^{ab}{}_{ab},\qquad \kappa_{\langle ab\rangle} = \tfrac{1}{d-1}\big(\bar{\mathscr R}{}^c{}_{\langle a|c|b\rangle} + \mathscr R^0{}_{\langle ab\rangle}\big),
\end{equation}
dividing into traceful and tracefree parts. 

In what follows, we shall mainly restrict our attention to the three-dimensional case $d=2$. Although the following discussions can be generalised to higher dimensions, there are a couple of reasons why the case of three spacetime dimensions is of particular interest to us. First, the Cotton tensor is completely determined from the Weyl tensor in the higher-dimensional Lorentzian case, see Eq. \eqref{eq:identity weyl cotton}. It is thus instrumental in quantifying deviations from conformal flatness in $d=2$ only, a feature that we aim at generalising to Carrollian realms. Second, three-dimensional Carroll--Cotton tensors have been shown to encode gravitational degrees of freedom at the boundary of four-dimensional asymptotically flat spacetime \cite{Campoleoni:2023fug}, see also \cite{Ciambelli:2017wou,Petkou:2022bmz,Mittal:2022ywl,Miskovic:2023zfz} for explicit computations in Robinson--Trautman spacetimes. The radiative degrees of freedom outgoing at future null infinity are understood there as part of the boundary connection's degrees of freedom \cite{Penrose:1965am,Geroch:1977big,Ashtekar:1981hw}, see \cite{Fiorucci:2025twa} for a fully intrinsic Carrollian derivation of this statement. In higher dimensions, outgoing radiation appears in more subleading orders in the radial expansion of the metric field near null infinity \cite{Campoleoni:2020ejn,Bekaert:2026cib}, making its Carrollian encoding and interpretation more obscure.

First, it is worth noticing that the traceless part of the spatial Ricci tensor vanishes identically in this case. Indeed, the purely spatial Riemann tensor has only one degree of freedom:
\begin{equation}
	\bar{\mathscr R}_{abcd} = \bar{\mathscr K}(\delta_{ac}\delta_{bd}-\delta_{ad}\delta_{bc}),\qquad \bar{\mathscr R} = \bar{\mathscr R}^{ab}{}_{ab} = 2\bar{\mathscr K},\qquad \bar{\mathscr K} = \mathring K + \mathring \nabla_a\alpha^a, \label{eq:riemann abcd 2D carroll}
\end{equation}
where $\mathring K$ is the Gauss mean curvature of the Levi-Civita connection $\mathring\nabla$ on the base space $\mathscr S$. The Carroll--Schouten tensor thus simplifies to
\begin{equation}
	\boxed{
		\kappa^a{}_a = \tfrac12 \bar{\mathscr R}{}^{ab}{}_{ab},\quad \kappa_{\langle ab\rangle} = \mathscr R^0{}_{\langle ab\rangle},\quad \kappa^0{}_0 = \tfrac12\mathscr R^{0a}{}_a - \tfrac14\bar{\mathscr R}{}^{ab}{}_{ab},\quad \kappa^a{}_0 = \bar{\mathscr R}{}^a,\quad \kappa^0{}_a = \mathscr R^{0b}{}_{ab}.
	} \label{eq:schouten carr in terms of Riemann}
\end{equation}
One can check that the above formulas are directly obtained from a $c\to 0$ limit of Eqs. \eqref{eq:kappa for rheo relat}. By computing the relevant contractions of the Carroll--Riemann tensors in order to nourish the components of the Carroll--Schouten tensors, we find
\begin{subequations} \label{eq:pieces of curvature carroll}
\begin{align}
		\mathscr R^{0a}{}_a &= \bar{\mathscr D}_\bot \beta - \big(\bar{\mathscr D}_a + \beta_a\big)\beta^a, \\
		\mathscr R^{0a}{}_{ab} &= \tfrac12 \bar{\mathscr D}_a\beta + \tfrac12 \bar{\mathscr D}{}^b\mathscr C_{ab} - \varepsilon_{ab}\big(\bar{\mathscr D}{}^b + 2 \beta^b\big)(\ast\varpi), \\
		\mathscr R^0{}_{\langle ab\rangle} &= -\tfrac12\mathscr N_{ab} - (\bar{\mathscr D}_{\langle a}+\beta_{\langle a})\beta_{b\rangle},
\end{align}
\end{subequations}
where $\bar{\mathscr D}$ denotes the Weyl-covariant derivative built from the transverse connection, and we have defined the following symmetric transverse traceless tensor
\begin{equation}
	\boxed{
		\mathscr N_{ab} \coloneqq \bar{\mathscr D}_\bot \mathscr C_{ab}.
	} \label{eq:news tensor}
\end{equation}
In gravitational applications where $\mathscr C_{ab}$ is identified with the Bondi shear, $\mathscr N_{ab}$ is the natural candidate for the Weyl-covariant instance of \textit{Bondi news tensor} \cite{Campoleoni:2023fug,Fiorucci:2025twa}, hence the choice of notation. In the traceless rheotactic gauge fixing, both $\beta$ and $\beta_a$ vanish and Eqs. \eqref{eq:pieces of curvature carroll} become
\begin{equation}
	\boxed{\mathscr R^{0a}{}_a = 0,\qquad \mathscr R^{0b}{}_{ab} = \tfrac12 \bar{\mathscr D}{}^b\mathscr C_{ab} - \varepsilon_{ab}\bar{\mathscr D}{}^b(\ast\varpi),\qquad \mathscr R^0{}_{\langle ab\rangle} = -\tfrac12 \mathscr N_{ab}.} \label{eq:riemann carroll 2D suite}
\end{equation}
In this case, looking again at holographic applications, it corresponds both to the boundary value of the bulk Schouten tensor at null infinity \cite{Geroch:1977big,Ashtekar:1981hw} and to the transverse symmetric traceless part of the boundary Schouten tensor \cite{Fiorucci:2025twa}. Stated here as a mere definition, Eq. \eqref{eq:news tensor} equates in fact the derivative of the transverse shear to null infinity ($\mathscr C_{ab}$) along the generators on the flow of $\bfupsilon$ to the longitudinal shear at first non-trivial leading order ($\mathscr N_{ab}$) in the holographic coordinate.\footnote{In the conformal compactification picture, future null infinity (for definiteness) is located at the geometric locus where the defining function $\Omega$ vanishes. The shear of its normal vector $\bfupsilon$ can therefore be expanded as $\xi_{ab} + \Omega\mathscr N_{ab} + \mathscr O(\Omega^2)$, and the Carrollian intrinsic shear $\xi_{ab}$ vanishes on account of bulk Einstein equations. In passing, at the black hole horizon, $\mathscr C_{ab}$ and $\mathscr N_{ab}$ exchange their roles and conformality is lost in general.} As an extra check, one can prove that Eqs. \eqref{eq:riemann abcd 2D carroll}--\eqref{eq:riemann carroll 2D suite} derive from the $c\to 0$ limit of Eqs. \eqref{eq:contractions riemann rheotactic} under the crucial hypotheses \eqref{eq:hypotheses limit}.

\subsection{Carroll--Cotton tensors}
\label{sec:CarrollCotton}

Having the Carroll instance of the Schouten tensor, we now define the components of the Carroll--Cotton tensors as the $\gen K$-curvatures:
\begin{equation}
	\vect R[\gen K]^0 = \tfrac12 C^0{}_{ab}\htheta^a\wedge\htheta^b + C^0{}_{0a}\bftau\wedge\htheta^a,\qquad \vect R[\gen K]^a = \tfrac12 C^a{}_{bc}\htheta^b\wedge\htheta^c + C^a{}_{0b}\bftau\wedge\htheta^b,
\end{equation}
where, by virtue of Eqs. \eqref{eq:RKa carroll} and \eqref{eq:RK0 carroll}:
\begin{equation}
	C^0{}_{ab} = 2\mathscr D_{[a}\kappa^0{}_{b]},\quad C^0{}_{0a} = \mathscr D_0\kappa^0{}_a - \mathscr D_a\kappa^0{}_0, \quad C^a{}_{bc} = 2\mathscr D_{[b}\kappa^a{}_{c]},\quad C^a{}_{0b} = \mathscr D_0\kappa^a{}_b - \mathscr D_b\kappa^a{}_0,
\end{equation}
and we recall that ${\mathscr D}$ denotes the Weyl-covariant derivative built up from a given choice of Weyl--Carroll--Levi-Civita connection, Eqs. \eqref{eq:WCLC omegaAB} and \eqref{eq:WCLC detailed}. Since $d=2$, the components of the Carroll--Cotton tensors can be dualised into a spacetime two-index object thanks to the three-dimensional Levi-Civita symbol $\varepsilon_{ABC}$ where indices $0$ correspond to the longitudinal direction and $\varepsilon_{0ab} \equiv -\varepsilon_{ab}$ in order to match with the $c\to 0$ limit of the Lorentz-relativistic case. We set $C^A{}_{BC} = C^{AD}\varepsilon_{BCD}$ as before, with the proviso that now the $0$-indices can no longer be raised in a meaningful way. We have
\begin{equation}
	\vect R[\gen K]^0 = \varepsilon_{ab} \big(C^{0a} \htheta^b\wedge\bftau + \tfrac12 C^{00} \htheta^a \wedge \htheta^b\big),\qquad \vect R[\gen K]^a = \varepsilon_{bc} \big(C^{ab}\htheta^c\wedge \bftau + \tfrac12 C^{a0} \htheta^b \wedge \htheta^c \big), \label{eq:cotton carroll in comp}
\end{equation}
where the remaining two-dimensional Levi-Civita symbols implement transverse dualisation. On account of Bianchi identities \eqref{eq:bianchi carroll RJab}, \eqref{eq:bianchi carroll RBa} and \eqref{eq:bianchi carroll RD}, the transverse part of the Carroll--Cotton tensors is symmetric and tracefree and the mixed time-space components are equal. Indeed, in three dimensions, requiring that the Ricci tensor vanishes is equivalent to cancel the whole Riemann tensor, and this statement holds irrespective of the Lorentzian or Carrollian nature of the background manifold. Therefore, the conditions \eqref{eq:EOM carroll} are equivalent to
\begin{equation}
	\vect R[\gen P]^a = \bm 0,\quad \vect R[\gen H] = \bm 0,\quad \vect R[\gen D] = \bm 0,\quad \vect R[\gen J]^{ab} = \bm 0,\quad \vect R[\gen B]^a = \bm 0,
\end{equation}
and the quoted Bianchi identities degenerate into
\begin{subequations}
	\begin{alignat}{3}
		\vect R[\gen K]^{[a}\wedge\htheta^{b]} = \bm 0 &\qquad\Rightarrow\qquad& &C^{[ab]} = 0, \\
		\vect R[\gen K]^a \wedge \bftau = \vect R[\gen K]^0 \wedge\htheta^a &\qquad\Rightarrow\qquad& &C^{0a} = C^{a0}, \\
		\vect R[\gen K]^a\wedge\htheta_a = \bm 0 &\qquad\Rightarrow\qquad& &C^{ab}\delta_{ab}=0.
	\end{alignat}
\end{subequations}
The Carroll--Cotton tensors thus separate into a Carrollian scalar $C \coloneqq C^{00}$, a Carrollian vector $C^a \coloneqq C^{a0} = C^{0a}$ and a Carrollian symmetric traceless rank-two tensor $C^{ab} = C^{\langle ab\rangle}$, summing up to five independent components. In terms of the symmetric tracefree part of the Carroll--Schouten tensor, they are given in full generality by
\begin{subequations}
	\begin{align}
		C &= \big(\bar{\mathscr D}^a\bar{\mathscr D}_a + 2\bar{\mathscr K}\big)(\ast\varpi) -\tfrac12 \bar{\mathscr D}_a\bar{\mathscr D}_b (\ast\mathscr C^{ab}) - \tfrac12 (\ast\mathscr C^{ab})\kappa_{\langle ab\rangle} \nonumber \\
		&\quad + (\ast\varpi)(\bar{\mathscr D}_a+\beta_a\big)\beta^a + 2 \bar{\mathscr D}_a\big(\!\ast\!\varpi\beta^a\big)  , \label{eq:Cgen}\\
		C^a &= -\bar{\mathscr D}_b\big(\!\ast\!\kappa^{\langle ab\rangle}\big) + \tfrac12\big(\!\ast\!\bar{\mathscr D}{}^a \bar{\mathscr K}\big) + \tfrac12 \bar{\mathscr D}^a\bar{\mathscr A} - 2(\ast\varpi)\bar{\mathscr R}^a,\label{eq:Cagen} \\
		C^{ab} &= \ast\bar{\mathscr D}_\bot \kappa^{\langle ab\rangle} - \ast\big(\bar{\mathscr D}^{\langle a}+\beta^{\langle a}\big)\bar{\mathscr R}^{b\rangle}. \label{eq:Cab gen}
	\end{align}
\end{subequations}
Moreover, the scalar and vector components obey Carrollian `evolution' equations on the account of the Bianchi identities \eqref{eq:bianchi carroll RKa} and \eqref{eq:bianchi carroll RK0}:
\begin{subequations}\label{eq:carroll cotton evolution}
	\begin{align}
		\pmb{\mathscr D}\wedge \vect R[\gen K]^0 &= \bm 0\qquad\Rightarrow\qquad \bar{\mathscr D}_\bot C + \big(\bar{\mathscr D}_a+\beta_a\big) C^a - \tfrac12\mathscr C_{ab} C^{ab} = 0, \label{eq:Dbot C} \\
		\pmb{\mathscr D}\wedge \vect R[\gen K]^a &= \bm 0 \qquad\Rightarrow\qquad \bar{\mathscr D}_\bot C^a + \bar{\mathscr D}_b C^{ab} = 0.\label{eq:Dbot Ca}
	\end{align}
\end{subequations}
We now select again the traceless rheotactic gauge, $\beta = 0 = \beta_a$, to display explicit expressions for the independent pieces of information encompassed by the Carroll--Cotton tensors. As already mentioned, the related Weyl--Carroll--Levi-Civita connection coincides with the $c\to 0$ limit of the Lorentzian Weyl--Levi-Civita in rheotactic gauge, and, in particular, the transverse projected connection is such that $\bar{\mathscr D}\equiv \hat{\mathscr D}$. The components of the Weyl curvature are
\begin{equation}
	\hat{\mathscr R}_a = \hat{\nabla}_\bot \varphi_a - \tfrac12\hat\nabla_a\theta,\quad
	\Omega_{ab} = -\varepsilon_{ab}\hat{\mathscr A},\quad \hat{\mathscr A} = \ast\varpi\,\theta - \varepsilon^{ab}\hat{\nabla}_a\varphi_b = -2\hat{\mathscr D}_\bot(\ast\varpi),
\end{equation}
which corresponds to the $c\to 0$ limit of Eqs. \eqref{eq:weyl curv rheotactic}--\eqref{eq:weyl curv rheo 2}. A straightforward computation on the basis of Eqs. \eqref{eq:riemann abcd 2D carroll} and \eqref{eq:riemann carroll 2D suite} allows us to derive:
\begin{equation}
\boxed{
\begin{aligned}
	C &= \big(\hat{\mathscr D}_a \hat{\mathscr D}{}^a + 2\hat{\mathscr K} \big)(\ast\varpi) - \tfrac12 \hat{\mathscr D}_a \hat{\mathscr D}_b (\ast \mathscr C^{ab}) + \tfrac14 (\ast \mathscr C_{ab}) {\mathscr N}^{ab} , \\
	C^{a} &= \tfrac12 \hat{\mathscr  D}_b (\ast\mathscr N^{ab}) + \tfrac12 \hat{\mathscr D}{}^a \hat{\mathscr A} + \tfrac12\! \ast\! \hat{\mathscr D}{}^a \hat{\mathscr K} - 2 (\ast \varpi) \hat{\mathscr R}{}^a,\\
	C^{ab} &= -\tfrac12 \hat{\mathscr D}_\bot (\ast\mathscr N^{ab}) + \tfrac12 \!\ast\!\hat{\mathscr D}{}^b\hat{\mathscr R}{}^a - \tfrac12 \hat{\mathscr D}{}^a(\ast\hat{\mathscr R}{}^b).
\end{aligned}
} \label{eq:carroll cotton rheo}
\end{equation}
In the gravitational/holographic setting, $C$ is twice the \textit{magnetic-mass aspect} of the dual gravitational solution, see Eq. (3.38) of \cite{Campoleoni:2023fug}, also referred to as \textit{dual mass aspect} \cite{Freidel:2021fxf,Freidel:2021qpz}. In terms of the variables $C_{(0)}$, $\chi^a$ and $X^{ab}$ respectively introduced in Eqs. \eqref{eq:Cotton Cs}, \eqref{eq:Cotton psi chi z} and \eqref{eq:Cotton Psi Chi Z}, the Carroll--Cotton tensors \eqref{eq:carroll cotton rheo} read
\begin{equation}
	\begin{split}
		C &= C_{(0)} - \tfrac12 \hat{\mathscr D}_a \hat{\mathscr D}_b (\ast \mathscr C^{ab}) + \tfrac14 (\ast \mathscr C_{ab}) {\mathscr N}^{ab}, \\
		C^a &= \chi^a\big|_{\xi_{ab} = 0} + \tfrac12 \hat{\mathscr  D}_b (\ast\mathscr N^{ab}), \\
		C^{ab} &= X^{ab}\big|_{\xi_{ab}=0} -\tfrac12 \hat{\mathscr D}_\bot (\ast\mathscr N^{ab}).
	\end{split}	
\end{equation}
More importantly, assuming that the limit is taken while keeping $\mathscr C_{ab}$ finite in the constraints \eqref{eq:hypotheses limit}, we can check that the Carroll--Cotton tensors derive from the $c\to 0$ limit of the components of the Lorentz-relativistic Cotton tensor, \eqref{eq:cotton in c2}, reduced with respect to a congruence $\bfupsilon$ of timelike observers that becomes the Carrollian field of observers in the limit, \textit{i.e.},
\begin{equation}
	C \equiv \lim_{c\to 0} C^{00},\quad C^a\equiv \lim_{c\to 0} C^{a0},\quad C^{ab} \equiv \lim_{c\to 0} C^{\langle ab\rangle}.
\end{equation}
The constraints \eqref{eq:hypotheses limit} have the net effect of implementing the vanishing of the Carrollian shear $\xi_{ab}$ by trading it for $\mathscr C_{ab}$ up to $c^2$ factors that mix the various terms in the expansion \eqref{eq:cotton in c2} and allow for crucial contributions of $\mathscr C_{ab}$ and $\mathscr N_{ab}$ to appear in the Carroll--Cotton tensors. Therefore, our expressions in Eq. \eqref{eq:carroll cotton rheo} offer the generalisation in the presence of a non-trivial Carroll-boost connection of the Carroll--Cotton tensors previously introduced via the $c\to 0$ limit.

To conclude this discussion, let us display the expressions of the Carroll--Cotton tensors in the conventional Bondi--Sachs gauge \cite{Bondi:1962px,Sachs:1962wk} and match its components with well-known bulk gravitational quantities. From a Carroll-geometric point of view, and as reviewed in \cite{Fiorucci:2025twa}, the choice of gauge coined as Bondi--Sachs amounts to embed null infinity in the bulk as a leaf of a Gaussian null normal foliation. This supposes that the induced boundary Ehresmann connection $\bftau$ is integrable, and can therefore be written as
\begin{equation}
	\bftau = \D u,\quad u\in\mathscr C^\infty(\mathscr C). \label{eq:tau bondi}
\end{equation}
In this context, $u$ is treated as a coordinate and extends into the bulk as a timelike coordinate representing retarded (resp. advanced) time of asymptotic observers near future (resp. past) null infinity. Owing to \eqref{eq:tau bondi}, both acceleration $\varphi_a$ and vorticity $\varpi_{ab}$ vanish, hence $\alpha_a = 0$ in rheotactic gauge. Furthermore, one can use the local gauge freedom, \textit{i.e.} Weyl transformations and rotations, to reach a frame in which $\theta=0$ and $c_{[ab]}=0$, which is referred to as a \textit{Bondi frame} \cite{Ashtekar:1981hw}. This restricts the residual gauge parameters $B$ and $\lambda_{ab}$ to be invariant on the flow of $\bfupsilon$ and makes the full Weyl connection vanish, $\bfalpha=\bm 0$. Therefore, the Bondi--Sachs gauge actually allows us to identify $\hat{\mathscr D}_\bot \equiv \hat\nabla_\bot \equiv \partial_u$, $\hat{\mathscr D}_a \equiv \hat\nabla_a$ and to set $\hat{\mathscr R}^a = 0 = \hat{\mathscr A}$.
\begin{description}
 	\item[Remark.] Looking at total transformations --- that is the combination of infinitesimal general diffeomorphisms and local gauge transformations generated by rotations, boosts and dilations --- that preserve the particular choice of a Bondi frame allows one to express all local gauge parameters in terms of the components of the diffeomorphism $\bm \upxi = f \bfupsilon + Y^a \hframe_a$
\begin{equation}
 	\lambda_a(\bm\upxi) = \hframe_a(f),\quad B(\bm\upxi) = \bfupsilon(f), \quad \lambda_{ab}(\bm\upxi) = \delta_{c[a} \htheta^c(\mathscr L_{\bm\upxi}\hframe_{b]}),
 \end{equation}
and imposes the following constraints on the diffeomorphism
\begin{equation}
 	\htheta^a(\mathscr L_\bfupsilon \bm\upxi) = 0, \quad \delta_{c(a} \htheta^c\big(\mathscr L_{\bm\upxi}\hframe_{b)}\big) + \delta_{ab} B(\bm\upxi) = 0.
\end{equation}
As was explained, \textit{e.g.}, in \cite{Fiorucci:2025twa}, these conditions imply that the $Y^a$ are invariant along the longitudinal direction and satisfy the conformal Killing equation on $\mathscr S$, further reducing general boundary diffeomorphisms to the well-known BMS transformations $f = T + \frac{u}{2} \hat\nabla_a Y^a$, where $T$ is an arbitrary function on $\mathscr S$ known as a \textit{super-translation}. Along with the choice of the tracefree rheotactic gauge, this implies that all the local gauge parameters are fixed in terms of $T$ and $Y^a$, and it can be checked that the behaviour of the various fields, for instance $\mathscr C_{ab}$, reproduce the ones arising from a bulk analysis \cite{Barnich:2010eb} under BMS transformations.

\end{description}
Our Eqs. \eqref{eq:carroll cotton rheo} thus reduce to the well-known expressions
\begin{subequations}
\begin{align}
	C &= - \tfrac12 \hat\nabla_a\hat\nabla_b (\ast\mathscr C^{ab}) + \tfrac14\!\ast\!\mathscr C_{ab}\mathscr N^{ab}, \label{eq:C00 bondi}\\
	C^{a} &=  \tfrac12 \hat\nabla_b (\ast\mathscr N^{ab}) + \tfrac14\! \ast\!\hat\nabla{}^a \hat R ,\\
	C^{ab} &= \tfrac{1}{2} \partial_u (\ast\mathscr N^{ab}),
\end{align}
\end{subequations}
in the context of asymptotically flat gravity, where ${\mathscr N}_{ab} = \partial_u\mathscr C_{ab}$ is now the original Bondi news tensor introduced in \cite{Bondi:1962px} and $\hat R$ is the Ricci scalar on $\mathscr S$. Projecting these five components onto the Newman--Penrose double-null basis \cite{Newman:1961qr,Newman:1962cia}, we have the following correspondence
\begin{equation}
	C \leftrightarrow \text{Im}\Psi_2^0,\quad C^{a} \leftrightarrow \text i\Psi_3^0,\quad C^{ab} \leftrightarrow \text i\Psi_4^0 \label{eq:link with NP}
\end{equation} 
with the leading components of the near-zone and Coulombic Newman--Penrose coefficients. In particular, $C$ is sensitive to the Taub-NUT parameter of the solution under scrutiny. Following \cite{10.1063/1.525274,Campoleoni:2023fug}, the NUT aspect $N$ describing a solution to four-dimensional asymptotically flat Einstein gravity is given by the following combination
\begin{equation}
	N \coloneqq \tfrac12 C + \tfrac18 \mathscr C_{ab}(\ast\mathscr N^{ab})
\end{equation}
of the magnetic-mass aspect $C$ and the Bondi news tensor $\mathscr N_{ab}$. Evolution equations of the Carroll--Cotton tensors, Eq. \eqref{eq:carroll cotton evolution}, then take the following familiar form \cite{Freidel:2021qpz,Mittal:2022ywl,Campoleoni:2023fug}
\begin{equation}
	\partial_u C^{a} + \hat{\nabla}_b C^{ab} = 0, \quad \partial_u C + \hat{\nabla}_a C^{a} - \tfrac12 \mathscr C_{ab} C^{ab} = 0.
\end{equation}
The second one in fact implies the conservation of the Taub-NUT charge \cite{10.1063/1.525274} obtained from integrating $N$ on each cut of null infinity, homeomorphic to the transverse Riemannian space $\mathscr S$.\footnote{Indeed, the $u$-derivative of the NUT aspect $N$ reduces to a divergence on $\mathscr S$, which cancels whenever $\mathscr S$ is assumed to have no boundaries. Since $\mathscr N_{ab}(\ast \mathscr N^{ab})=0$, the evolution equation for $C$ becomes $\partial_u N + \tfrac12 \hat{\nabla}_aC^a =0$.} Finally, the evolution equation for $C^a$ does not provide any non-trivial information on the gravitational side as it boils down to the usual bulk consistency equation between the time derivative of $\Psi^0_3$ and the space derivative of $\Psi^0_4$ in the Newman--Penrose viewpoint.

\section{Radiative fields at null infinity and covariance} 
\label{sec:News}

Thanks to the gauging of the conformal Carroll algebra, we have new tools to discuss the encoding of gravitational radiation in the flat-space holographic setting, focusing here on the physically relevant bulk four-dimensional case, hence setting $d=2$ everywhere. In particular, we discuss more specifically the instrumental role of the Carroll--Cotton tensors in that respect, while pondering the quest for a covariant notion of Bondi news tensor under the relevant symmetries. The latter endeavour is rather ancient and dates back to Geroch \cite{Geroch:1977big}, who addressed the question in a coordinate-free way but assuming the Bondi condition $\theta=0$. It has been pursued more recently \cite{Barnich:2010eb,Compere:2018ylh,Campoleoni:2023fug,Rignon-Bret:2024gcx,Rignon-Bret:2024wlu,Rignon-Bret:2024mef} by relaxing the asymptotic structure at null infinity namely under the action of the so-called superrotations, \textit{i.e.}, either local conformal transformations \cite{Barnich:2010eb,Barnich:2011ct,Barnich:2011mi} or general diffeomorphisms on the celestial sphere \cite{Campiglia:2014yka,Campiglia:2015yka,Compere:2018ylh,Fiorucci:2021pha}; and under the action of arbitrary boundary Weyl rescalings \cite{Barnich:2010eb,Campoleoni:2023fug,Fiorucci:2025twa}. We show here that a definition of the news tensor that is covariant under all the local symmetries\footnote{Ensuring that a tensorial quantity tensor is covariant under all the local symmetries arising from the gauging of the conformal Carroll algebra is sufficient to ensure its covariance under residual diffeomorphisms arising from boundary gauge fixings, because general covariance is already guaranteed. This is the case, \textit{e.g.}, for the news tensor, see \cite{Fiorucci:2025twa} for more details.} cannot be constructed from the curvatures related to a particular bulk gravitational solution, but requires to compare them with those of a background structure, that is, by design, recognised as non-radiative. To discuss all these concepts in what follows, we consider a Weyl--Carroll structure endowed with a Weyl--Carroll--Levi-Civita connection, hence with $\xi_{ab} = 0$, whose prototypical example is null infinity. Since we do not want to introduce too much nomenclature, we shall frequently assimilate the two notions by slight abuse of terminology. We first discuss non-radiative configurations, that shall be assimilated with conformally flat geometries at null infinity; next, we discuss the construction of the compensator to reach a definition of physical news tensor in the most general context, \textit{i.e.}, without assuming any local-symmetry or coordinate gauge fixing.

\subsection{Hierarchy of conformally flat Carroll geometries}
\label{sec:confflat}

In the following lines, we introduce three hierarchised notions of conformal flatness adapted to three-dimensional Carroll geometries and discuss their respective holographic interpretation. Inspired by the original definition of flat Carrollian geometry by Henneaux \cite{Henneaux:1979vn}, we qualify as \textit{conformally flat} a Carrollian manifold $(\mathscr C,\bfg,\bfupsilon)$ for which there exists a non-vanishing function $\mathscr B\in\mathscr C^\infty(\mathscr C)$ such that the metric and the volume form obey $\bfg = \mathscr B^{-2}\bfg_0$ and $\bfmu = \mathscr B^{-3}\bfmu_0$, where $\bfg_0$ and $\bfmu_0$ are constant in at least one coordinate system. In what follows, we find criteria to test conformal flatness in a coordinate-free way and also discuss the effect on the degrees of freedom in the connection.

As we have reviewed in Section \ref{sec:gauging Lorentzian}, a Lorentzian manifold is recognised as locally conformally flat\footnote{A three-dimensional Lorentzian manifold $\mathscr M$ is globally conformally flat if it is locally conformally flat and admits a developing map from its universal covering onto the three-dimensional Einstein space, \textit{i.e.}, the conformal compactification of the three-dimensional Minkowski space, that is a local conformal immersion. One also requires that this map  is equivariant under a holonomy representation $\rho : \pi_1(\mathscr M)\to SO(3,2)$ of its conformal group. The Carrollian instance of this statement is immediate to find, \textit{mutatis mutandis}. We shall not enter into these global issues in this paper and therefore stick to the local notion of conformal flatness provided by the Carroll--Cotton tensors.} if the Cotton tensor $C^{AB}$ derived from the (Weyl--)Levi-Civita connection related to the equipped (Weyl-)metric structure vanishes. Contrary to the Lorentzian case, we have a bit more refinement at our disposal in the Carrollian case. Indeed, owing to the transformation laws of the curvatures \eqref{eq: transfo Carrollian curvatures} we deduce that the components $C$, $C^a$ and $C^{ab}$ of the Carroll--Cotton tensors transform into the Weyl-weight-three irreducible representations of $\mathfrak{so}(2)$, respectively scalar, vector and symmetric traceless tensor representations:
\begin{equation}
	\delta_{(\lambda,B)}C = 3BC,\qquad \delta_{(\lambda,B)} C^a = \lambda^a{}_b C^b + 3BC^a,\qquad \delta_{(\lambda,B)} C^{ab} = 2\lambda^{\langle a}{}_c C^{b\rangle c} + 3 B C^{ab}, 
\end{equation}
but into an irreducible indecomposable representation (multiplet) under Carroll boosts:
\begin{equation}
	\delta_{\bm{\uplambda}} C = -2\lambda_a C^a,\qquad \delta_{\bm{\uplambda}} C^a = -\lambda_b C^{ab},\qquad \delta_{\bm{\uplambda}} C^{ab} = 0.
\end{equation}
Therefore, one can require that only parts of the Carroll--Cotton tensors vanish without breaking the local symmetries at work, leading to a hierarchy of concepts of conformal flatness on which we briefly comment below:
\begin{enumerate}
	\item \textit{Quadrupolar conformal flatness:} $C^{ab} = 0$. The evolution by the flow of $\bfupsilon$ of the degrees of freedom $\mathscr C_{ab}$ is constrained by a second-order differential equation. All gauge curvatures vanish, except $\vect R[\gen K]^0$ and the transverse projection of $\vect R[\gen K]^a$. At future null infinity, this corresponds to the absence of outgoing gravitational radiation as the boundary Noether charges are conserved by the flow of $\bfupsilon$.
	\item \textit{Dipolar conformal flatness:} $C^{ab} = 0$ and $C^a=0$. Since $\mathscr D_\bot C^a=0$ when $C^{ab}$ vanishes by virtue of Eq. \eqref{eq:Dbot Ca}, the second condition can be imposed on one element of the transverse distribution only. This partially constrains the initial data for the evolution problem of $\mathscr C_{ab}$. All gauge curvatures vanish except the transverse components of $\vect R[\gen K]^0$. At future null infinity, this allows one to fix completely one of the two `integration constants' in $\mathscr C_{ab}$.
	\item \textit{Total conformal flatness:} $C^{ab}=0$, $C^a=0$ and $C=0$. When the transverse parts of the Carroll--Cotton tensors vanish, $C$ is conserved by the flow of $\bfupsilon$ by virtue of Eq. \eqref{eq:Dbot C}. The condition $C=0$ can then again be imposed on one element of the transverse distribution only. This spoils the residual freedom in the initial data in the evolution problem of $\mathscr C_{ab}$. All gauge curvatures vanish. At future null infinity, this corresponds to geometries that are diffeomorphic to Minkowski spacetime, \textit{i.e.}, gravitational vacua. The shear $\mathscr C_{ab}$ reduces to its electric, \textit{i.e.} parity-even, modes and Taub--NUT charges are absent.
\end{enumerate}
An important observation is in order before proceeding. Carroll conformal flatness in three dimensions does not restrict the background geometry as much as the degrees of freedom of the gauge connection. This is rooted again into the fibre-bundle structure of such geometries. Indeed, deviations from conformal flatness in the usual sense can only appear on the base space, $\mathscr S$, and parallel transport and curvature along the fibre driven by $\bfupsilon$ is either absent by design or left to the degrees of freedom $\mathscr C_{ab}$. Since $\mathscr S$ is Riemannian and two-dimensional, it is automatically conformally flat. Conclusively, the only constraints a requirement of conformal flatness can impose shall involve the degrees of freedom in the gauge connection. This feature offers again a striking difference compared to three-dimensional (pseudo-)Riemannian geometry.

In this Section, we shall often exemplify our results by setting the Bondi--Sachs gauge, but \textit{without} necessarily restricting ourselves to a Bondi frame. We thus have $\varphi_a = 0$, $\varpi_{ab} = 0$, $c_{[ab]}=0$, but $\theta\neq 0$, hence rheotactic and transverse metric gauge fixings are equivalent and boil down to $\alpha_a = 0 = \beta_a$. We also assume that $\beta = 0$, which restricts the gauge parameters as \cite{Fiorucci:2025twa}
\begin{equation}
	\lambda_a = \hframe_a(f),\quad \bfupsilon(\lambda_{ab}) = 0,\quad B = \bfupsilon(f),\quad Z_a = \bfupsilon(\lambda_a) = \hframe_a(B),\quad Z^0 = -\tfrac12\hat\nabla_a\lambda^a, \label{eq:gaugefixed gauge param}
\end{equation}
where $\bm\upxi = f\bfupsilon+Y^a\hframe_a$ is an arbitrary boundary diffeomorphism.

\subsection{Quadrupolar conformal flatness} 
One can start by requiring $C^{ab}=0$, which we coin as the condition of \textit{quadrupolar conformal flatness}, since only the spin-two Carroll--Cotton tensor vanishes. Owing to Eq. \eqref{eq:Cab gen}, the symmetric tracefree part of the Carroll--Schouten tensor therefore obeys the following first-order differential equation along the fibre:
\begin{equation}
	\bar{\mathscr D}_\bot\kappa_{\langle ab\rangle} = \big(\bar{\mathscr D}_{\langle a}+\beta_{\langle a}\big)\bar{\mathscr R}_{b\rangle}. \label{eq:Cab=0 gen}
\end{equation}
We denote by $\kappa_{\langle ab\rangle} = -\tfrac12\rho_{\langle ab\rangle}$ the general solution to this equation. To comply with Eq. \eqref{eq:schouten carr in terms of Riemann}, we set
\begin{equation}
	\rho{}^a{}_a = -\bar{\mathscr R} \label{eq:geroch1}
\end{equation}
to be able to write $\kappa_{(ab)} = -\tfrac12\rho_{ab}$, defining $\rho_{ab}$ as a symmetric transverse tensor. In the presence of a non-trivial Weyl curvature, we recall that $\kappa_{ab}$ is not directly symmetric, but can be canonically brought back to its symmetric part thanks to Bianchi identities without losing any physical information, see Appendix \ref{sec:ids} for more details. By virtue of Eq. \eqref{eq:Cab=0 gen}, the tensor $\rho_{ab}$ contains an `integration constant' in the form of a symmetric rank-two tensor, $\hat\rho_{ab}$, whose tracefree part is parallel transported along the fibre, $\bar{\mathscr D}_\bot \hat\rho_{\langle ab\rangle} = 0$. The latter provides initial data for the Cauchy problem controlling the evolution of the Carroll--Schouten tensor on the flow of $\bfupsilon$. The right-hand side of \eqref{eq:Cab=0 gen} only depends on the Weyl connection, and therefore is left to gauge freedom. In the Weyl-flat gauge, we find $\kappa_{(ab)} = -\tfrac12\hat{\rho}_{ab}$. At the level of the degrees of freedom $\mathscr C_{ab}$, the requirement that $C^{ab}=0$ is a second-order differential equation along the fibre. Still in Weyl-flat gauge, choosing further the Bondi--Sachs gauge, this differential equation is solved by the particularly simple expression \cite{Compere:2016jwb,Compere:2018ylh} 
\begin{equation}
	\mathscr C_{ab} = \hat{\mathscr C}_{ab} + u\, \hat\rho_{\langle ab\rangle}, \label{eq:Cab linear in u}
\end{equation}
where $\hat{\mathscr C}_{ab}$ and $\hat\rho_{\langle ab\rangle}$ are both $u$-independent. 

In holographic applications, the condition $C^{ab} = 0$ has been recognised, from the early times of asymptotic analysis by Pirani \cite{1957PhRv..105.1089P}, Sachs \cite{Sachs:1961zz}, Newman and Penrose \cite{Newman:1962cia,Newman:1961qr}, and revisited recently in gauge-fixing approaches \cite{Compere:2018ylh,Fiorucci:2021pha,Freidel:2021fxf,Freidel:2021cjp,Kmec:2024nmu}, as the condition of \textit{absence of outgoing radiation} at future null infinity. More precisely, it is sufficient to ensure that the \textit{improved mass aspect}, defined as
\begin{equation}
	\mathscr M \coloneqq M + \tfrac18 \mathscr C_{ab}\mathscr N^{ab} = \text{Re} \Psi^0_2
\end{equation}
from the Bondi mass aspect $M$, leads to a definition of gravitational energy that is conserved in Bondi time. Indeed, on account of bulk Einstein equations, the improved mass aspect evolves as prescribed by the Bondi--Trautman mass loss formula \cite{Trautman:1958zdi,Bondi:1962px,Tamburino:1966zz,Barnich:2010eb}, conveniently written as
\begin{equation}
	\partial_u \mathscr M = \tfrac14 (\ast\mathscr C_{ab})C^{ab} - \tfrac12 \hat{\nabla}_a (\ast C^a) . \label{eq:bondi mass loss}
\end{equation}
Importantly, this equation establishes by its own the central role of the boundary Carroll--Cotton tensors in driving radiative fluxes at null infinity \cite{Ciambelli:2018wre,Mittal:2022ywl,Campoleoni:2023fug,Ciambelli:2024kre,Fiorucci:2025twa}. If $C^{ab} = 0$, the right-hand side reduces to a divergence on a cut of null infinity, which vanishes upon integration on a closed manifold like the celestial sphere. It is important to stress that in spite of its merits regarding BMS covariance, link with Newman-Penrose coefficients, and simplicity of the related flux-balance law, the improved mass aspect does not lead to a positivity theorem for the energy flux, in the sense that the first term in the right-hand side of Eq. \eqref{eq:bondi mass loss} is not negative definite.\footnote{It has been argued that it is the case up to a total derivative in $u$: hence, after integration by parts, one can discard the resulting boundary term if one imposes stronger early- and late-time boundary conditions on the Bondi shear $\mathscr C_{ab}$. These conditions are however unphysical beyond the linearised/scattering regime of the bulk theory of gravity.} Furthermore, the at-most-linear dependence on $u$ in the tensor $\mathscr C_{ab}$ is the typical behaviour of gravitational vacua \cite{Compere:2016jwb,Compere:2018ylh,Freidel:2021fxf} obtained by exponentiating the action of BMS vector fields around Minkowski spacetime. In this context, $\hat\rho_{\langle ab\rangle}$ has been coined as the \textit{vacuum news tensor}. We shall revisit this construction in a more general setting in Section \ref{sec:covariantnews}.

\subsection{Dipolar conformal flatness} When $C^{ab}=0$, one can impose $C^a = 0$ in agreement with all the local symmetries, to reach a configuration of \textit{dipolar conformal flatness}, since the spin-one component is now affected. Since $\bar{\mathscr D}_\bot C^a = 0$ when the transverse Carroll--Cotton tensor vanish, the condition $C^a=0$ can be imposed on any transverse section, modelled on $\mathscr S$, and therefore reduces to a constraints on the initial data for the Cauchy problem \eqref{eq:Cab=0 gen}. From Eqs. \eqref{eq:riemann abcd 2D carroll}, \eqref{eq:schouten carr in terms of Riemann} and \eqref{eq:Cagen}, we get
\begin{equation}
	\bar{\mathscr D}_{[b}\rho_{c]a} = \tfrac12 \varepsilon_{bc}\bar{\mathscr D}_a\bar{\mathscr A} - 2\varpi_{bc}\bar{\mathscr R}_a , \label{eq:Ca=0 gen}
\end{equation}
using the two-dimensional identity $\bar{\mathscr D}_{[b}\rho_{c]a} = \tfrac12\varepsilon_{bc} \varepsilon^{de}\bar{\mathscr D}_d \rho_{ea}$ to trade divergences for curls of two-dimensional symmetric tensors. The condition $C^a=0$ thus fixes the curl of $\rho_{ab}$ in terms of the Weyl curvature. When the Weyl connection is flat, we recover the following conditions \cite{Geroch:1977big}
\begin{equation}
	\bar{\mathscr D}_{[b}\hat\rho_{c]a} = 0,\qquad \hat\rho{}^a{}_a = -\bar{\mathscr R}, \label{eq:conditions Geroch}
\end{equation}
which have a deep geometric meaning as the flatness conditions of the projected gauge connection on $\mathscr S$, as we discuss in Section \ref{sec:geroch}. When the base $\mathscr S$ has the topology of a two-sphere, a theorem by Geroch states that the constraints \eqref{eq:conditions Geroch} admit a unique solution.\footnote{One can show that the set of `integration constants' in the solution to Eq. \eqref{eq:conditions Geroch} is finite-dimensional and grows with the genus $g$ of $\mathscr S$, as it is given by $3(g-1)$ for any $g\geq 2$ and $1$ for tori (with $g=1$) \cite{Burstall2010ConformalSG}.\label{foot:moebius}} This unique tensor $\hat\rho_{ab}$ has then been used by Geroch to improve the boundary Schouten tensor of asymptotically flat spacetimes and define a Weyl-covariant notion of the news tensor. The joint solution to Eqs. \eqref{eq:Cab=0 gen} and \eqref{eq:Ca=0 gen} offers a generalisation of this procedure to arbitrary geometries and gauge fixings at null infinity: for this reason, we shall refer to $\rho_{ab}$ as the \textit{Geroch tensor} in the general case.

Dipolar conformal flatness imposes that not only the charge $\int_\mathscr{S}\mathscr M$, but the momentum $\mathscr M$ itself, is conserved on the flow of Bondi retarded time, by virtue of Eq. \eqref{eq:bondi mass loss}.\footnote{For this reason, the conditions $C^{ab}=0$, $C^a=0$ of dipolar conformal flatness have also been coined to determine a \textit{strongly} non-radiative configuration \cite{Freidel:2021qpz}.} It still allows for an arbitrary $\hat{\mathscr C}_{ab}$ as well as an arbitrary `nut' charge, if topology is non-trivial. Therefore, this situation models the boundary geometry of gravitational solutions with no outgoing radiation at future null infinity and possible non-trivial topology/Taub-NUT charge at null infinity. However, it allows for a complete determination of the Geroch tensor, at least in the important case of topologically spherical celestial spaces. In geometrical terms, this makes the dipolar conformal flatness a stronger and handier non-radiation criterion.

\begin{description}
	\item[Remark.] In Bondi--Sachs gauge, the constraints \eqref{eq:conditions Geroch} are equivalent to $\hat\nabla^b \hat\rho_{\langle ab\rangle } +\tfrac12 \hat\nabla_a \hat R = 0$ and $\hat\rho^a{}_a = -\hat R$ \cite{Compere:2018ylh,Campiglia:2020qvc}. Setting conformal coordinates $x^\pm$ on $\mathscr S$, \textit{i.e.}, $\delta_{ab}\htheta^a\otimes\htheta^b = 2 e^{-\phi}\D x^+\D x^-$ for an arbitrary scalar field $\phi(x^+,x^-)$ on $\mathscr S$, these equations are solved by
\begin{equation}
	\hat\rho_{++} = \tfrac12 (\partial_+\phi)^2 - \partial_+^2\phi,\quad \hat\rho_{--} = \tfrac12 (\partial_-\phi)^2 - \partial_-^2\phi,\quad \hat\rho_{+-} = \partial_+\partial_-\phi = -\tfrac12\hat{R} e^{\phi}.
\end{equation}
On the unit-round sphere, $e^{-\phi} = \tfrac12(1+x^+x^-)^2$ and $\hat\rho_{++} = 0 = \hat\rho_{--}$. The Geroch tensor is therefore a pure trace. These considerations reproduce the heuristic computations of \cite{Compere:2018ylh}. 
\end{description}

\subsection{Total conformal flatness}  Finally, when both $C^{ab}$ and $C^a$ vanish, one can require that the complete Carroll--Cotton tensor vanishes, and thus $C=0$, in a local-conformal-Carroll-covariant way. This is the strongest acceptance of conformal flatness in Carroll geometries, which we refer to as \textit{total conformal flatness}.\footnote{The study of the most general $C=0$ condition, on the basis of Eq. \eqref{eq:Cgen} can be performed as an interesting problem of partial differential equations but is far beyond the scope of the present analysis.} Since $\bar{\mathscr D}_\bot C = 0$ on the account of the Bianchi identity \eqref{eq:Dbot C}, the last condition $C=0$ spoils the residual freedom in the initial data, namely imposing non-trivial constraints on $\hat{\mathscr C}_{ab}$. To disclose them, and to fix ideas, we work directly in the Bondi--Sachs gauge. On account of Eqs. \eqref{eq:C00 bondi}, \eqref{eq:Cab linear in u} and \eqref{eq:conditions Geroch}, this implies that
\begin{equation}
	\varepsilon^{ab} \left(\hat\nabla_a\hat\nabla_c -\tfrac12 \hat\rho_{ac}\right)\hat{\mathscr C}^c{}_b = 0. \label{eq:elec}
\end{equation}
The particular differential operator that appears in this condition has a deep geometrical origin. Indeed, when working in Bondi--Sachs coordinates within the rheotactic gauge fixing for the Weyl connection, Weyl transformations on $\mathscr S$ are locked in terms of local Lorentz boosts, \textit{i.e.}, purely-conformal transformations. In this case, $\hat\rho_{\langle ab\rangle}$ plays the role of Lorentz-boost connection \cite{Compere:2018ylh,Campiglia:2020qvc,Barnich:2021dta,Donnay:2021wrk} and the combination $\hat{\nabla}_{\langle a}\hat{\nabla}_{b\rangle} - \tfrac12 \hat\rho_{\langle ab\rangle}$ is in fact covariant under local Lorentz boosts. To solve Eq. \eqref{eq:elec}, we recall that $\hat{\mathscr C}_{ab}$ transforms into an irreducible representation of an $\mathfrak{so}(3)$ subalgebra of $\mathfrak{so}(3,1) \simeq \mathfrak{confcarr}_1(2,1)$. It is indeed symmetric traceless and can be decomposed as
\begin{equation}
	\hat{\mathscr C}_{ab} = -2\big(\hat{\nabla}_{\langle a}\hat{\nabla}_{b\rangle}-\tfrac12 \hat\rho_{\langle ab\rangle}\big) \hat{\mathscr C} + \varepsilon_{c(a}\hat{\nabla}_{b)}\hat\nabla^c \hat\Psi
\end{equation}
where $\hat{\mathscr C}$ and $\hat\Psi$ are arbitrary functions on $\mathscr S$ that parameterise the two independent degrees of freedom in $\hat{\mathscr C}_{ab}$. Owing to Eq. \eqref{eq:elec}, $\Psi$ is committed to obey the quartic differential equation $\hat{\nabla}^2(\hat{\nabla}^2+\hat R)\hat\Psi = 0$. If $\mathscr S$ has the topology of $S^2$, one can expand $\Psi$ in spherical harmonics. The quartic differential equation then implies that $\Psi$ vanishes up to the lowest $\ell = 0,1$ modes. However, these are annihilated by the differential operator $\varepsilon_{c(a}\hat{\nabla}_{b)}\hat\nabla^c$ and thus do not appear in the final tensor:
\begin{equation}
	\boxed{\mathscr C_{ab} = - 2\big(\hat{\nabla}_{\langle a}\hat{\nabla}_{b\rangle}-\tfrac12 \hat\rho_{\langle ab\rangle}\big) \hat{\mathscr C} + u\,\hat{\rho}_{\langle ab\rangle}.}
\end{equation}
On the round sphere where $\hat{\rho}_{\langle ab\rangle}$ disappears, $\hat{\mathscr C}$ is also defined up to the lowest $\ell = 0,1$ spherical-harmonic modes that encode the translation frame. Otherwise, the second-order differential operator is accompanied with a non-linear term $\hat{\mathscr C}\hat\rho_{\langle ab\rangle}$ and the sum of both terms do no longer annihilate translations.\footnote{This feature has been discussed in \cite{Compere:2018ylh} and coined as a non-linear memory effect, that can appear in elementary configurations like impulsive Robinson--Trautman gravitational waves.} The resulting tensor corresponds to the Bondi shear obtained from a finite supertranslation of intensity $\hat{\mathscr C}$ on Minkowski spacetime. This is therefore a pure-gauge tensor under the transformation \eqref{eq:delta Cab} when $\lambda_a = \hframe_a(f)$ \cite{Fiorucci:2025twa}, assuming that $\delta_f \hat{\mathscr C} = f$. As a conclusion, totally conformally flat Carroll manifolds are therefore meant to holographically encode stationary configurations of the gravitational field.

\subsection{Canonical nature of the Geroch tensor}
\label{sec:geroch}
We discuss here the geometrical nature of the Geroch tensor $\rho_{ab}$ through our gauging methods. We namely clarify the assertion according to which $\rho_{ab}$ lives on $\mathscr S$ rather than on the whole manifold. This property arises from the following algebraic statement.

We first recall that the three-dimensional isotropic conformal Carroll algebra, that we denote $\mathfrak{g}$ for short, is isomorphic to the Poincaré algebra, $\mathfrak{iso}(3,1) = \mathfrak{h}\loplus\mathfrak{t}$, where $\mathfrak{h}=\mathfrak{so}(3,1)\simeq \mathfrak{conf}(2)$, the Lorentz group in four dimensions.\footnote{See Eqs. \eqref{eq:decomposotion poincare} and \eqref{eq:carroll to poin gens} for the relation between Carroll and Poincaré generators. The isomorphism $\mathfrak{conf}(2)\simeq \mathfrak{so}(3,1)$ is an \textit{algebraic statement}, insensitive to the topology of $\mathscr S$, and our gauging methods only ever require it locally. In the context of asymptotically flat spacetime, $\mathfrak{conf}(2)$ is differentially realised by vector fields tangent to the celestial sphere, and this realisation depends on the global definiteness of the vectors. If this is the case, they form a six-dimensional sets of Lorentz rotations and boosts. If they are allowed to have one pole on the sphere, then the algebra is extended into two copies of the Witt algebra, leading to superrotations \cite{Barnich:2010ojg,Barnich:2010eb}. Global definiteness can be reached back if one trades the topology of the sphere for that of the plane, for instance. See, \textit{e.g.}, \cite{Barnich:2021dta} for a recent discussion.} The projection $\pi : \mathfrak g\to\mathfrak h$ is a Lie-algebra morphism since the Poincaré translations $\mathfrak t$ constitute an Abelian ideal of $\mathfrak g$. Hence, the projected gauge connection
\begin{equation}
	\bm\Gamma_\mathfrak{h} \coloneqq \pi(\bm\Gamma) = \gen P_a\htheta^a + \tfrac12\gen J_{ab}\bfgamma^{ab} - \gen D\bfalpha + \tfrac12\gen K_a \bfkappa^a \label{eq:A}
\end{equation}
is a rightful connection on $T\mathscr C$ with values into the Lorentz algebra. This is already clear at the level of the symmetries, since Eqs. \eqref{eq:transfo theta a}, \eqref{eq:transfo gamma carr}, \eqref{eq:transfo alpha carr} and \eqref{eq:transfo kappa carr} do not involve neither Carroll boosts nor vertical special conformal transformations, and the Lorentz sector $(\htheta^a,\bfgamma^{ab},\bfalpha,\bfkappa^a)$ of the gauge connection decouples. Moreover, $\bm\Gamma_\mathfrak{h}$ canonically descends onto $T\mathscr S$ if and only if it is projectable, \textit{i.e.}, transverse and invariant by $\bfupsilon$ up to gauge transformations. In equations:
\begin{equation}
	\bm\Gamma_\mathfrak{h}\in \Omega^1\big(\mathscr S,\mathfrak{conf}(2)\big)\quad\Leftrightarrow\quad \bm\Gamma_\mathfrak{h}(\bfupsilon) = 0,\quad \mathscr L_\bfupsilon\bm\Gamma_\mathfrak{h} = \bm 0 \label{eq:conditions projectable}
\end{equation}
in at least one gauge frame. Since
\begin{equation}
	\bm\Gamma_\mathfrak{h}(\bfupsilon) = -\tfrac12 \gen J_{ab}c^{[ab]} - \tfrac12 \gen D \theta + \tfrac12 \gen K_a \bar{\mathscr R}^a
\end{equation}
can always be eliminated by a choice of how the gauge parameters $\lambda_{ab}$, $B$ and $Z_a$ behave along the fibre, the requirement that $\bm\Gamma_\mathfrak{h}$ is projectable boils down to demanding that $\vect R[\bm\Gamma_\mathfrak{h}](\bfupsilon,\cdot) = \bm 0$ owing to Cartan's magic formula.\footnote{Indeed, choosing a gauge frame the conditions \eqref{eq:conditions projectable} are obeyed, one has $\mathscr L_\bfupsilon \bm\Gamma_\mathfrak{h} = \D\bm\Gamma_\mathfrak{h}(\bfupsilon,\cdot) + \D\big(\bm\Gamma_\mathfrak{h}(\bfupsilon)\big) = \vect R[\bm\Gamma_\mathfrak{h}](\bfupsilon,\cdot) = \bm 0$. However, the requirement that a piece of curvature vanishes is a gauge-invariant requirement, hence it holds in all gauge frames.} Since we work with a Weyl--Carroll--Levi-Civita connection by hypothesis, all curvatures vanish except $\vect R[\gen K]^a$, this condition amounts in fact to requiring that $\vect R[\gen K]^a(\bfupsilon,\cdot) = \bm 0$, or equivalently, $C^{ab} = 0$. Therefore, in the gauging viewpoint, the projected connection \eqref{eq:A} descends onto $\mathscr S$ if and only if the Carroll manifold is quadrupolar-conformally flat. When it is the case, $\bm\Gamma_\mathfrak{h}$ can be assimilated with a Lorentz connection on $T\mathscr S$.

Denoting by $\bfgamma =\tfrac12\varepsilon_{ab}\bfgamma^{ab} = \bfgamma^{12}$ the only independent spin-connection one-form, the aforementioned gauging condition reads as
\begin{equation}
	\D\htheta^a + \varepsilon^a{}_b\bfgamma\wedge\htheta^b - \bfalpha\wedge\htheta^a = \bm 0,\quad \D\bfgamma = \varepsilon_{ab}\bfkappa^a\wedge\htheta^b,\quad \bfkappa_a\wedge\htheta^a = \D\bfalpha.
\end{equation}
The first equation is satisfied if $\bfgamma$ is the Weyl--Levi-Civita connection on $\mathscr S$ for the particular choice of Weyl connection $\bfalpha = \alpha_a\htheta^a$. The second equation has only one independent component and only fixes the trace $\kappa^a{}_a = \tfrac12\bar{\mathscr R}$ of the special-conformal connection. The tracefree part of it, $\kappa_{\langle ab\rangle}$ is completely free. It has been coined by Calderbank as a \textit{M\"obius structure} \cite{Calderbank1998MOBIUSSA,2000math......1150C,Burstall2010ConformalSG}. Requiring further that $C^a = 0$ implies that the projected connection $\bm\Gamma_\mathfrak{h}$ is \textit{flat}, which constitutes a canonical choice that uniquely fixes $\kappa_{\langle ab\rangle}$ in terms of the background geometry, of course, up to topological obstruction, see Footnote \ref{foot:moebius}, and irrelevant gauge freedom. Indeed, 
\begin{equation}
	\vect R[\gen K]^a = \bar{\pmb{\mathscr D}}\wedge\bfkappa^a = (\bar{\pmb{\mathscr D}}\kappa^a{}_b)\wedge\htheta^b + (\bar{\pmb{\mathscr D}}\kappa^a{}_0)\wedge\bftau + \bar{\mathscr R}^a\big(\bar{\pmb{\mathscr D}}\wedge\bftau) = \bm 0,
\end{equation}
which reproduces Eqs. \eqref{eq:Cab=0 gen} and \eqref{eq:Ca=0 gen} after splitting into longitudinal and transverse components. As a conclusion, the Geroch tensor emerges from the gauging of the Lorentz algebra on $\mathscr S$, demanding that the projected connection $\bm\Gamma_\mathfrak{h}$ is flat.

\subsection{Physical news tensor}
\label{sec:covariantnews}

We now conclude by constructing a suitable improved definition of the Bondi news tensor that is covariant under all the symmetries. We first work in a general setting without imposing any gauge, and we proceed by illustrating our result in some gauge fixings and match with existing literature.

An instrumental remark is in order: the Bondi news tensor $\mathscr N_{ab}$, Eq. \eqref{eq:news tensor}, although defined through a Weyl-covariant derivative from the Bondi shear, is never a covariant object, as it is obviously sensitive to Carroll boosts. A Carroll covariant object is obtained from the Carroll--Schouten tensor, see \eqref{eq:schouten carr in terms of Riemann} and \eqref{eq:pieces of curvature carroll}, as
\begin{equation}
	\mathscr N_{ab}' \coloneqq -2\kappa_{\langle ab\rangle} = \mathscr N_{ab} + 2 (\bar{\mathscr D}_{\langle a}+\beta_{\langle a})\beta_{b\rangle}.
\end{equation} 
Its transformation law is computed as
\begin{equation}
	\delta_\Sigma\mathscr N_{ab}' = 2\lambda_{\langle a}{}^c \mathscr N'_{b\rangle c} + 2 B\mathscr N_{ab}' - 2 \bar{\mathscr D}_{\langle a}Z_{b\rangle} - 2 \lambda_{\langle a}\bar{\mathscr R}_{b\rangle} \label{eq:delta news prime}
\end{equation}
from Eq. \eqref{eq:transfo kappa carr}. The modified news tensor does no longer transform as a connection under Carroll boost. The last term is present whenever Weyl curvature is non-vanishing, which shows that, on general grounds, the news tensor is not a Carrollian tensor but only the fully transverse part of it. However, since it now involves derivatives of the field $\beta_a$ that transforms algebraically under transverse special conformal transformations, the modified news tensor behaves like a connection for these transformations. Since the difference of two connections is automatically a tensor, what should rather be considered as the \textit{physical news tensor} is the following difference
\begin{equation}
	\mathscr N_{ab}^{\text{phys}} \coloneqq \mathscr N_{ab}' - \mathscr N_{ab}'^{(0)},
\end{equation}
where $\mathscr N_{ab}'^{(0)}$ is the modified news of a given gravitational solution. This is the core of Geroch's procedure, here exploited to its full generality and geometrical meaning: the correct instance of the news tensor is nothing but the difference of the transverse projection of two special-conformal connections at null infinity, up to a numerical factor. By construction, $\mathscr N_{ab}'^{(0)}$ transforms exactly as $\mathscr N_{ab}'$ and we find that
\begin{equation}
	\delta_\Sigma \mathscr N_{ab}^{\text{phys}} = 2\lambda_{\langle a}{}^c\mathscr N_{b\rangle c}^{\text{phys}} + 2B \mathscr N_{ab}^{\text{phys}}.
\end{equation}
There exists a \textit{canonical choice} of origin to construct the physical news tensor, which consists in choosing for $\mathscr N_{ab}'^{(0)}\equiv \rho_{\langle ab\rangle}$ a flat special-conformal connection, \textit{i.e.}, a configuration where the special-conformal curvature vanishes: $\vect R[\gen K]^a = \bm 0$. This configuration is equivalent to a dipolar conformally flat Carroll manifold, with $C^{ab} = 0$ and $C^a=0$. It reduces to the usual Geroch tensor in a gauge where the Weyl connection is flat. As a result,
\begin{equation}
	\boxed{\mathscr N_{ab}^{\text{phys}} = \mathscr N_{ab} + 2(\bar{\mathscr D}_{\langle a}+\beta_{\langle a})\beta_{b\rangle} - \rho_{\langle ab\rangle }} \label{eq:nphys final}
\end{equation}
is the definitive instance of the fully coordinate-independent gauge-covariant news tensor.

Let us now exemplify the solution in Bondi--Sachs gauge, see Eq. \eqref{eq:gaugefixed gauge param}. From Eq. \eqref{eq:delta news prime}, we recover the celebrated anomalous transformation of the Bondi news tensor:
\begin{equation}
\begin{split}
	\delta_\Sigma \mathscr N'_{ab} &=  2\lambda_{\langle a}{}^c \mathscr N'_{b\rangle c} + 2 B\mathscr N'_{ab} - 2 \hat{\nabla}_{\langle a}\hat\nabla_\bot \lambda_{b\rangle} + \hframe_{\langle a}(\theta)\hframe_{b\rangle}(f) \\
	&= 2\lambda_{\langle a}{}^c \mathscr N'_{b\rangle c} + 2 B\mathscr N'_{ab} - 2 \hat{\nabla}_{\langle a}\hat\nabla_{b\rangle} B + \hframe_{\langle a}(\theta)\hframe_{b\rangle}(f),
\end{split} \label{eq:transfo bondi news}
\end{equation} 
see, \textit{e.g.}, \cite{Barnich:2010eb,Compere:2018ylh,Rignon-Bret:2024gcx}. After choosing Bondi--Sachs gauge, Carroll boosts and Weyl rescalings are no longer independent, which explains that the above inhomogeneous transformation can be interpreted either as a Carroll-boost anomaly or as a Weyl anomaly. In his seminal work, Geroch has shown how to repair the transformation of the news by adding a tensor transforming as a Weyl connection \cite{Geroch:1977big}, which was in fact, prior to any gauge fixing, a flat special-conformal connection. He also chose the Bondi frame, $\theta=0$, which makes the transformation law blind to the fact that, as we have already mentioned, the Bondi news tensor transforms into a Carroll-boost multiplet whenever the Weyl connection is not strictly flat. This is the role of the last term in Eq. \eqref{eq:transfo bondi news}. In Bondi--Sachs gauge, the expression of the physical news tensor is found to be\footnote{We thank Simone Speziale for discussions on that point.}
\begin{equation}
	\mathscr N_{ab}^\text{phys} = \big(\partial_u + \tfrac12\theta\big)\mathscr C_{ab} - 2(\bar{\nabla}_{\langle a}+\alpha_{\langle a})\alpha_{b\rangle} - \rho_{\langle ab\rangle}.
\end{equation}
The first term corresponds to the refined Bondi news tensor in the presence of a non-trivial $u$-dependent boundary metric \cite{Barnich:2010eb,Compere:2019bua,Geiller:2022vto}. The second term indicates that the physical news differs from the Bondi news tensor in the presence of a non-trivial Weyl connection, even on the round sphere where $\rho_{\langle ab\rangle}$ vanishes. This confirms the heuristic derivations of \cite{Campiglia:2020qvc,Freidel:2021dfs}. Finally, it is worth noting that, in the presence of a non-trivial expansion scalar, the Geroch tensor has a non-trivial $u$-dependence, governed by Eqs. \eqref{eq:Cab=0 gen} as
\begin{equation}
	(\partial_u+\theta)\rho_{\langle ab\rangle} = \hat{\nabla}_{\langle a}\hat{\nabla}_{b\rangle}\theta.
\end{equation}
However, it still obeys \eqref{eq:conditions Geroch} on each constant-$u$ section of null infinity, since $\hat{\mathscr A} = 0$ in Bondi--Sachs gauge. The solution to this equation reproduces the news tensor describing the strength of Robinson--Trautman waves, as derived in \cite{Barnich:2026wpw}. In conclusion, since gravitational radiation is holographically encoded in the deviation from conformal flatness at null infinity, the Carroll--Cotton tensors are naturally preferred over the news tensor as the geometrical objects carrying the boundary values of the radiative degrees of freedom. Indeed, the Carroll--Cotton tensors vanish identically in conformally flat configurations, whereas the Bondi news tensor, which involves one derivative fewer, remains non-trivial there; the Geroch improvement then has to be understood as measuring the departure from such configurations. This concludes our gateway into the gravitational applications of Carroll--Cotton tensors and their multiple declinations.

\section{Carroll--Chern--Simons actions}
\label{sec:ChernCarroll}

As we reviewed in Section \ref{sec:relat chern simons}, an elegant way to derive the Cotton tensor is to compute the covariant energy--momentum tensor related to the gravitational Chern--Simons action \eqref{eq:grav CS}. In this last section, we study the formulation of Carrollian avatars of the Chern--Simons action \eqref{eq:CS} and show how they encode information on the Carroll--Cotton tensors.

\subsection{Chern--Simons basics}

First, let us recall that a Chern--Simons action is not attached to any Lie algebra, but to a Lie algebra equipped with an invariant, symmetric, non-degenerate bilinear form $G$. Such an object exists for the three-dimensional conformal algebra and the non-vanishing pairings are given by Eq. \eqref{eq:GforConformal}. In the Carroll limit, this is no longer the case \cite{Figueroa-OFarrill:1995opp,Matulich:2019cdo}.\footnote{The complete statement is that the conformal Carroll algebra $\mathfrak{confcarr}_1(d,1)\simeq \mathfrak{iso}(d+1,1)$ admits a non-degenerate bilinear symmetric form in two dimensions only, \textit{i.e.}, for $d=1$ \cite{Witten:1988hc}.} This can be seen as follows. To ease the notation, we write $\mathfrak g=\mathfrak h\loplus \mathfrak t$, where $\mathfrak g=\mathfrak{iso}(3,1)$ is the four-dimensional Poincaré algebra, $\mathfrak h$ is the Lorentz subalgebra and $\mathfrak t$ the Abelian ideal of translations. The Carrollian generators from which each subalgebra is spanned are given by Eq. \eqref{eq:decomposotion poincare}. Let us assume that $G_\mathfrak{g}$ is a symmetric bilinear invariant form on $\mathfrak g$. Denoting elements of $\mathfrak h$ by $\gen J'{}_{MN}$ and elements of $\mathfrak t$ by $\gen P'{}_M$, where $M=0,1,2,3$ as usual, we get
\begin{equation}
\begin{split}
	G_\mathfrak{g}(\gen J'{}_{MN},\gen J'{}_{PQ}) &= G_\mathfrak{h}(\gen J'{}_{MN},\gen J'{}_{PQ}) = 2(\hat\eta_{MQ}\hat\eta_{NP}-\hat\eta_{MP}\hat\eta_{NQ}),\\
	G_\mathfrak{g}(\gen J'{}_{MN},\gen P'{}_P) &= 0,\\
	G_\mathfrak{g}(\gen P'{}_M,\gen P'{}_N) &= G_{\mathfrak{t}}(\gen P'{}_M,\gen P'{}_N) = 0.
\end{split} \label{eq:G for carroll}
\end{equation}
where $\hat\eta_{MN} = \text{diag}(-1,1,1,1)$ is the bulk Minkowski metric with the speed of light set to one. Indeed, by $\mathfrak{h}$-invariance, $G_\mathfrak{g}(\gen{J}'{}_{MN},\gen{P}'{}_P)$ would have to be an $\mathfrak{so}(3,1)$-invariant tensor with three indices that is skew-symmetric in the last pair. No such invariant exists, hence $G_\mathfrak{g}(\mathfrak{h},\mathfrak{t})=0$. Furthermore, $\mathfrak{h}$-invariance forces $G_\mathfrak{g}(\gen P'{}_M,\gen P'{}_N)=k\hat\eta_{MN}$ by Schur's lemma,
$\mathfrak{t}$ being $\mathfrak{h}$-irreducible. Requiring $\mathfrak{t}$-invariance, \textit{i.e.}, using Eq. \eqref{eq:invariance of G} with $\gen X\in\mathfrak{t}$, imposes $k=0$, hence $G_\mathfrak{g}(\mathfrak t,\mathfrak t)=0$. We conclude that any invariant bilinear form on $\mathfrak{confcarr}_1(2,1)$ is supported on the Lorentz subalgebra $\mathfrak{h}\simeq\mathfrak{so}(3,1)$. In other words, strictly conformal-Carroll-invariant
Chern--Simons action cannot depend on $\bftau$, $\bfbeta^a$ and $\bfkappa^0$, since the latter are the gauge fields conjugated to elements of the Abelian ideal $\mathfrak t$.

Consequently, it seems too naive to directly applying the defining formula \eqref{eq:CS} to the conformal Carroll algebra. We therefore take a step back, and consider an expansion in powers of $c^2$ of this action. Choosing a timelike congruence $\bfupsilon$ and splitting in longitudinal ($0$) and transverse ($a=1,2$) components, the special-relativistic Chern--Simons action is expanded as
\begin{equation}
	\begin{split}
		S_{\text{CS}} &= \frac{1}{2c}\int_{\mathscr C} \left[ \bfgamma^a{}_b \wedge \D \bfgamma^b{}_a + 2 \bfalpha \wedge \D \bfalpha - 4 \bfkappa_a \wedge \left(\D \htheta^a + \bfgamma^a{}_b \wedge \htheta^b - \bfalpha \wedge \htheta^a \right) \right] \\
		& + \frac{c}{2}\int_{\mathscr C} \left[ 2 \bfgamma^0{}_a \wedge \left(\D \bfgamma^{0a} + \bfgamma^a{}_b \wedge \bfgamma^{0b} \right) + 4 \bfkappa^0 \wedge \left(\D \bftau + \bfgamma^0{}_a \wedge \htheta^a - \bfalpha \wedge \bftau \right) + 4 \bfkappa^a \wedge \bftau \wedge \bfgamma^0{}_a \right] . \label{eq:CS in c2}
	\end{split}
\end{equation}
We have therefore two\footnote{This has to be contrasted with the four replicas derived by $c\to 0$ limit in Ref. \cite{Miskovic:2023zfz}. There, the question was to obtain the replicas of the \textit{gravitational} Chern--Simons action \eqref{eq:grav CS}, and the limit has been taken partially on-shell, \textit{i.e.}, assuming that composite torsion and Riemann curvature vanish. In this case, the additional orders originate not from the algebra but from the hidden powers of $c^2$ carried by $\xi_{ab}$ and $\varpi_{ab}$ in the geometric data, see Eqs. \eqref{eq:projconn} and \eqref{eq:hypotheses limit}. Contrariwise, we perform here the $c$-expansion completely off-shell.} Carrollian replicas of the Chern--Simons action, on which we comment in what follows. Note that we are only interested in the derivation of the actions and their link with the Carroll--Cotton tensors. Their relevance to three-dimensional Carroll gravity and applications thereof is an interesting question that shall be discussed elsewhere.

\subsection{The electric action}
The leading-order action in the exact small-$c$ expansion \eqref{eq:CS in c2} provides, in the widely accepted terminology, the \textit{electric action}:
\begin{equation}
	S_{\text{CCS}}^{\text{e}} \coloneqq - \int_{\mathscr C} \left[ \bfgamma \wedge \D \bfgamma - \bfalpha \wedge \D \bfalpha + 2 \bfkappa_a \wedge \left(\D \htheta^a + \bfgamma^a{}_b \wedge \htheta^b - \bfalpha \wedge \htheta^a \right) \right].
\end{equation}
This is the Chern--Simons action supported by the non-degenerate part $G_\mathfrak{h}$ of the invariant bilinear symmetric form on $\mathfrak{g}$. The cubic term in $\bfgamma^a{}_b$ is absent, $\mathfrak{so}(2)$ being Abelian and $\bfgamma^{ab} = \varepsilon_{ab}\bfgamma$ as before. Since
\begin{equation}
	G_\mathfrak{h}(\gen J_{ab},\gen J_{cd}) = 2(\delta_{ad}\delta_{bc} - \delta_{ac}\delta_{bd}),\quad G_\mathfrak{h}(\gen P_a,\gen K_b) = -4\delta_{ab}, \quad G_\mathfrak{h}(\gen D,\gen D) = 2
\end{equation}
from Eqs. \eqref{eq:carroll to poin gens} and \eqref{eq:G for carroll}, one can check that the electric action is in fact a Chern--Simons action for the projected Lorentz connection \eqref{eq:A}, which can be written as
\begin{equation}
	\boxed{S_{\text{CCS}}^{\text{e}} = \frac{1}{2}\int_\mathscr{C}\text{CS}[\bm\Gamma_\mathfrak{h}] = \frac{1}{2}\int_\mathscr{C}\text{Tr}_{G_\mathfrak{h}}\left(\bm\Gamma_\mathfrak{h}\wedge\D \bm\Gamma_\mathfrak{h} + \tfrac23 \bm\Gamma_\mathfrak{h}\wedge\bm\Gamma_\mathfrak{h}\wedge\bm\Gamma_\mathfrak{h}\right).} \label{eq:SeCCS}
\end{equation}
It is directly invariant under local conformal Carroll symmetries. Indeed, since it only involves the projected connection, it is trivially invariant by Carroll boosts and longitudinal special-conformal transformations. Next, it is rotation invariant as all indices are contracted. Finally, it is invariant under Weyl and transverse special-conformal transformations as one can show that the related variation of $S^\text{e}_{\text{CCS}}$ reduces to a boundary term. As in the special-relativistic case, there is a distinction between dynamical and background fields in order to extract the Cotton tensor from the action. In this case, natural dynamical fields are the connections $\bfgamma^a{}_b$, $\bfalpha$ and $\bfkappa^a$, while the background is the transverse basis covectors $\htheta^a$. Equations of motion are therefore
\begin{equation}
	\frac{\delta S^{\text e}_\text{CCS}}{\delta \bfgamma} \to \vect R[\gen J] = \bm 0,\quad \frac{\delta S^{\text e}_\text{CCS}}{\delta\bfalpha} \to \vect R[\gen D] = \bm 0,\quad \frac{\delta S^{\text e}_\text{CCS}}{\delta\bfkappa_a} \to \vect R[\gen P]^a = \bm 0,
\end{equation}
where $\vect R[\gen J] = \varepsilon_{ab}\vect R[\gen J]^{ab}$. On-shell, the electric Carroll--Chern--Simons action selects the Weyl--Carroll--Levi-Civita connection associated with $\bfalpha$. Performing the variation on-shell now yields
\begin{equation}
	\boxed{\delta S^{\text e}_\text{CCS} \approx \int_{\mathscr C}\delta\htheta^a\wedge\vect R[\gen K]_a.}
\end{equation}
Varying the electric Carroll--Chern--Simons action thus provides the expressions of the Carroll--Cotton vector $C^a$ and tensor $C^{ab}$, by virtue of Eq. \eqref{eq:cotton carroll in comp}. It is however blind to the longitudinal piece $C$, since it only involves the projected Lorentz connection. Making the coframe $\htheta^a$ dynamical and solving all the equations of motion imposes dipolar conformal flatness.\footnote{In a similar fashion than in the relativistic case, we note that the whole set of equations of motion deriving from the electric Carroll--Chern--Simons action do not impose any constraint on $\alpha_a$, in agreement with the fact that this field is pure-gauge under transverse special conformal transformations: see Eq. \eqref{eq:delta alpha a carroll}.} Therefore, equations of motion of the electric Carroll--Chern--Simons action are weaker than their special-relativistic ascendant, which justifies the presence of a second action that is worth to be studied.

\subsection{The magnetic action}
The subleading term in Eq. \eqref{eq:CS in c2} provides another possible action for Carrollian Chern--Simons dynamics. By analogy with usual Carrollian analyses, we refer to it as the magnetic action:
\begin{equation}
	\tilde S_{\text{CCS}}^{\text m} = \int_{\mathscr C}\left[\bfbeta_a\wedge \big(\bar{\bm\nabla}\wedge\bfbeta^a\big)+2\bfkappa^0\wedge\big(\D\bftau + \bfbeta_a\wedge\htheta^a -\bfalpha\wedge\bftau\big) - 2 \bfkappa^a\wedge\bfbeta_a\wedge\bftau \right], \label{eq:Stildem}
\end{equation}
identifying $\bfgamma^0{}_a = \bfbeta_a$. The geometric origin of this action is more subtle than the electric one, but can be traced back to the \.In\"on\"u--Wigner contraction of the conformal algebra into the Poincaré algebra. Prior to taking the $c\to 0$ limit, the Killing form of $\mathfrak{so}(3,2)$ naturally separates as $G(\mathfrak h,\mathfrak h) = G_\mathfrak{h}$, $G(\mathfrak h,\mathfrak t) = 0$ and $G(\mathfrak t,\mathfrak t) = c^2 G_\mathfrak{t}$, where
\begin{equation}
	G_\mathfrak{t}(\gen B_a,\gen B_b) = 2\delta_{ab},\quad G_\mathfrak{t}(\gen H,\gen K_0) = 4,
\end{equation}
all the other pairings are vanishing. Armed with this definition, we can massage Eq. \eqref{eq:Stildem} to get the particularly simple form\footnote{The trace with respect to the reduced bilinear form $G_\mathfrak{t}$ is well-defined because the quadratic term $[\bm\Gamma_\mathfrak{h},\bm\Gamma_\mathfrak{t}]$ is a two-form taking values in $\mathfrak t$ only, by virtue of the fact that $\mathfrak t$ is an ideal of $\mathfrak{g}$.}
\begin{equation}
	\boxed{
		\tilde S^\text{m}_{\text{CCS}} = \frac12 \int_\mathscr{C} \text{Tr}_{G_\mathfrak{t}}\left(\bm\Gamma_\mathfrak{t}\wedge\left(\D\bm\Gamma_\mathfrak{t}+\big[\bm\Gamma_\mathfrak{h},\bm\Gamma_\mathfrak{t}\big]\right)\right) - \oint_{\partial\mathscr C}\bftau\wedge\bfkappa^0 ,
	}\label{eq:StildemTr}
\end{equation}
with $\bm\Gamma_\mathfrak{t} \coloneqq \bm\Gamma - \bm\Gamma_\mathfrak{h} = \gen \bftau \gen H + \bfbeta^a\gen B_a + \tfrac12\bfkappa^0\gen K_0$ is the complementary piece of connection of the projected Lorentz connection, that we can refer to as vertical gauge connection. The magnetic action therefore couples the vertical gauge connection to its derivative by the Lorentz connection.

Contrary to the electric action, the magnetic action \eqref{eq:Stildem} is not Carroll-invariant, as it involves explicitly the Carroll-boost connection. It is neither invariant under longitudinal special-conformal transformations. More precisely, we find that
\begin{subequations}
	\begin{align}
		\delta_{\bm\uplambda}\tilde S^\text{m}_{\text{CCS}} &= -2\int_\mathscr{C} \lambda_a\big(\bfbeta_b\wedge\vect R[\gen J]^{ab} + \bftau\wedge\vect R[\gen K]^a + \bfkappa^0\wedge\vect R[\gen P]^a\big), \\
		\delta_{Z^0}\tilde S^\text{m}_{\text{CCS}} &= 2\int_\mathscr{C} Z^0 \big(\bfbeta_a\wedge\vect R[\gen P]^a + \bftau\wedge\vect R[\gen D]\big).
	\end{align}
\end{subequations}
This is a well-known phenomenon: in a small-$c$ expansion, only the leading term is committed to be Carroll-invariant; the subleading pieces are not automatically invariant, unless they are supplemented by constraints implying that, at least, the leading-order theory is on-shell \cite{Hansen:2020pqs,deBoer:2021jej,Henneaux:2021yzg,Rivera-Betancour:2022lkc}. We recover this feature in the present case. An invariant improvement of the magnetic action is therefore
\begin{equation}
	S^\text{m}_{\text{CCS}} =\tilde S^\text{m}_{\text{CCS}}  - \int_{\mathscr C}\left(2\htheta_a^{(1)}\wedge\vect R[\gen K]^a + 2\bfkappa_a^{(1)}\wedge\vect R[\gen P]^a + \bfgamma_{ab}^{(1)}\wedge\vect R[\gen J]^{ab} - 2\bfalpha_{(1)}\wedge\vect R[\gen D]\right). \label{eq:improved magn action}
\end{equation}
The Lagrange multipliers that introduce the electric equations of motion as constraints can easily be interpreted as an auxiliary Lorentz gauge connection
\begin{equation}
	\bm\Gamma^{(1)}_\mathfrak{h} \coloneqq \htheta^a_{(1)}\gen P_a + \tfrac12 \bfgamma^{ab}_{(1)}\gen J_{ab} - \bfalpha_{(1)}\gen D + \tfrac12 \bfkappa^a_{(1)}\gen K_a,
\end{equation}
such that, at any finite $c$, the Lorentz connection reads $\bm\Gamma_{\mathfrak h}^{(c\neq 0)} = \bm\Gamma_\mathfrak{h} + c^2 \bm\Gamma^{(1)}_\mathfrak{h}$.\footnote{In a holographic perspective, the boundary value of the speed of light is related to the cosmological constant. Hence, the coefficient $\bm\Gamma^{(1)}_\mathfrak{h}$ in this expansion is the first subleading correction to the Carroll gauge connection when one performs the asymptotically flat limit \cite{Campoleoni:2023fug}. The invariant improvement of the magnetic action is therefore compatible with the approach through limits \cite{Ahlouche:2026tba}. The extension of the aforementioned expansion to higher-order terms in $c^2$ might play a role in the emergence of Chthonian degrees of freedom in the asymptotically flat limit.} With these notations, the improved magnetic action can be compactly presented as
\begin{equation}
\boxed{
	S^\text{m}_\text{CCS} = \frac12 \int_\mathscr{C} \text{Tr}_{G_\mathfrak{t}}\left(\bm\Gamma_\mathfrak{t}\wedge\left(\D\bm\Gamma_\mathfrak{t}+\big[\bm\Gamma_\mathfrak{h},\bm\Gamma_\mathfrak{t}\big]\right)\right) - \int_\mathscr{C}\text{Tr}_{G_\mathfrak{h}}\big(\bm\Gamma_\mathfrak{h}^{(1)}\wedge \vect R[\bm\Gamma_\mathfrak{h}]\big),
} \label{eq:magnetic final}
\end{equation}
up to a neglected boundary term, see Eq. \eqref{eq:StildemTr}. Therefore, the transformation laws of the Lagrange multiplier need not be postulated: they are simply derived from Eqs. \eqref{eq:delta gauge relat}, keeping the $c^2$-terms in the transformation:
\begin{subequations}
\begin{align}
	\delta_\Sigma \htheta^a_{(1)} &= \lambda^a{}_b \htheta^b_{(1)} - B\htheta^a_{(1)} - (\lambda^a\bftau), \\
	\delta_\Sigma \bfgamma^{ab}_{(1)} &= - 2\bfgamma^{c[a}_{(1)}\lambda^{b]}{}_c + 2 Z^{[a}\htheta^{b]}_{(1)} + (2 \bfbeta^{[a}\lambda^{b]}), \\
	\delta_\Sigma \bfalpha_{(1)} &= Z_a\htheta^a_{(1)} -(Z^0\bftau),\\
	\delta_\Sigma \bfkappa^a_{(1)} &= \lambda^a{}_b \bfkappa^b_{(1)} + B \bfkappa^a_{(1)} + \bfgamma^{ab}_{(1)}Z_b + (Z^0\bfbeta^a +\bfalpha_{(1)}Z^a - \lambda^a\bfkappa^0).
\end{align}
\end{subequations}
We have bracketed the terms that supplement the expected terms from \eqref{eq:transfo gauge fields carroll} to highlight them: they are instrumental in proving that the improved magnetic action is now fully invariant under local conformal Carroll transformations: $\delta_\Sigma S^\text{m}_\text{CCS} = 0$.

The equations of motion derived from the magnetic Carroll--Chern--Simons action \eqref{eq:magnetic final} are
\begin{equation}
	\frac{\delta S^{\text m}_\text{CCS}}{\delta \bfbeta_a} \to \vect R[\gen B]^a = \bm 0,\quad \frac{\delta S^{\text m}_\text{CCS}}{\delta\bfkappa^0} \to \vect R[\gen H] = \bm 0,\quad \frac{\delta S^{\text m}_\text{CCS}}{\delta\bm\Gamma_\mathfrak{h}^{(1)}} \to \vect R[\bm\Gamma_\mathfrak{h}] = \bm 0.
\end{equation}
Again, we have considered that the coframe is a background structure and, accordingly, have not varied with respect to the Ehresmann connection $\bftau$. Should we have varied with the respect to $\bftau$, we would have obtained the vertical component of the Carroll--Cotton tensor, $\vect R[\gen K]^0$, providing a variational origin of the scalar $C$, or equivalently, the boundary NUT aspect. Importantly, the last set of equations of motion, $\vect R[\bm\Gamma_\mathfrak{h}] = \bm 0$, indicates that the magnetic theory exists, by construction, on dipolar conformally flat Carroll geometries, in the sense that they team up with a flat Lorentz connection on the base manifold. In gravitational terms, the magnetic Carroll--Chern--Simons theory is defined on non-radiative configurations. Evaluating the action with the assumption that the other equations of motion are obeyed, we find
\begin{equation}
	\boxed{
		S^\text{m}_\text{CCS} \approx \frac12 \int_\mathscr{C} C\,\varepsilon_{ab}\htheta^a\wedge\htheta^b\wedge\bftau = \int_\mathscr{C} \bfmu\, C.
	}
\end{equation}
While the electric action gives access to the Carroll--Cotton vector and tensor pieces upon variation with respect to the geometry, the magnetic action only encodes the last independent piece that is the scalar field $C$. Consistently, the
evolution equation \eqref{eq:Dbot C} degenerates to $\bar{\mathscr{D}}_\bot C = 0$, so that the on-shell magnetic action is conserved by the flow of $\bfupsilon$. Therefore, it represents the topological functional computing the gravitational magnetic charge of the underlying asymptotically flat solution.

\begin{description}
	\item[Remark.] From our analysis, it seems impossible to derive expressions of the Carroll--Cotton scalar $C$ on more general backgrounds than dipolar conformally flat configurations from a functional that is invariant under conformal Carroll symmetries. This is rooted to the fact that the conformal Carroll algebra does not admit a non-degenerate invariant bilinear symmetric form. Of course, such a derivation is always possible from non-invariant functionals, like the bare magnetic action \eqref{eq:Stildem} obtained through a polynomial expansion of \eqref{eq:CS} in $c^2$.
\end{description}

\subsection{A dual electric action}
As a side remark, we recall that the Lorentz algebra admits a second invariant bilinear symmetric form, that we denote $G^*_\mathfrak{h}$. Adopting the same normalisation convention as before, it reads as
\begin{equation}
  {G}^*_{\mathfrak{h}}(\mathbb{P}_a,\mathbb{K}_b) = -4\varepsilon_{ab},\qquad
  {G}^*_{\mathfrak{h}}(\mathbb{J}_{ab},\mathbb{D}) = 2\varepsilon_{ab},\qquad
  {G}^*_{\mathfrak{h}}(\mathbb{J}_{ab},\mathbb{J}_{cd}) = {G}^*_{\mathfrak{h}}(\mathbb{D},\mathbb{D}) = 0.
  \label{eq:Gexotic}
\end{equation}
We can then form the following contraction:
\begin{equation}
	\boxed{S_{\text{CCS}}^{\text{e},*} = \frac{1}{2}\int_\mathscr{C}\text{CS}^*[\bm\Gamma_\mathfrak{h}] = \frac{1}{2}\int_\mathscr{C}\text{Tr}_{G^*_\mathfrak{h}}\left(\bm\Gamma_\mathfrak{h}\wedge\D \bm\Gamma_\mathfrak{h} + \tfrac23 \bm\Gamma_\mathfrak{h}\wedge\bm\Gamma_\mathfrak{h}\wedge\bm\Gamma_\mathfrak{h}\right).}
\end{equation}
Expanding in components, we find
\begin{equation}
	S^{\text e,*}_{\text{CCS}} = -2\int_{\mathscr C}\big(\bfgamma\wedge\D\bfalpha + (\ast\bfkappa_a)\wedge\vect R[\gen P]^a\big) + \oint_{\partial\mathscr C} \big(\bfalpha\wedge\bfgamma + \varepsilon_{ab}\htheta^a\wedge\bfkappa^b\big). \label{eq:Sexotic}
\end{equation}
This action is also rightfully invariant under conformal Carroll symmetries. It corresponds to the electric action, Eq. \eqref{eq:SeCCS}, where pairings through $\delta_{ab}$ are traded for $\varepsilon_{ab}$. In particular, diagonal terms $\bfgamma\wedge\D\bfgamma$ and $\bfalpha\wedge\D\bfalpha$ are replaced by the crossed-term $\bfgamma\wedge\D\bfalpha$, and the Carroll--Cotton tensors $C^a$ and $C^{ab}$ by their transverse dual $\ast C^a$ and $\ast C^{ab}$ in front of the torsion-free constraint. We therefore propose to baptise the action \eqref{eq:Sexotic} the \textit{dual electric action}. Considering again all fields as dynamical except the coframe itself, the equations of motion are identical, though redistributed, to the ones deriving from the electric action:
\begin{equation}
	\frac{\delta S^{\text e,*}_\text{CCS}}{\delta \bfgamma} \to \vect R[\gen D] = \bm 0,\quad \frac{\delta S^{\text e,*}_\text{CCS}}{\delta\bfalpha} \to \vect R[\gen J] = \bm 0,\quad \frac{\delta S^{\text e,*}_\text{CCS}}{\delta\bfkappa_a} \to \vect R[\gen P]^a = \bm 0,
\end{equation}
and the response to variations with respect to the coframe yields both $\ast C^a$ and $\ast C^{ab}$. The role of the dual electric action in holographic applications is at the moment unclear to us but more than likely it might play a role in addressing issues in self-dual gravity and the twistor approach to it, see, \textit{e.g.}, \cite{Adamo:2013tja,Freidel:2021ytz,Kmec:2024nmu,Cresto:2024fhd,Kmec:2026dis}.

To conclude, let us stress that, group-theoretically, the three actions \eqref{eq:SeCCS}, \eqref{eq:magnetic final} and \eqref{eq:Sexotic} exhaust the range of possibilities to construct Carroll--Chern--Simons actions. Indeed, the expansion of the Killing form for $\mathfrak{so}(3,2)$ terminates at order $c^2$ and adding higher-order terms would amount to extend the conformal algebra, either in a trivial or non-trivial way. Second, the dual electric action admits no magnetic partner. Indeed, $\mathfrak{t}$ being an irreducible $\mathfrak{h}$-module, its invariant symmetric form is unique up to scale, by virtue of Schur's lemma. Moreover, demanding invariance of the Killing form with the replacement $G_\mathfrak{h}\to G^*_\mathfrak{h}$ and splitting again into Lorentz and translation components would force $G_\mathfrak{t}\to G_\mathfrak{t}^*$ with $G_\mathfrak{t}^*(\gen B_a,\gen B_b) \propto \varepsilon_{ab}$ for instance, which is incompatible with the fact that the $G_\mathfrak{t}^*$ must be symmetric. Therefore, it has to vanish identically, and we are left with only three Carroll instances of the Chern--Simons action.

\section{Summary \& outlook}

In this work we have given a first-principles, intrinsically Carrollian definition of Carroll--Cotton tensors and discussed their importance for the encoding of gravitational radiative modes at null infinity. To this end, we began by revisiting conformal geometry in general $d+1$ dimensions from the point of view of gauging methods, recovering the Cotton tensor as the curvature associated with local special conformal transformations, and, in three dimensions, relating it to the variation of the gravitational Chern--Simons action for $\mathfrak{so}(3,2)$ \cite{Horne:1988jf}. We then repeated the construction for the isotropic conformal Carroll algebra. This produced the Weyl--Carroll--Levi-Civita connection, torsionless and compatible with the Carrollian conformal class in accordance with \cite{Fiorucci:2025twa}, together with two features that have no Lorentzian counterpart: the vanishing of the intrinsic shear $\xi_{ab}$, which is a genuine restriction on the background geometry, and the persistence of degrees of freedom $\mathscr C_{ab}$ in the connection, which no compatibility requirement can remove. From this connection we constructed the Carroll--Schouten and Carroll--Cotton tensors and derived their evolution equations from the Bianchi identities. 

The holographic reading of these objects is straightforward. The degrees of freedom in the connection are identified with the Bondi shear, that is, the boundary value of gravitational radiative fields at null infinity. The transverse rank-two Carroll--Schouten tensor carries the news together with the remaining geometric data of the cut; and the components of the Carroll--Cotton tensors reproduce the radiative and Coulombic Newman--Penrose Weyl scalars, Eq. \eqref{eq:link with NP}, the scalar part controlling the NUT aspect. Building upon these findings, we have established three important results for flat-space holography. First, we discuss how the concept of conformal flatness is hierarchised in three-dimensional Carrollian manifolds due to the particular structure of the underlying symmetry group, and the physical interpretation of each of them at null infinity. The absence of outgoing radiation is represented by quadrupolar conformally flat geometry at null infinity, while the stronger condition of total conformal flatness singles out gravitational vacua. Second, we showed how our gauging methods can clarify the outcomes of a long-standing quest for a fully-covariant notion of news tensor. We show that the Geroch approach was nothing but the construction of a rightful tensor out of the difference of two transverse special-conformal connections. Finally, we showed how to define Carrollian avatars of the Chern--Simons actions from which the Carroll--Cotton tensors derive upon variations with respect to the background geometry.

Among the objects considered here, the Cotton tensor plays a distinguished role: it is the only one that is simultaneously local, intrinsic and covariant under the full set of boundary symmetries, namely general diffeomorphisms, local rotations, Carroll boosts, Weyl rescalings and special conformal transformations. The Carroll--Schouten tensor, hence the news tensor from which it derives immediately, is not: being the gauge field of special conformal transformations, its symmetric traceless part transforms inhomogeneously, and this inhomogeneity is precisely the anomalous transformation under conformal symmetry of the news tensor, written in Weyl-covariant variables. Having a covariant version of the news at our disposal is desirable, in particular when dealing with graviton scattering processes on general background geometries, since it provides the entry into the so-called Carrollian amplitudes \cite{Donnay:2022wvx,Mason:2023mti} that encode such processes in a holographic manner. We have given a fundamental geometric origin to the heuristic procedure of \cite{Compere:2018ylh} by proving that the physical news tensor, unlike the Carroll--Cotton tensor, cannot be entirely determined by the conformal structure, and must be supplied as additional data: a flat M\"obius structure \cite{Calderbank1998MOBIUSSA,Burstall2010ConformalSG}, which possesses a canonical meaning and is moreover unique if the topology of null infinity is $\mathbb R\times S^2$. In plain words, the news tensor is not a genuine Carrollian field but measures the deviation of the boundary geometry from a flat background of reference. The Cotton tensor evades this discussion entirely, since it is the curvature of that geometry and is therefore insensitive to the choice of origin.

This distinction matters a lot for Carrollian holography. In a relativistic bulk, the Newman--Penrose formalism is adapted to a choice of null frame, and the individual Weyl scalars are not invariant under arbitrary changes of frame. At null infinity one faces an analogous issue, compounded by the degeneracy of the boundary geometry and by the additional Carrollian gauge freedom. The Cotton tensor isolates the conformal information
in a robust way, whereas the Newman--Penrose-like decomposition should be understood as a convenient parameterisation, once the relevant geometric structure has been fixed. The classification obtained here, which separates fully covariant quantities from those requiring supplementary geometric input, should prove useful in any systematic treatment of the intrinsic geometry of null infinity.

Let us now conclude by offering some future perspectives. The results presented here call for a systematic treatment of holographic renormalisation directly at null infinity. It would be interesting to formulate the renormalised variational principle for asymptotically flat gravity in a way adapted to the Carrollian boundary geometry, and to determine which combinations of Carrollian data play the role of renormalised sources and responses. The relation between this construction and the boundary terms required for a well-defined variational principle should provide a bridge between the geometric analysis developed here and the holographic framework that we advocated in \cite{Fiorucci:2025twa} (see also \cite{Hartong:2025jpp,Hartong:2026rbr}). We have deliberately kept this aspect general, since a complete treatment requires a systematic analysis of counterterms, of the symplectic potential and of the admissible phase space at null infinity. The present results nevertheless indicate that the Cotton tensor should play a distinguished role, precisely because of its conformal covariance and its intrinsic character. Indeed, unlike the Fefferman--Graham gauge, Einstein gravity in AlAdS$_4$ written in a Bondi or Newman--Unti gauge forces some components of the boundary Lorentzian Cotton tensor to appear directly in the asymptotic expansion; these components mix with the holographic stress tensor and become the source of gravitational radiation in the flat, Carrollian limit \cite{Campoleoni:2023fug}. Whether the components of the Carroll--Cotton tensors defined here appear in the asymptotic expansion of four-dimensional asymptotically flat spacetimes, in a form similar to the one proposed in \cite{Fiorucci:2025twa,Hartong:2025jpp,Hartong:2026rbr}, is an interesting question that deserves further investigation.

Having constructed Carrollian instances of the Cotton tensor raises the possibility of importing into asymptotically flat holography structures familiar from self-dual solutions in AdS$_4$ gravity. In holographic descriptions of four-dimensional self-dual gravity, the bulk self-duality condition translates into a relation between the holographic stress tensor and the Cotton tensor of the boundary geometry, schematically $T_{AB}\propto C_{AB}$, with a factor of $\mathrm{i}$ in Lorentzian signature \cite{Petkou:2015fvh,Mittal:2022ywl}. This gives an appealing geometric characterisation of self-duality purely in terms of boundary data: the response encoded in the holographic stress tensor is constrained by the conformal geometry of the boundary itself. Whether an analogous statement survives in asymptotically flat spacetimes is an open question. The natural candidate is a relation between the appropriate Carrollian stress tensor at null infinity and the Carrollian Cotton tensor identified here, which would provide a boundary characterisation of asymptotically flat self-dual gravity and,
more broadly, a concrete way of embedding self-dual sectors into Carrollian holography \cite{Adamo:2013tja,Kmec:2024nmu}.

Finally, the analysis of Carrollian conformal flatness raises a more global question about the space of solutions. The supertranslated, superrotated and superboosted configurations provide a particularly interesting family of conformally flat Carrollian geometries, and suggest a direct relation between large Carrollian transformations and non-trivial representatives of the vacuum orbit \cite{Compere:2016jwb,Compere:2018ylh,Barnich:2021dta}. It is natural to ask whether these configurations exhaust the physically relevant solutions of the conformal-flatness conditions, or whether the equations admit genuinely distinct branches. Put differently, one would like to understand the full orbit structure of the conformally flat Carrollian vacua: is every configuration with vanishing Carroll--Cotton tensor related to the standard vacuum by a combination of Carrollian diffeomorphisms, Weyl rescalings and superboosts, and what global or regularity conditions are needed for such a statement to hold? Our analysis already suggests \textit{where} the answer must be topological. The integration of the conformal-flatness conditions is hierarchical: $C^{ab}=0$ integrates along the generators, $C^{a}=0$ is a constraint on a single cut which is then propagated by the Bianchi identities, and $C=0$ constrains the remaining $u$-independent shear. The second step is Geroch's equation, whose solution space is crucially dependent upon the topology of the cuts. On the sphere, the orbit is  therefore expected to be a single one, whereas in higher genus additional branches should appear, with possible consequences for the interpretation of memory effects, asymptotic charges and soft sectors in Carrollian holography.


\subsection*{Acknowledgements}
We would like to thank Daniel Grumiller, Anastasios Petkou, Konstantinos Siampos and Tom Wetstein for useful discussions, and in particular Isma\"el Ahlouche and Mathieu Beauvillain for discussions and collaboration on closely related topics. The work of AF is partially supported by the \textit{Fondation de l'\'Ecole polytechnique}, by the \textit{Fonds de la Recherche Scientifique} -- FNRS Belgium (under the convention IISN4.4503.15), as well as by research funds from the Solvay Family. The work of SP is supported by the \textit{European Research Council} (ERC) Project 101076737 -- CeleBH, and partially supported by \textit{INFN Iniziativa Specifica ST \& FI}. Views and opinions expressed are however those of the authors only and do not necessarily reflect those of the European Union or the European Research Council. Neither the European Union nor the granting authority can be held responsible for them. The work of MV is supported by the \textit{Fonds de la Recherche Scientifique} -- FNRS Belgium under the Grant No. T.0047.24.

\appendix

\section{Two-dimensional conformal manifolds}
\label{sec:2d}

In this Appendix, we give precisions on the gauging of the conformal algebra in two dimensions ($d=1$), both on special-relativistic and Carrollian backgrounds, which is of direct relevance for holographic dualities in three-dimensional gravity. 

\subsection*{Lorentzian case}

When $d=1$, the composite Riemann tensor has only one independent component,
\begin{equation}
	\vect R[\gen J]^{01} = R[\gen J]^{01}{}_{01}\bfmu,
\end{equation}
where $\bfmu = \htheta^0\wedge\htheta^1$ is the volume form. Solving Eq. \eqref{eq:RJdef} determines only the trace $\kappa^A{}_A$ of the special-conformal gauge field in terms of data coming from the spin connection $\bfgamma = \tfrac12\varepsilon_{AB}\gamma^{AB}$. Indeed, since the Riemann curvature of the spin connection reduces to $\vect R[\bfgamma] = \D\bfgamma = \tfrac12 R \bfmu$, we have
\begin{equation}
	\D\bfgamma - \kappa^A{}_A\bfmu = \bm 0\quad\Rightarrow\quad \kappa^A{}_A = \tfrac12 R, \label{eq:KAA}
\end{equation}
The Bianchi identities involve three forms and are thus trivial: in particular, Eq. \eqref{eq:RD vanishes auto} does no longer constrain the composite Weyl curvature, which can be required to vanish independently. In this particular case, \textit{i.e.}, $\kappa_{[AB]} = -\tfrac12 (\D\bfalpha)_{AB}$, one can show that there exists no set of equations of motion that is compatible with all the symmetries and do not put any constraint on the geometry that are able to determine the symmetric tracefree part $\kappa_{\langle AB\rangle}$ of the special-conformal gauge field, leaving two independent degrees of freedom. Defining
\begin{equation}
	T_{AB} \coloneqq \kappa_{(AB)} - \tfrac12 R \eta_{AB},
\end{equation}
and demanding that the gauge curvature $\vect R[\bm\Gamma]$ is flat, \textit{i.e.} $\vect R[\gen K]^A= \bm 0$ on top of the others, we find that $T^A{}_B$ is Weyl-covariantly conserved:
\begin{equation}
	\bar{\mathscr D}_{[A}\kappa_{B]C} = 0\quad\Rightarrow\quad \bar{\mathscr D}_A T^A{}_B = 0. \label{eq:DT}
\end{equation}
To reach this implication, we have used the two-dimensional identity $\bar{\mathscr D}_{[B}\kappa_{C]A} = \tfrac12\varepsilon_{BC} \varepsilon^{DE}\bar{\mathscr D}_D \kappa_{EA}$. Furthermore, $T^A{}_A = -\tfrac12 R$ from Eq. \eqref{eq:KAA}. In a holographic setting, $T_{AB}$ is then a candidate for the holographic energy--momentum tensor, whose conservation equation \eqref{eq:DT} emanates from bulk Einstein equations. Moreover, the non-vanishing trace of $T_{AB}$ is the signature of the Weyl anomaly in three-dimensional gravity \cite{Brown:1986nw,Coussaert:1995zp,Henningson:1998gx,Balasubramanian:1999re,deHaro:2000vlm}. Note that there always exists a special-conformal gauge choice in which the Schouten tensor is symmetric, the metric gauge for instance. There is therefore no Lorentz anomaly. 

Alternatively, one can trade the Weyl anomaly for a Lorentz anomaly \cite{Campoleoni:2022wmf} thanks to a particular choice of special-conformal frame. Indeed, using Eq. \eqref{eq:delta sigma gammaAB}, we can set $\bfgamma = \bm 0$. This is a particular feature of the two-dimensional case. Then, $\vect R[\gen P]^A = \bm 0$ imposes
\begin{equation}
	\D\htheta^A - \bfalpha\wedge\htheta^A =\bm 0\quad \Rightarrow \quad \bfalpha = \theta\htheta^0+\varphi_1\htheta^1,
\end{equation}
where $[\hframe_0,\hframe_1] = \varphi_1\hframe_0 - \theta\hframe_1$. Imposing the vanishing of the composite Weyl curvature yields
\begin{equation}
	\D\bfalpha = c^2\bfkappa^0\wedge\htheta^0 - \bfkappa^1\wedge\htheta^1\quad\Rightarrow\quad
	s = \kappa_{01} - \kappa_{10} = \hframe_0(\varphi_1) - \hframe_1(\theta).
\end{equation}
Since the special-conformal gauge choice has already been spoiled by the condition $\bfgamma=\bm 0$, the presence of a non-vanishing $s$ is the signal of a Lorentz anomaly. Indeed, the vanishing of the composite Riemann curvature implies
\begin{equation}
	\bfkappa^{[A}\wedge\htheta^{B]} = \bm 0\quad\Rightarrow\quad \kappa^A{}_A = 0.
\end{equation}
Therefore, $T_{AB} \equiv \kappa_{AB}$ in this case, which is also conserved if one requires that $\vect R[\gen K]^A = \bm 0$. For this second choice of frame, there is no Weyl anomaly since $T^A{}_A = 0$.

\subsection*{Carrollian case}

Let us now focus on the Carroll case. The only spacelike index is $a=1$, hence $\bfgamma^{ab} = \bm 0$. The only non-trivial parts of the gauge connection are the Carroll-boost connection,
\begin{equation}
	\bfbeta^1 = \beta^1\bftau + \beta\htheta^1,
\end{equation}
the Weyl connection $\bfalpha = \theta\bftau + \alpha_1\htheta^1$ and the special-conformal connection $\bfkappa^0$ and $\bfkappa^1$. Again, Bianchi identities \eqref{eq: Bianchi Carroll} are all trivial since there are no three-forms in two dimensions. Moreover, assuming no torsion, demanding that the composite Ricci tensor vanishes is equivalent to demanding that the only non-trivial independent component $\vect R[\gen B]^1(\bfupsilon,\hframe_1)$ of the Riemann tensor vanishes. As in the Lorentz-relativistic case, the vanishing of the composite Weyl curvature is no longer automatic and can be required on top of the aforementioned conditions: then, Eqs. \eqref{eq:RH carroll}, \eqref{eq:RBa carroll} and \eqref{eq:RD carroll} imply
\begin{equation}
	\beta_1 = \varphi - \alpha_1,\quad \D\bfbeta^1 = (\kappa^0{}_0+\kappa^1{}_1)\bftau\wedge\htheta^1,\quad \D\bfalpha + s\htheta^1\wedge\bftau=\bm 0,
\end{equation}
and nothing else. We write $s = \kappa_{10}$ by compatibility with the Lorentz case. Therefore, there exist only two scalar constraints on the gauge fields $\bfkappa^0$ and $\bfkappa^1$, which are in turn not fully determined: more specifically, their symmetric tracefree part is again left unconstrained, which amounts to two degrees of freedom. The interpretation of these degrees of freedom as the components of the holographic energy--momentum tensor in three-dimensional gravity is again possible, and one can posit
\begin{equation}
	\Pi = \kappa^0{}_0,\quad \Pi^1 = \kappa^1{}_0,\quad P_1 = \kappa^0{}_1,\quad \Pi^1{}_1 = \kappa^1{}_1,
\end{equation}
for the Carrollian momenta, in the notation of \cite{Fiorucci:2025twa}. One can show that $\vect R[\gen K]^0=\bm 0$ and $\vect R[\gen K]^1 = \bm 0$ reproduce the expected evolution equations \cite{Campoleoni:2018ltl,Ciambelli:2020ftk,Ciambelli:2020eba,Campoleoni:2022wmf} of the Carrollian momenta induced by bulk Einstein equations:
\begin{equation}
	\bar{\mathscr D}_\bot \Pi + \bar{\mathscr D}_1 \Pi^1 = 0,\quad \bar{\mathscr D}_\bot P_1 + \bar{\mathscr D}_1\Pi^1{}_1 = 0.
\end{equation}
To be explicit, there are various choices of special-conformal gauge frames that can be discussed. 

First, the transverse metric gauge assumes $\alpha_1 = 0$, and one can further set $\beta = 0$ to completely fix the special-conformal gauge frame. In this case, the vanishing of all gauge curvatures yields
\begin{equation}
	\Pi + \Pi^1{}_1 = \bfupsilon(\theta) -(\hframe_1+\varphi)\varphi,\quad \Pi^1 = s,
\end{equation}
which is the signal of both boost and Weyl anomalies. In fact, the former is related to the presence of degrees of freedom in the Carroll-boost connection, namely $\beta$, and a non-vanishing $\Pi^1$ can be recovered from the Noether identity related to Carroll boosts by introducing a hypermomentum as reaction to a variation with respect to $\beta$, see \cite{Fiorucci:2025twa}. However, the Weyl anomaly is not essential and can be traded for a Carroll-boost anomaly by changing the transverse special-conformal frame.

Second, in the rheotactic gauge $\bfbeta^1=\bm 0$, since $\beta$ is frozen, the presence of $s\neq 0$ signals a true Carroll-boost anomaly, which is the mechanism that has been investigated in \cite{Campoleoni:2022wmf}. In that case, we have
\begin{equation}
	\Pi + \Pi^1{}_1 = 0,\quad \Pi^1 = s,
\end{equation}
and the same equations of motion. The identification of the momenta is now straightforward: $\Pi$ and $P_1$ are the energy and angular momentum densities respectively.

\section{Invariant bilinear form on the 3-dimensional conformal algebra}
\label{sec:invariantbilinear}
In this Appendix, we review the computation of the invariant bilinear form over the conformal algebra in three dimensions, $\mathfrak{so}(3,2)$ \cite{Horne:1988jf}. First, we recall that this algebra is simple: it therefore admits one and only one invariant bilinear form only, the Killing form $G : \mathfrak{so}(3,2)\times\mathfrak{so}(3,2)\to\mathbb R$, up to normalisation. It is defined as
\begin{equation}
	G(\gen X_1,\gen X_2) \coloneqq \tfrac12 \text{tr}_{\text{ad}}(\text{ad}_{\gen X_1},\text{ad}_{\gen X_2}),
\end{equation}
where $\text{ad}$ is the adjoint representation of $\mathfrak{so}(3,2)$, \textit{i.e.}, $\text{ad}_\gen{X_1}\gen{X}_2 = [\gen X_1,\gen X_2]$. Second, from the invariance condition of $G$:
\begin{equation}
	G\big([\gen X_1,\gen X_2],\gen X_3\big) + G\big(\gen X_2,[\gen X_1,\gen X_3]\big) = 0,\qquad\forall\gen X_1,\gen X_2,\gen X_3\in\mathfrak{so}(3,2), \label{eq:invariance of G}
\end{equation}
one can show that only a few pairings through $G$ are non-vanishing. Indeed, assuming that elements $\gen X_1,\gen X_2$ of $\mathfrak{so}(3,2)$ have conformal weight $w_1,w_2$, \textit{i.e.}, $[\gen D,\gen X_1] = w_1\gen X_1$ and $[\gen D,\gen X_2] = w_2\gen X_2$, one has
\begin{equation}
	G\big([\gen D,\gen X_1],\gen X_2\big) + G\big(\gen X_1,[\gen D,\gen X_2]\big) = (w_1+w_2) G(\gen X_1,\gen X_2) = 0,
\end{equation}
which implies that $G(\gen X_1,\gen X_2) = 0$ if the conformal weights sum up to zero. Taking a careful look at Eq. \eqref{eq:conformal alg}, all pairings through $G$ are vanishing, except $G(\gen P_A,\gen K_B)$, $G(\gen J_{AB},\gen J_{CD})$, $G(\gen J_{AB},\gen D)$ and $G(\gen D,\gen D)$, up to symmetry. By Lorentz invariance, these pairings can only depend on Lorentz-invariant tensors, and therefore
\begin{equation}
	G(\gen J_{AB},\gen J_{CD}) = a_1 (\eta_{AC}\eta_{BD} - \eta_{AD}\eta_{BC}),\qquad G(\gen P_A,\gen K_B) = a_2 \eta_{AB},\qquad G(\gen D,\gen D) = a_3,
\end{equation}
where $a_1,a_2,a_3\in\mathbb R$. Furthermore, since there is no such invariant that has two skew-symmetric indices, $G(\gen J_{AB},\gen D)$ can only be vanishing. Using the invariance condition of $G$, Eq. \eqref{eq:invariance of G} allows to relate the unknown numbers as $a_2 = 2a_1$ and $a_3=-a_1$. This derivation is easier and more direct than the computation of the Killing form, but it does not provide a canonical normalisation for, say, $a_1$. To this end, one can compute the value of the Killing form on $\gen D$ only. Since $\gen D$ acts diagonally on each generator through the Lie bracket, it is represented by the diagonal matrix
\begin{equation}
	\text{ad}_{\gen D} = \begin{pmatrix}
		\delta^A{}_B & 0 & 0 & 0 \\
		0 & 0 & 0 & 0 \\
		0 & 0 & 0 & 0 \\
		0 & 0 & 0 & -\delta^A{}_B 
	\end{pmatrix}
\end{equation}
in the adjoint representation of $\mathfrak{so}(3,2)$ when reported to the basis $\{\gen P_A,\gen J_{AB},\gen K_A\}$. Therefore, 
\begin{equation}
	a_3 = G(\gen D,\gen D) = \tfrac12\text{Tr}(\text{ad}_\gen{D}^2) = d = 2\qquad\Rightarrow\qquad a_1 = -2,\quad a_2 = -4,\quad a_3 = 2.
\end{equation}
This canonical normalisation has been used to derive the Chern--Simons action in Eq. \eqref{eq: Chern-Simons conformal}.

\section{Useful identities involving the Weyl-covariant curvatures}
\label{sec:ids}

In this Appendix, we give some useful identities on the curvatures that arise from gauging the conformal algebra, either in the pseudo-Riemannian or Carrollian case, when a non-zero Weyl curvature is present. In particular, this allows us to show that the usual symmetries of the Riemann tensor can be recovered by a suitable projection onto the vector space of rank-four tensors that are skew-symmetric by pairs of indices in a canonical way.

To stress that the construction is not \textit{ad hoc} and is in fact fairly general, we shall start by paying attention back to the Lorentz-relativistic case. By definition, a spin connection $\bfgamma^{AB}$ is always skew-symmetric on $(A,B)$, hence compatible with a particular representative of the Weyl class of the metric. Therefore, the components $R_{ABCD}$ of the related Riemann tensor $\vect R[\bfgamma]^{AB}$ are always skew-symmetric by pairs, \textit{i.e.}, $R_{AB(CD)} = 0 = R_{(AB)CD}$, and the rank-four tensor built upon them therefore belongs to $\Omega^2(\mathscr M)\otimes\Omega^2(\mathscr M)$. Under the action of the linear group on the tangent bundle to $\mathscr M$, this space separates into three independent pieces as
\begin{equation}
	\Omega^2(\mathscr M)\otimes\Omega^2(\mathscr M) = \mathcal K(\mathscr M) \oplus \Omega^4(\mathscr M) \oplus \Omega^2(\Omega^2(\mathscr M)), \label{eq:decomp}
\end{equation}
where $\mathcal K(\mathscr M)$ is the vector space of \textit{algebraic curvature tensors} which obey the standard algebraic Bianchi identity $R_{A[BCD]}=0$, and therefore $R_{ABCD} = R_{CDAB}$, $\Omega^4(\mathscr M)$ is, as the notation implies, the set of full skew-symmetric rank-four tensors, and 
$\Omega^2(\Omega^2(\mathscr M))$ gathers the rank-four twice-skew-symmetric tensors that are also skew-symmetric under the exchange of pairs of indices, \textit{i.e.}, $R_{ABCD} = -R_{CDAB}$.\footnote{If $n=d+1$ denotes the spacetime dimension, each vector space has the following dimensions: $\dim \mathcal K(\mathscr M) = \frac{n^2(n^2-1)}{12}$, $\dim \Omega^4(\mathscr M) = C^4_n$ and $\dim \Omega^2(\Omega^2(\mathscr M)) = C^2_{n(n-1)/2}$. The sum is correctly given by $(C^2_n)^2$, in accordance with Eq. \eqref{eq:decomp}.} A necessary condition for a curvature tensor related to a spin connection to evade $\mathcal K(\mathscr M)$ is the presence of torsion $\vect T^A \coloneqq \tfrac12 T^A{}_{BC}\htheta^B\wedge\htheta^C$, since the algebraic Bianchi identity becomes
\begin{equation}
	R^A{}_{[BCD]} = \nabla_{[B}T^A{}_{CD]} + T^A{}_{E[B}T^E{}_{CD]}
\end{equation}
in this case. It is however always possible to construct an auxiliary curvature tensor $\tilde R^A{}_{BCD}$ that behaves like an algebraic curvature tensor, simply by projecting any element of $\Omega^2(\mathscr M)\otimes\Omega^2(\mathscr M)$ onto $\mathcal K(\mathscr M)$ as follows:
\begin{equation}
	\tilde R^A{}_{BCD}\coloneqq S^A{}_{BCD} - S{}^A{}_{[BCD]},\qquad S_{ABCD} \coloneqq \tfrac12\big(R_{ABCD}+R_{CDAB}\big).
\end{equation}
Indeed, the first term eliminates any component lying in $\Omega^2(\Omega^2(\mathscr M))$, while the second eliminates the fully skew-symmetric part, as it is not difficult to show that $S_{A[BCD]}\in\Omega^4(\mathscr M)$ by design. It is important to stress that the above algebraic construction does not guarantee, \textit{a priori}, that the auxiliary Riemann tensor can be interpreted as the curvature tensor for a certain connection.

This construction specialises to the Weyl--Levi-Civita connection. We recall that, regardless of the dimensions of the background manifold, this connection is defined by spin coefficients $\gamma^A{}_{BC}$ given by Eq. \eqref{eq:gamma for WLC} where $\alpha_A$ is, at the moment, arbitrary and responsible for a non-trivial torsion, $T^A{}_{BC} = 2\alpha_{[B}\delta^A{}_{C]}$. We can check that $R^A{}_{[BCD]} = \delta^A{}_{[B}\Psi_{CD]}$, where the right-hand side involves the components of the Weyl curvature, $\D\bfalpha\coloneqq \bm\Psi = \tfrac12\Psi_{AB}\htheta^A\wedge\htheta^B$. As a result, 
\begin{equation}
	S_{ABCD} = R_{ABCD} + \Psi_{A[C}\eta_{D]B} - \Psi_{B[C}\eta_{D]A},\qquad S_{A[BCD]} = 0.
\end{equation}
Then, the components of the auxiliary Riemann tensor are simply found as $\tilde R^A{}_{BCD} = S^A{}_{BCD}$. 

In the Carroll case, the curvatures of the Weyl--Carroll--Levi-Civita connection satisfy the Bianchi identities \eqref{eq:Bianchi carroll detail}, where the skew-symmetric part of the Carroll--Schouten tensors is governed by Eq. \eqref{eq:antisym schouten carroll}. Upon combining these identities using cyclic permutations yields the definition of Carroll--Riemann curvatures as
\begin{equation}
\begin{aligned}
	\tilde{\mathscr R}_{abcd} &\coloneqq \bar{\mathscr R}_{abcd} + \Omega_{a[c}\delta_{d]b} - \Omega_{b[c}\delta_{d]a},\qquad &\tilde{\mathscr R}^0{}_{abc} &\coloneqq \mathscr R^0{}_{abc}, \\
	\tilde{\mathscr R}_{abc} &\coloneqq \bar{\mathscr R}_{abc} - 2\bar{\mathscr R}_{[a}\delta_{b]c},\qquad &\tilde{\mathscr R}^0{}_{ab} &\coloneqq \mathscr R^0{}_{ab} + \tfrac12\Omega_{ab}.
\end{aligned} \label{eq:tilde curvature carroll}
\end{equation}
By construction, they obey the following Bianchi identities:
\begin{equation}
	\tilde{\mathscr R}^a{}_{[bcd]} = 0,\qquad \tilde{\mathscr R}^a{}_{[bc]} = 0,\qquad \tilde{\mathscr R}^0{}_{[abc]} = 0,\qquad \tilde{\mathscr R}^0{}_{[ab]} = 0.
\end{equation}
In particular, the vertical components of the Carroll-boost curvature are now symmetric, and the horizontal curvature tensor has now the same symmetries as the usual Riemann tensor, \textit{i.e.}, $\tilde{\mathscr R}_{abcd}\in\mathcal K(\mathscr S)$. Furthermore, owing to the geometric condition $\xi_{ab} = 0$, one can show that $\tilde{\mathscr R}_{abc} = 0$. This is consistent with the $c\to 0$ limit of Eq. \eqref{eq:Rabc weyl}.

\bibliographystyle{style}
\bibliography{csrefs}

\end{document}